\documentclass[twocolumn,showpacs,amsmath,amssymb,prd,floatfix,preprintnumbers]{revtex4}
\usepackage[dvipdfmx]{graphicx}
\usepackage{bm}
\usepackage{amsmath}
\usepackage{color}
\usepackage{xcolor}
\usepackage{hyperref}
\usepackage{comment}
\usepackage{multirow}
\usepackage[varg]{txfonts}

\hypersetup{
  bookmarksnumbered=true,
  linkcolor=red,
  citecolor=green,
  urlcolor=cyan
}
\newcommand{\bx}{{\bf x}}

\newcommand{\bs}{{\bf s}}
\newcommand{\bk}{{\bf k}}

\newcommand{\bp}{{\bf p}}
\newcommand{\bn}{{\bf n}}

\def\avrg#1{\left\langle #1 \right\rangle}

\newcommand{\simgt}{\lower.5ex\hbox{$\; \buildrel > \over \sim \;$}}
\newcommand{\simlt}{\lower.5ex\hbox{$\; \buildrel < \over \sim \;$}}

\begin{document}

\title[]{Accurate emulator for the redshift-space power spectrum of dark matter halos
and its application to galaxy power spectrum}

\author{Yosuke Kobayashi${}^{1,2}$}
\email{yosuke.kobayashi@ipmu.jp}
\author{Takahiro Nishimichi${}^{3,1}$}
\author{Masahiro Takada${}^{1}$}
\author{Ryuichi Takahashi${}^{4}$}
\author{Ken Osato${}^{5}$}
\affiliation{
${}^{1}$Kavli Institute for the Physics and Mathematics of the Universe
(WPI), The University of Tokyo Institutes for Advanced Study (UTIAS),
The University of Tokyo, Chiba 277-8583, Japan\\
${}^{2}$ Physics Department,
The University of Tokyo, Bunkyo, Tokyo 113-0031, Japan\\
${}^3$Center for Gravitational Physics, Yukawa Institute for Theoretical Physics, Kyoto University, Kyoto 606-8502, Japan\\
${}^4$Faculty of Science and Technology, Hirosaki University, 3 Bunkyo-cho, Hirosaki, Aomori 036-8561, Japan\\
${}^5$Institut d’Astrophysique de Paris, Sorbonne Universit\'{e}, CNRS, UMR 7095, 75014 Paris, France
}

% \date{\today}

\begin{abstract}
An accurate theoretical template of the redshift-space galaxy power spectrum, if applicable out to
nonlinear scales, enables us to extract more stringent and robust constraints on cosmological parameters
from the measured galaxy clustering. In this work, we develop a simulation-based template, so-called
\textit{emulator}, for the redshift-space power spectrum of dark matter halos.
Using the redshift-space halo power spectra measured from the \textsc{Dark Quest} $N$-body simulation suite that covers $101$ flat-geometry $w$-cold dark matter ($w$CDM) cosmologies around the {\it Planck} $\Lambda$CDM model,
we feed these data into a feed-forward neural network to build the fast and accurate emulation of the power spectrum from the linear to nonlinear scales up to $k \simeq 0.6 \, h \, {\rm Mpc}^{-1}$.
Our emulator achieves about 1\% and 5\% fractional accuracies in predicting
the monopole and quadrupole moments of the power spectrum, respectively, for halos of $\sim 10^{13}h^{-1}M_\odot$ that correspond to host halos of the Sloan Digital Sky Survey (SDSS) LOWZ- and CMASS (constant mass)-like galaxies, where the achieved accuracies are sufficient compared to the statistical errors of SDSS volume.
The validation and performance of the emulator are given by the comparison of the emulator predictions with the power spectra directly measured from the simulations for validation sets that are not used in the training. 
We demonstrate that the emulator outputs can be used to make model predictions for the redshift-space power spectrum of galaxies by employing user-fed models for the halo-galaxy connection, such as the halo occupation distribution. 
The emulator allows us to easily incorporate the Finger-of-God effect due to the virial motions of galaxies and the Alcock-Paczy\'{n}ski distortions.
Our code can compute the redshift-space galaxy power spectrum in a CPU subseconds and is ready to perform the emulator-based cosmological analysis for the exiting and upcoming galaxy redshift surveys.
\end{abstract}

\preprint{IPMU20-0046}
\preprint{YITP-20-41}

\maketitle

\section{Introduction}
\label{sec:intro}

The three-dimensional distribution of galaxies, measured from wide-area spectroscopic surveys of galaxies, is a powerful probe of cosmology, e.g.
for constraining cosmological parameters such as parameters characterizing the nature of dark energy and for testing gravity theory on cosmological scales.
To attain the fundamental cosmology, there are various exiting, ongoing and planned galaxy redshift surveys: 
the SDSS-III Baryon Oscillation Spectroscopic Survey [BOSS] \cite{2013AJ....145...10D}, 
the SDSS-IV extended Baryon Oscillation Spectroscopic Survey \cite{2016AJ....151...44D}, the Subaru Prime Focus Spectrograph  \cite{2014PASJ...66R...1T}, the Dark Energy Spectroscopic Instrument \cite{Aghamousa:2016zmz}, the ESA \textit{Euclid} satellite mission \cite{Laureijs:2011gra}, and the NASA Wide Field Infrared Survey Telescope \cite{Gehrels:2014spa}.

The galaxy distribution observed by spectroscopic surveys is modulated by
the Doppler effect due to the line-of-sight peculiar velocities of galaxies, and exhibits characteristic anisotropies,
called the redshift-space distortion (RSD) \cite{kaiser87,Hamilton92,Hamilton:1997zq}.
The RSD effect is useful to improve cosmological constraints by 
breaking degeneracies between the cosmological parameters and uncertainties in galaxy bias relative to the underlying matter distribution \citep{PhysRevD.101.023510}.
In addition, since the RSD effect is a gravitational effect, it 
can be used, if precisely measured, to probe the strength of gravitational field in large-scale structure, which can be in
turn used to test gravity theory on cosmological scales.

In order to exploit the full information from galaxy redshift surveys,
we need a sufficiently accurate theoretical template that enables a high-fidelity comparison with the measured clustering statistics of galaxies to obtain robust constraints on cosmological parameters.
The standard approach has been analytic prescriptions based on the perturbation theory of large-scale structure \cite{bernardeau02,Desjacques18}. 
This approach describes the distribution of galaxies in terms of a series expansion of both the matter density and velocity fields with a set of free coefficients/terms including bias parameters, under the single-stream approximation \citep{taruya10,nishimichi11}.
A further refined model enforcing the mass and momentum conservations, the so-called Effective Field Theory of Large-Scale Structure, has also been developed \citep{baumann12}.
These models have been applied to actual datasets to obtain cosmological constraints \citep{2014MNRAS.444..476R,10.1093/mnras/stu1051,Beutler:2016arn,2017MNRAS.470.2617A,Ivanov_2020,d_Amico_2020}.
While these perturbation theory-based templates give useful predictions at linear and quasi-nonlinear scales up to 
$k \sim 0.2 \, h \, {\rm Mpc}^{-1}$,
% \mtrv{$k \sim 0.1 \, h \, {\rm Mpc}^{-1}$,}
the application of these models to on even smaller scales is still disturbed by even higher-order contributions of both the density and velocity fields as well as nonperturbative effects arising from the dynamics beyond shell crossing, i.e., formation of galaxies (or dark matter halos) \citep[e.g.,][]{Pueblas_2009,blas14,bernardeau:2014lr,nishimichi2016,taruya2017,saga2018,halle2020}.
Consequently, the cosmological analysis on the galaxy power spectrum has been typically limited to the wave number 
$k \lesssim 0.15~h\,{\rm Mpc}^{-1}$
%\text{--}0.2\,h\,{\rm Mpc}^{-1}$ 
\cite{Beutler:2016arn,10.1093/mnras/stu1051}.
In other words, the clustering information on the higher-$k$ scales does not seem useful for cosmology in this method, because the information is used to basically constrain higher-order bias parameters and other nuisance parameters that need to be introduced for the theoretical consistency of models.

In this paper, we take an alternative approach to the galaxy clustering cosmology.
We develop a simulation-based theoretical template, called emulator, with the aim to obtain accurate model predictions for redshift-space galaxy power spectrum.
There have been several previous works on the emulator approach for the large-scale structure probes.
As a pioneering work on the emulator construction, the \textsc{Coyote Universe} \cite{Coyote1,Coyote2,Coyote3} employed the Gaussian process regression \cite{10.5555/1162254} on 1000 $N$-body simulations covering 38 $w$CDM cosmologies to construct an emulator for the nonlinear matter power spectrum in the redshift range
$0 \le z \le 1$, which can predict the matter power spectrum at $k \lesssim 1\,h\,{\rm Mpc}^{-1}$ to within about 1\% accuracy.
In the context of the galaxy clustering, the \textsc{Aemulus} Project \cite{DeRose_2019,McClintock_2019,Zhai_2019} constructed an emulator for the monopole and quadrupole moments of the redshift-space galaxy correlation function, as well as the halo mass function.
It used 47 $w$CDM cosmologies and a specific form of the halo occupation distribution (HOD) to produce the mock galaxy catalogs, and constructed an emulation of the galaxy correlation function, which has about 1\% accuracy in the redshift-space separations of $1 \lesssim s/(h^{-1}\,{\rm Mpc}) \lesssim 10$.

In this work we develop an emulator for the redshift-space power spectrum of halos, instead of galaxies.
Our basic philosophy is similar to that of \textsc{Dark Quest} \citep{Nishimichi_2019}; it is based on the fact that the redshift-space power spectrum of halos can be accurately modeled by using $N$-body simulations. 
Then, a model of the halo-galaxy connection, which a user adopts, can be combined with the emulator outputs to compute the redshift-space power spectrum of galaxies for a galaxy sample of interest.
Given uncertainties in physics of galaxy formation and evolution, a large number of nuisance parameters to model the halo-galaxy connection need to be introduced and then be marginalized over to obtain unbiased constraints on cosmological parameters (also see \cite{2017JCAP...10..009H,PhysRevD.101.023510} for the study based on a similar motivation).
Using an ensemble of the cosmological $N$-body simulations 
for 101 cosmological models in the six-dimensional parameter space of the flat $w$CDM cosmology around the best-fit model to the {\it Planck} CMB data \cite{planck-collaboration:2015fj}, we construct an emulator for the redshift-space halo power spectrum by utilizing a deep feed-forward neural network with a simple architecture.
Among the wide variety of machine learning techniques, the neural network is suitable to the multi-dimensional interpolation of the multi-output functions, such as the redshift-space power spectrum that depends on the wave number $k$ and the cosine angle between the wave vector and the line-of-sight direction, $\mu$, for two halo samples of masses $M_1$ and $M_2$ for a given cosmology.
The emulator of halo power spectrum, developed in this way, includes all the complicated effects on nonlinear scales: nonlinear clustering, nonlinear bias, nonlinear redshift-space distortion, the exclusion effect and so on.
We carefully assess the performance and validation of the emulator outputs by comparing with the redshift-space power spectra directly measured from the $N$-body simulations in validation sets that are not used in the training.
As demonstration, we combine the emulator outputs with the halo occupation distribution to compute the model predictions for the galaxy power spectrum for the SDSS LOWZ- and CMASS-like galaxies \citep{2013AJ....145...10D}.
Our emulator easily enables one to include the Finger-of-God (FoG) effects due to 
the virial motions of galaxies in the host halos and the Alcock-Paczy\'{n}ski (AP) distortion effect \citep{alcock79}. 
In the end, our emulator allows for a computation of the galaxy power spectrum in a CPU subseconds with a laptop computer.
This is a huge reduction in computational time compared to a brute-force approach, where high-resolution $N$-body simulations are run, galaxies are populated into halos, and then the galaxy power spectrum is measured from the mocks. 
This work is a preparation study for the emulator-based cosmology analysis of the SDSS galaxy data.

This paper is organized as follows:
in Sec.~\ref{sec:haloclustering}, we first give the rationale of why we focus on the redshift-space power spectrum of halos, and then define the redshift-space power spectrum and the multipole moments. 
In Sec.~\ref{sec:simulation}, we briefly review the \textsc{Dark Quest} simulation suite including a description of the halo catalogs we use to construct our emulator.
Section~\ref{sec:emulator} describes in detail the machine learning-based scheme to develop the emulator of the redshift-space halo power spectrum and show the main results, i.e., the performance of the emulator.
In Sec.~\ref{sec:discussion}, we demonstrate how to combine the emulator outputs with the halo occupation distribution to make model predictions for the redshift-space power spectrum of galaxies for the SDSS-like galaxies.
Finally, Sec.~\ref{sec:conclusion} gives summary and conclusion.

\section{Emulation Design}
\label{sec:haloclustering}

We first describe the overall design of our emulator.
In particular we give a rationale of why we want to develop an ``emulator'' of redshift-space power spectrum of ``halos'', and describe 
the relation of the halo power spectrum to the galaxy power spectrum in redshift space, which is a direct observable from galaxy surveys. 

\subsection{Cosmological information content of redshift-space galaxy power spectrum}
\label{subsec:cosmo_info_galaxy_power_spectrum}

Under the halo model picture \citep{seljak:2000uq,peacock:2000qy,ma:2000lr}, without loss of generality, the redshift-space power spectrum of galaxies is generally given by the sum of the one- and two-halo terms, 
\begin{align}
P_\mathrm{gg}^\mathrm{S}(\bk)&=P^\mathrm{S}_{\rm gg, 1h}(\bk)+P^\mathrm{S}_{\rm gg, 2h}(\bk)\nonumber\\
&= P^\mathrm{S}_{\rm gg, 1h}(\bk)\nonumber\\
&\hspace{1em}+\int\!\mathrm{d}M_1{\cal G}(\bk; M_1)\int\!\mathrm{d}M_2{\cal G}(\bk; M_2)
P^\mathrm{S}_\mathrm{hh}(\bk; M_1, M_2).
\label{eq:psgg_general}
\end{align}
Throughout this paper we often omit the redshift dependence ``$z$'' in a function for notational simplicity. 
The first term is the one-halo term arising from the contribution of correlations between galaxies inside the same halo, while the second term is the two-halo term arising from those between galaxies that reside in different halos.
Due to the redshift-space distortion, the redshift-space galaxy power spectrum is given as a function of the wave vector $\bk$; that is, it depends on the direction $\bk$ in addition to the length, $|\bk|$.
In the above equation $P_\mathrm{hh}^\mathrm{S}(\bk; M_1,M_2)$ is the redshift-space power spectrum for halos of masses $M_1$ and $M_2$. 
Other functions, $P^\mathrm{S}_{\rm gg, 1h}$ and ${\cal G}$, are needed to model the relation of halos to galaxies and therefore depend on galaxy physics --- often referred to as galaxy bias uncertainties.
The halo emulator approach in this study is motivated by the fact that the redshift-space power spectrum of halos can be accurately modeled using $N$-body simulations, as done in Ref.~\cite{Nishimichi_2019}.
On the other hand, since it is still quite challenging to model the formation and evolution of galaxies from the first principles, one has to employ an empirical prescription to describe characteristics of a target galaxy sample, by employing
a sufficient number of nuisance parameters to model the effects due to properties and physics of galaxies.
Then the nuisance parameters have to be marginalized over to 
obtain unbiased and robust constraints on cosmological parameters at the price of
conservative confidence intervals. 
In summary, we assume that $P_\mathrm{hh}^\mathrm{S}$ carries cosmological information, while the galaxy-related functions are treated as theoretical errors/uncertainties that lead to degradation of the cosmological parameter constraints, 
\begin{align}
\mbox{Halos (cosmology):}\ &\ P^\mathrm{S}_\mathrm{hh}(\bk; z, M_1,M_2, \bp_{\rm cosmo}) \nonumber \\
\mbox{Galaxies (errors and nuisance):} \ & \  \left\{P^\mathrm{S}_{\rm gg, 1h}(\bk), {\cal G}(\bk; M)\right\},
\end{align}
where ${\bf p}_{\rm cosmo}$ is a set of cosmological parameters. 

Hence we use an ensemble of high-resolution $N$-body simulations to develop an emulator that allows for fast and accurate computation of the redshift-space halo power spectrum, $P_\mathrm{hh}^\mathrm{S}(\bk; z, M_1,M_2, {\bf p}_{\rm cosmo})$, as a function of redshift, halo masses ($M_1$ and $M_2$), and cosmological models (${\bf p}_{\rm cosmo}$) for the $w$CDM cosmology.
Since we use $N$-body simulations, the halo power spectrum we emulate includes all complicated effects in the nonlinear regime: nonlinear clustering, nonlinear redshift-space distortion, nonlinear bias, the exclusion effect, and so on.
This is complementary to perturbation theory-based approaches.
On the other hand, galaxy-related functions ($P^\mathrm{S}_{\rm gg, 1h}$, ${\cal G}$) need to be provided by a user.
In this paper, as a working example, we use the HOD prescription to model the relation between halos and galaxies and further introduce the functions to model the spatial and velocity distributions of galaxies inside halos to model the RSD effect due to galaxies in the host halos. 

If the formation and evolution of galaxies arise from local physics, which is relevant for scales below some scale $\lambda \lesssim R_{\ast}$ or $k \gtrsim k_{\ast}$ in Fourier space with $k_{\ast} \sim 1/R_{\ast}$, the clustering properties of galaxies on larger scales are governed purely by gravitational interaction or properties of the primordial perturbations.
For example, the nonlinear scale to divide scales of galaxy physics would be around a virial scale of massive halos at most, i.e., $R_{\ast} \sim$ a few Mpc, 
in the standard CDM dominated structure formation scenario.
Under this consideration, in the limit of $k \ll k_{\ast}$, the galaxy power spectrum [Eq.~(\ref{eq:psgg_general})] can be expressed as
\begin{align}
P_\mathrm{gg}^\mathrm{S}(\bk)&\xrightarrow[k\ll k_{\ast}]{} 
\frac{1}{\bar{n}_{\rm g}^2}\int\!\mathrm{d}M_1\frac{\mathrm{d}n}{\mathrm{d}M} (M_1) \langle N_{\rm g}\rangle(M_1)\nonumber\\
& \hspace{4em}\times \int\!\mathrm{d}M_2\frac{\mathrm{d}n}{\mathrm{d}M} (M_2) \langle N_{\rm g}\rangle(M_2)P_\mathrm{hh}^\mathrm{S}(\bk; M_1,M_2),
\label{eq:pgg_asymptotic}
\end{align}
where 
$\langle N_{\rm g}\rangle(M)$ is the HOD that models the average number of galaxies in halos of mass $M$, 
$\mathrm{d}n/\mathrm{d}M$ is the halo mass function, 
and $\bar{n}_{\rm g}$ is the 
mean number density of galaxies, defined as
\begin{align}
\bar{n}_{\rm g}&=\int\!\!\mathrm{d}M\frac{\mathrm{d}n}{\mathrm{d}M} (M) \langle N_{\rm g}\rangle(M). 
\end{align}
In Eq.~(\ref{eq:pgg_asymptotic}), we assumed that, due to the spatial locality of galaxy physics and/or gas physics, the galaxy-related functions have asymptotic behaviors of $P^\mathrm{S}_{\rm gg, 1h}\rightarrow 0$ and ${\cal G}\rightarrow k^0$ at the limit of $k\ll k_{\ast}$, respectively. 
Alternatively, in the quasi-nonlinear regime $k\lesssim k_{\ast}$, the galaxy-related functions would be well behaved in the sense that the functions can be approximately expanded by a series of polynomials of $k$, e.g. $P^\mathrm{S}_{\rm gg, 1h}\sim \sum_n c_n k^n$ \cite{2014MNRAS.445.3382M}. 
Thus, as explicitly described by Eq.~(\ref{eq:pgg_asymptotic}), the galaxy power spectrum at large scales ($k\ll k_{\ast}$) is directly related to the halo power spectrum, with constant coefficients.
In particular the anisotropy ($\bk$ dependence) in the redshift-space galaxy power spectrum arises from the redshift-space halo power spectrum. 
In our approach, we accurately model a part of the galaxy power spectrum that arises from gravitational effects in the large-scale structure, via the distribution of halos. 
Then the question arises; up to which nonlinear scale ($k$) can we use the redshift-space galaxy power spectrum to extract the cosmological information in a robust manner?
Can we use the galaxy power spectrum up to $k\simeq 0.2$ or even 0.3$~h\,{\rm Mpc}^{-1}$, without any significant bias in derived cosmological parameters?
This is not fully understood yet, and we have to carefully address this question (also see \cite{2020arXiv200308277N} for the study based on a similar motivation using the perturbation theory).

\subsection{Power spectrum in redshift space}

The redshift-space power spectrum of halos can be defined as a two-point correlation of number density field of halos in redshift 
space:
\begin{align}
    \avrg{\delta_{\rm h}^\mathrm{S}(\bk)\delta_{\rm h}^\mathrm{S}(\bk')} = (2\pi)^3 \delta_{\rm D}(\bk+\bk') P_\mathrm{hh}^\mathrm{S}(\bk),
\end{align}
where $\delta_\mathrm{D}(\bk)$ is the Dirac delta function.
Throughout this paper we employ the plane-parallel approximation or the distant observer approximation;
that is, we assume that the line-of-sight direction is parallel to one axis of the Cartesian coordinate, for which we take the $x_3$-axis direction, 
$\hat{\bn}\parallel x_3$. 
The power spectrum is expressed as a function of the wave number $k$ and the cosine angle between the wave vector and the line-of-sight direction, $\mu\equiv \hat{\bn}\cdot\hat{\bk}$. 
%\tnrv{[TN: if we write the line-of-sight direction by $\hat{\bn}$, the statement here (i.e., $P$ is a function of $k$ and $\mu$) is always true even without the plane-parallel approximation. So this is weird to put here.]} 
The power spectrum is symmetric with respect to the wave vector lying in the two-dimensional plane perpendicular to the line-of-sight, denoted as $\bk_\perp$. Hence the redshift-space power spectrum is given as a function of two variables that specify the wave vector $\bk$; $(k,\mu)$ or $(k_\perp, k_\parallel)$, where $k_\perp = k\sqrt{1-\mu^2}$ and $k_\parallel=k\mu$.

The redshift-space power spectrum given as a function of $(k,\mu)$ contains the full information in the halo distribution at the level of two-point statistics. 
However, the dimension of data vector can become readily huge, and the analysis would be computationally expensive; e.g., a calibration of the covariance matrix requires a large number of simulations that should be much larger than the dimension of data vector.
For this reason, a dimensional reduction of data vector is useful. 
The redshift-space power spectrum can be, without loss of generality, expressed as
\begin{align}
    P^\mathrm{S}(k,\mu) = \sum_{\ell=0}^\infty P^\mathrm{S}_\ell(k) \, {\cal L}_\ell(\mu),
\end{align}
where ${\cal L}_\ell(\mu)$ is the $\ell$th order Legendre polynomials. 
Here $P^\mathrm{S}_\ell(k)$ is the $\ell$th order multipole ``moments'' of the power spectrum, which is given as a function of $|\bk|$. 
Using the orthogonality of the Legendre polynomials, the multipole moments of the power spectrum can be estimated from the observed redshift-space power spectrum as
\begin{align}
    P^\mathrm{S}_\ell(k) \equiv \frac{2\ell+1}{2} \int_{-1}^1 \, {\rm d}\mu \, P^\mathrm{S}(k,\mu) \, {\cal L}_\ell(\mu).
\end{align}
Note that the odd-$\ell$ moments vanish due to the statistical isotropy.

At the level of Kaiser's linear theory \cite{kaiser87}, the redshift-space power spectrum has the nonvanishing moments of $\ell=0$ (monopole), $\ell=2$ (quadrupole) and $\ell=4$ (hexadecapole), and the higher-order moments vanish.

\section{Simulations}
\label{sec:simulation}

In this section we describe \textsc{Dark Quest}, a suite of cosmological $N$-body simulations that we use to develop the emulator. 
Detailed descriptions on this simulation suite can be found in Ref.~\cite{Nishimichi_2019}. 
Here we briefly describe the main properties of these simulations.

\subsection{$N$-body simulations}

All the simulations in \textsc{Dark Quest} \citep{Nishimichi_2019} were executed by using the Tree-Particle Mesh hybrid code \textsc{Gadget2} \cite{gadget2}. 
We generate the initial conditions of each simulation assuming the adiabatic Gaussian initial conditions based on the linear matter power spectrum for each cosmology.
\textsc{Dark Quest} consists of the simulations with two different particle resolutions: high-resolution (HR) and low-resolution (LR) runs.
The side lengths of the simulation boxes in the HR and LR runs are 1 and $2 \, h^{-1} \, \mathrm{Gpc}$, respectively, and both adopt $N_{\rm p} = 2048^3$ particles.
Thus the particle mass for the fiducial {\it Planck} cosmology is $m_{\rm p} = 1.02 \text{ and } 8.16 \times 10^{10}\,h^{-1}\,M_\odot$ for HR and LR runs, respectively.
In this work we utilize only the LR simulations to create the training and validation datasets to keep the sufficient statistics, since we need only the positions, velocities, and masses of halos to measure the redshift-space power spectrum of halos (for different mass thresholds) and we do not use $N$-body particles.
For LR simulations, the initial conditions are generated at redshift around $30$, using the second-order Lagrangian perturbation theory (2LPT; \cite{scoccimarro98,crocce06a}) based on the implementation by Refs.~\cite{nishimichi09,Valageas11a}.
Note that we slightly varied the initial redshifts depending on the input cosmology according to the criterion in Ref.~\cite{Nishimichi_2019}.
We also use the HR simulations to assess the effects of Fourier resolution on the power spectrum measurements in Appendix~\ref{sec:resolution_study}.

Throughout this paper we employ the flat-geometry $w$CDM cosmology framework that is characterized by the six cosmological parameters as follows.
The set of cosmological parameters for which the simulations are run is defined using the optimal maximin-distance sliced Latin hypercube design \cite{SLHD}, which enables an efficient sampling from a high-dimensional parameter space with a hierarchical structure among the samples.
Our purpose is to construct an emulator from the simulations each of which requires very high computational cost, and thus such an efficient simulation design is of great importance.
Following this scheme, we produce five disjoint subgroups of cosmological parameters (referred to as ``slice'' in the following), each of which satisfies a homogeneous sampling from the parameter space.
The cosmological parameters are sampled from the following ranges (also see Fig.~2 in \cite{Nishimichi_2019}):
\begin{align}
\label{eq:cosmo_param_range}
	&0.0211375 < \omega_{\rm b} < 0.0233625, \nonumber \\
	&0.10782 < \omega_{\rm c} < 0.13178, \nonumber \\
	&0.54752 < \Omega_{\rm de} < 0.82128, \nonumber \\
	&2.4752 < \ln (10^{10}A_{\rm s}) < 3.7128, \nonumber \\
	&0.916275 < n_\mathrm{s} < 1.012725, \nonumber \\
	&-1.2 < w < -0.8,
\end{align}
where $\omega_{\rm b} \equiv \Omega_{\rm b}h^2$ and $\omega_{\rm c} \equiv \Omega_{\rm c}h^2$ are the physical density parameters of baryon and cold dark matter, respectively; $h$ is the dimensionless Hubble constant defined as $h \equiv H_0 / (100\,{\rm km\,s^{-1}\,Mpc^{-1}})$; 
$\Omega_{\rm de}$ is the dark energy density parameter, $A_{\rm s}$ and $n_\mathrm{s}$ are the amplitude and spectral tilt of the power spectrum of primordial curvature perturbations, defined at the pivot scale, $k_\mathrm{p}=0.05\, \mathrm{Mpc}^{-1}$; $w$ denotes the equation-of-state parameter of dark energy.
The parameter range above Eq.~(\ref{eq:cosmo_param_range}) is defined to be centered at the best-fit $\Lambda$CDM model for the {\it Planck} 2015 data \cite{planck-collaboration:2015fj}, i.e., $\omega_{\rm b} = 0.02225, \omega_{\rm c} = 0.1198, \Omega_{\rm de} = 0.6844, \ln (10^{10} A_\mathrm{s}) = 3.094, n_\mathrm{s} = 0.9645, \text{ and } w = -1$.
For the neutrino abundance, we assume $\omega_\nu \equiv \Omega_\nu h^2 = 0.00064$ to include the effect on the initial linear power spectrum alone, and neglect the dynamical effect of massive neutrinos in the $N$-body simulations.
The Hubble constant is computed from the total energy budget condition assuming flatness, i.e., $\Omega_{\rm m} h^2 = \omega_\mathrm{b} + \omega_\mathrm{c} + \omega_\nu$ and $\Omega_{\rm m} + \Omega_\mathrm{de} = 1$.
Aside from the {\it Planck} cosmology, each of the five slices has 20 sets of cosmological parameters, and we have one realization for each of 100 cosmological models. 
Hence we sample $100+1$ models in total for our emulator construction. 
We note that the ranges of cosmological parameters are broad enough, e.g., to cover the current constraints of the large-scale structure probes such as those from the Subaru Hyper Suprime-Cam \citep{2019PASJ...71...43H}.

We stored outputs of each $N$-body simulation realization at 21 redshifts, given by $z =$ 1.48, 1.35, 1.23, 1.12, 1.02, 0.932, 0.846, 0.765, 0.689, 0.617, 0.549, 0.484, 0.422, 0.363, 0.306, 0.251, 0.198, 0.147, 0.0967, 0.0478, and 0.
These redshifts are evenly stepped by the linear growth factor for the fiducial {\it Planck} cosmology.

In addition we ran 15 random realizations for the fiducial {\it Planck} cosmology each of which has a volume of $8~(h^{-1}\,{\rm Gpc})^3$, which is larger than the volume of SDSS BOSS survey that has about $5.7~(h^{-1}\,{\rm Gpc})^3$. 
We will often use these simulations to estimate statistical errors such as the errors expected for the power spectrum measurements.

\subsection{Halo catalogs}

We construct halo catalogs from each simulation output based on the following procedure.
First, we identify halos using the friends-of-friends halo finder in six-dimensional phase space, \textsc{Rockstar}, developed in Ref.~\cite{Behroozi:2013}.
We define the center of each halo as the center-of-mass of the ``core particles,'' a subset of member particles in the inner part of that halo, which is considered as a proxy of the mass density maximum or the location of central galaxy if forms.
%\tnrv{position of the galaxy associated with the halo}.}
Similarly, the velocity of each halo is given as the center-of-mass velocity of the core particles.

In this paper we employ the halo mass definition given by
$M_{\rm 200} = (4\pi/3) 200 \bar{\rho}_{\rm m0} R_{\rm 200}^3$, where $R_{\rm 200}$ is the spherical halo boundary radius within which the interior mass is equal to 200 times the mean mass density $\bar{\rho}_{\rm m0}$. 
Note that the use of the mean mass density today $\bar{\rho}_{\rm m0}$ is due to our use of the comoving coordinates, meaning that $R_{\rm 200}$ is also in comoving length units.
Our definition of halo mass includes all the $N$-body particles within the boundary $R_{\rm 200}$ around the halo center, including those not gravitationally bound by the halo.
After identifying halo candidates, we determine whether they are central or satellite halos. When the separation between the centers of different halos is closer than $R_{\rm 200}$ of any other halo, we mark the most massive halo as a central halo, and the other halo(s) as a satellite subhalo(s).
We kept only the central halos with mass $M_{\rm 200} > 10^{12} \, h^{-1} \, M_\odot$.

\section{Construction of the emulator for the redshift-space halo power spectrum}
\label{sec:emulator}

In this section, we describe details of the emulator development.
The goal is to develop an emulator which allows for fast, accurate computation of the redshift-space power spectrum of halos for an input cosmological model within $w$CDM framework, given as a function of wave vector $(k,\mu)$, redshift ($z$), and halo masses ($M_1$ and $M_2$).

\subsection{Problem setting and emulation scheme}
\label{sec:overall_design}

Before going to details of the emulator development, we here describe several important aspects of our problem and discuss the machine learning scheme to meet the requirements from the problem setting.

The problem that we are dealing with falls in the category of regression, i.e., to find a reasonable function that reacts to the input parameters smoothly and predicts the outcome for new sets of inputs.
In particular, we construct a dataset of the redshift-space power spectrum corresponding to different cosmologies, redshifts and number densities.
Our goal is to implement the regression of the power spectrum dataset in the input parameter space.
As we will describe later, the number of combinations of the input parameter values is quite large (more than $10^5$), and for each of these inputs, a quite high-dimensional data vector of the redshift-space power spectrum, each component of which corresponds to each $(k,\mu)$ bin, needs to be output.
Such a regression of (i) a \textit{multi-output} (or equivalently, vector-valued) function in (ii) a \textit{multi-dimensional} input parameter space on (iii) a \textit{large dataset} is a highly nontrivial task, and furthermore we need to realize it in (iv) a small computational time.
To meet these requirements, we need to pay a special attention to select an efficient machine learning algorithm.

Traditionally, the Gaussian process (GP) regression \cite{10.5555/1162254} has been applied to the emulator constructions.
This is because of some advantages of the GP; it enables a nonparametric regression, i.e., we need not assume any specific function shape to fit the dataset, it works well even in a relatively high-dimensional input space, it is robust against the overfitting, and it can provide predictions in a probabilistic manner which accounts for the errors in the dataset (though we do not necessarily regard the last point as important in cosmology applications, since the physical interpretation of the predicted variance is somewhat unclear).
However, the GP has a drawback that it is difficult to apply to large datasets (specifically, the data size of order of $10^4$ or more) due to its high computational cost, unless we introduce some sparse approximation.
In addition, current typical GP applications mainly focus on the problems with a single scalar output, and the multi-output GP schemes are not straightforward to apply.
Therefore, in previous works on the emulator constructions using GPs, multiple single-output GPs were independently built so that each of them corresponds to each component of the multi-dimensional output data vector.
When the dimensionality of the output data vector is quite high (as in our problem), this can lead to a large computational time due to the call of each GP in the resultant emulator, as well as a quite large data size of the emulator code set.
Thus, in the application of GPs to the emulator construction, it is almost inevitable to employ some scheme to reduce the dimensionality of the output data, e.g., the principal component analysis.
However, to find a successful scheme of the dimensionality reduction for a given target quantity, i.e., the scheme by which we can precisely reconstruct the high-dimensional data vector from the reduced data encoded in a low-dimensional space, would be also a nontrivial task.
It is highly desirable, if possible, to build a learning pipeline that we can easily apply to almost any target quantities.

In this work, we choose a feed-forward neural network as a hopeful candidate for such a scheme of high versatility.
The neural network provides a smooth interpolation of the dataset in the multi-dimensional input space, and a relatively easy scheme for the multi-output regression without the necessity of dimensionality reduction, and it can be applied to a large dataset.
This approach was previously applied to cosmology \cite[e.g.,][]{2012MNRAS.424.1409A,2014MNRAS.439.2102A,Jennings_2018}, and its performance was shown to be competitive or sometimes better than other existing methods, when the network architecture was appropriately designed.
In addition, as we would extend the data size by additional simulation runs in the future, this scalability for larger datasets would be advantageous. 
This is opposed to the $\mathcal{O}(N^3)$ scaling of the popular Gaussian process regression with $N$ training samples.
On the other hand, the neural network has a drawback that it is more susceptible to the overfitting compared to GPs.
Hence we need to carefully check the emulation performance using the validation dataset and tune the network architecture so that the generalization error is successfully suppressed.

\subsection{Dataset}
\label{subsec:dataset}

Our aim is to develop a machine-learning pipeline that optimally finds the correspondence between the input parameters (cosmology, redshift, and two halo masses $M_1$ and $M_2$) and the output power spectrum $P^\mathrm{S}_\mathrm{hh}(k,\mu)$ from the training dataset.
Following the procedure in Ref.~\cite{Nishimichi_2019}, we set two distinct number density bins $n_1$ and $n_2$, instead of mass bins, to make the learning process easier.
This choice is intended to have similar error levels over the data vector among different cosmologies.
If we had specified a halo sample by its mass instead, the shot noise level could be quite different among different cosmologies especially in high-mass bins, reflecting the strong dependence of the halo mass function (i.e., the exponential damping as a function of the mass variance) on some of the cosmological parameters.
In Appendix~\ref{sec:cross_power_spectrum} we see that the two distinct number density bins, $n_1$ and $n_2$, are necessary to cover the full mass dependence of the halo power spectrum, and hence irreducible to a single set of bins.

For each number density bin $n$, we pick up the halos from top of the ranked list in which we sort all the halos in descending order of mass, to reach the given number density.
Such number density bins correspond to the halo mass threshold (minimum mass in the halo sample) derived from the halo mass function,
\begin{align}
\label{eq:hmf}
  n_{\rm h}(M>M_{\rm min}) = \int_{M_{\rm min}}^\infty\!{\rm d}M~ \frac{{\rm d}n}{{\rm d}M} (M),
\end{align}
where $M_{\rm min}$ denotes the halo mass threshold and ${\rm d}n / {\rm d}M$ is the halo mass function in the range $[M,M+\mathrm{d}M]$ which depends on the cosmology and redshift.
In the prediction stage, the emulator outputs the predictions (redshift-space power spectrum) as a function of mass by taking the numerical derivative at the target masses $M_1, M_2$,
\begin{align}
\label{eq:mass_deriv}
  P^\mathrm{S}_\mathrm{hh}(\bk; M_1,M_2) &= 
\frac{\frac{\partial^2}{\partial M \partial M'}\left.\left[n_\mathrm{h}(M) n_\mathrm{h}(M') P^\mathrm{S}_\mathrm{hh}(\bk;n(M),n(M'))\right]\right|_{\substack{M=M_1,\\M'=M_2}}}{ \frac{{\rm d}n}{{\rm d}M}(M_1) \frac{{\rm d}n}{{\rm d}M}(M_2)},
\end{align}
where $n(M)$ is the halo number density corresponding to the mass threshold $M$.

We employ nine logarithmic bins for the number density ranging from $10^{-7}$ to $10^{-3} \, (h^{-1} \, {\rm Mpc})^{-3}$, and add one \textit{mass} threshold bin $M_{\rm min} = 10^{12} \, h^{-1} \, M_\odot$ for each cosmology and redshift.
Since we adopt the (logarithm of) number density as the actual input argument to the neural network, the mass threshold $M_{\rm min} = 10^{12} \, h^{-1} \, M_\odot$ for each cosmology and redshift is converted to the number density through Eq.~(\ref{eq:hmf}), i.e., we have a slightly heterogeneous sampling of the number density over different cosmologies and redshifts.

We measure the halo power spectra from the halo catalogs by using the fast Fourier transform (FFT)-based method.
Each halo catalog contains the positions, velocities, and masses of halos.
We construct the number density field of halos in redshift space by shifting the halo positions along the $x_3$ axis according to their velocities and assigning each halo to the FFT grid using the cloud-in-cell (CIC) \cite{hockney81} interpolation kernel.
We then mitigate the aliasing contaminations in Fourier space by using the interlacing scheme in Ref.~\cite{Sefusatti:2015aex}.
We adopt $1024^3$ grids on the $2 \, h^{-1} \, {\rm Gpc}$ cubic box, which corresponds to the Nyquist wave number $k_{\rm Ny}=1.61 \, h \, {\rm Mpc}^{-1}$.
We use the $k$-bin width $\Delta k = 0.02\,h\,{\rm Mpc}^{-1}$ and define the $k$ value for each bin as its central value.
In Appendix~\ref{sec:resolution_study} we show a resolution study on this measurement procedure, where we find that our specific choice of the number of grids and the $k$-bin width has almost no significant impact on the measured power spectrum.

In the measurement, we assume the Poisson shot noise and subtract it from the measured power spectrum, but note that this procedure requires a slightly careful treatment as follows.
Suppose we measure the power spectrum between the halo samples of two different number densities, $n_1$ and $n_2$ ($n_1 < n_2$), defined by the procedure above (we call these two samples ``sample 1'' and ``sample 2,'' respectively).
By construction, sample 1 is a subsample of the sample 2, and hence the cross power spectrum between these two samples is decomposed into the auto and cross power spectra, respectively, for the overlapping and the exclusive subsamples:
\begin{align}
  P_{1,2}(\bk) = f P_{1,1}(\bk) + (1-f) P_{1,2\backslash1}(\bk),
\end{align}
where $f = n_1 / n_2$, $P_{1,2}(\bk)$ is the cross power spectrum between samples 1 and 2, 
$P_{1,1}(\bk)$ is the auto power spectrum of the sample 1, and $P_{1,2\backslash1}(\bk)$ is the cross power spectrum between the sample 1 and the subsample of the sample 2 which has no overlap with the sample 1. 
The Poisson noise that should be subtracted from the auto power spectrum $P_{1,1}(\bk)$ is simply $1/n_1$, and this is equivalent to the subtraction of $f / n_1 = 1 / n_2$ from $P_{1,2}(\bk)$. 
On the other hand, the second term in the right-hand side does not have a contribution from shot noise by construction.

\subsection{Data preprocessing}
\label{subsec:data_preprocessing}

\begin{figure*}[htpb]
\centering
  \begin{tabular}{c}
    \begin{minipage}{0.5\hsize}
    \centering
    \includegraphics[width=0.99\textwidth]{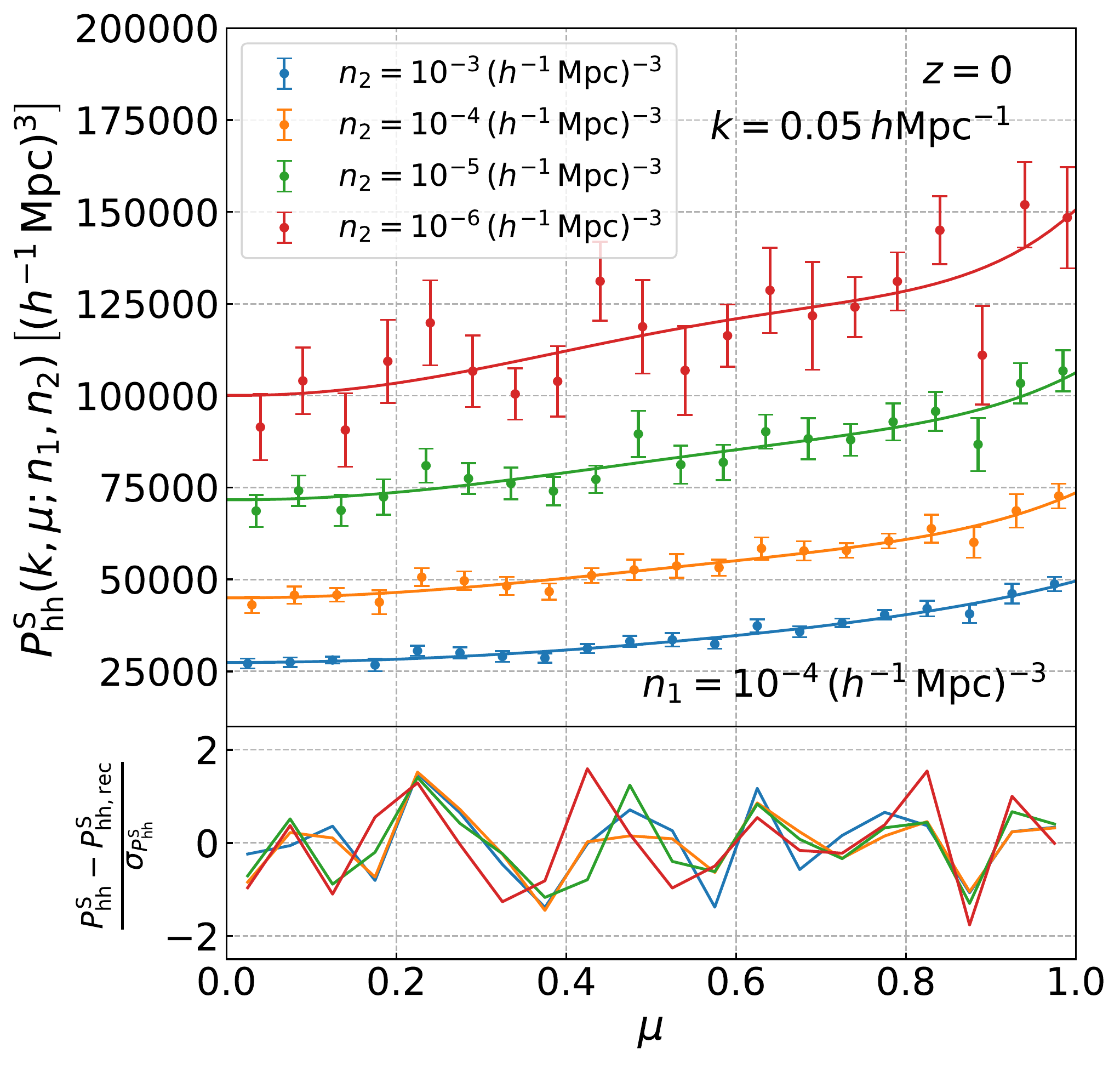}
    \end{minipage}
    
    \begin{minipage}{0.5\hsize}
    \centering
    \includegraphics[width=0.99\textwidth]{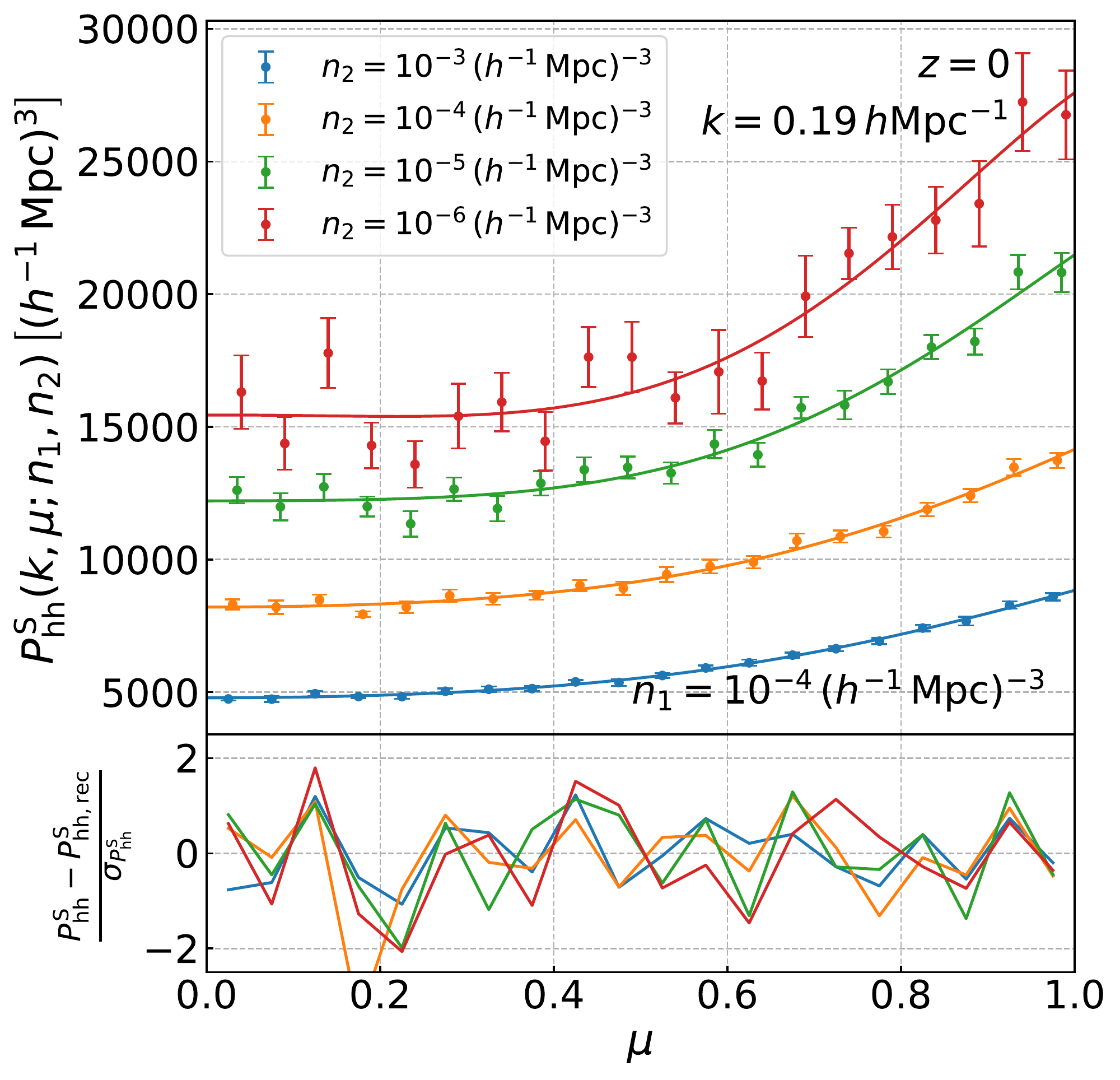}
    \end{minipage} \\

    \begin{minipage}{0.5\hsize}
    \centering
    \includegraphics[width=0.99\textwidth]{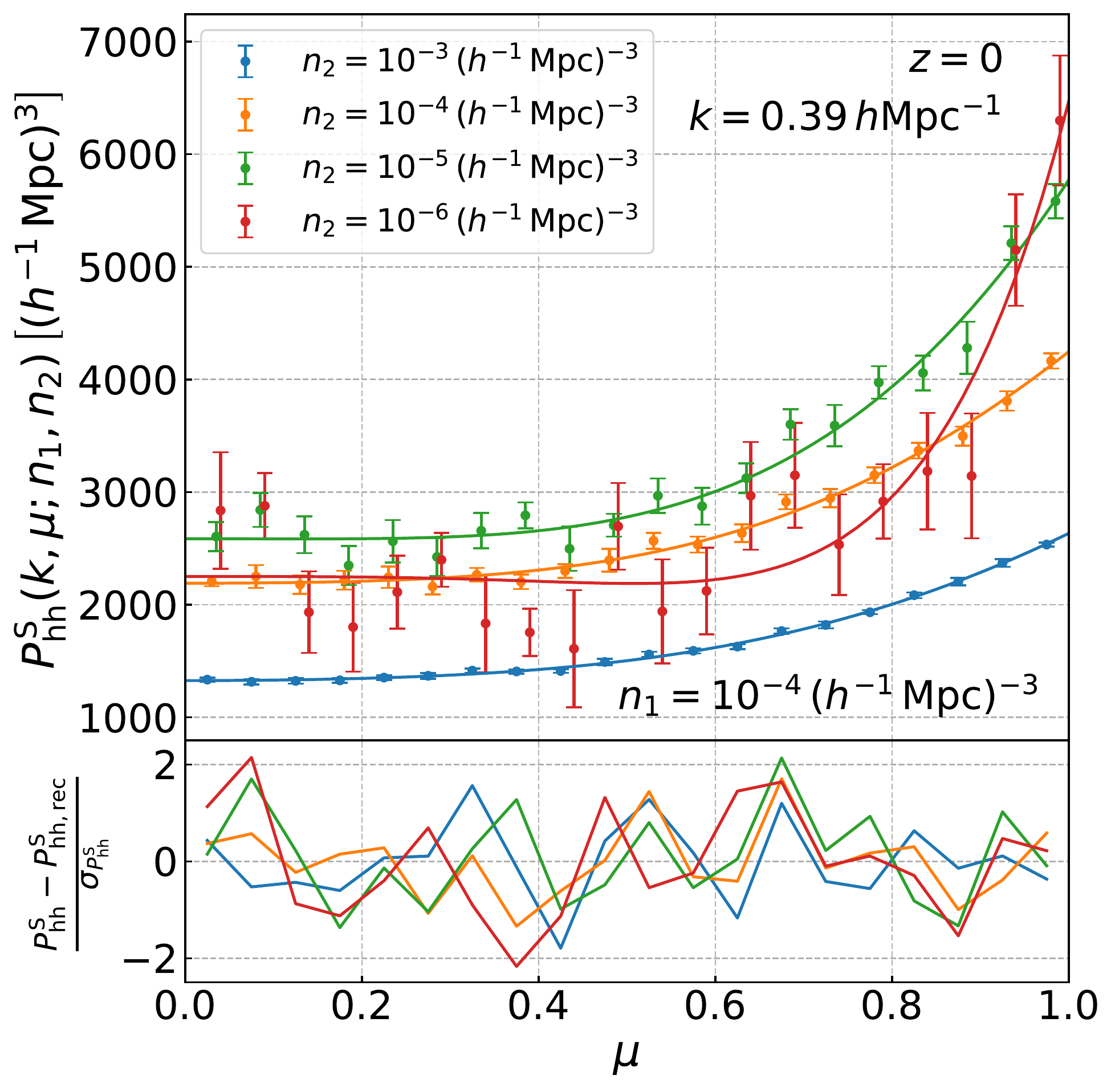}
    \end{minipage}

    \begin{minipage}{0.5\hsize}
    \centering
    \includegraphics[width=0.99\textwidth]{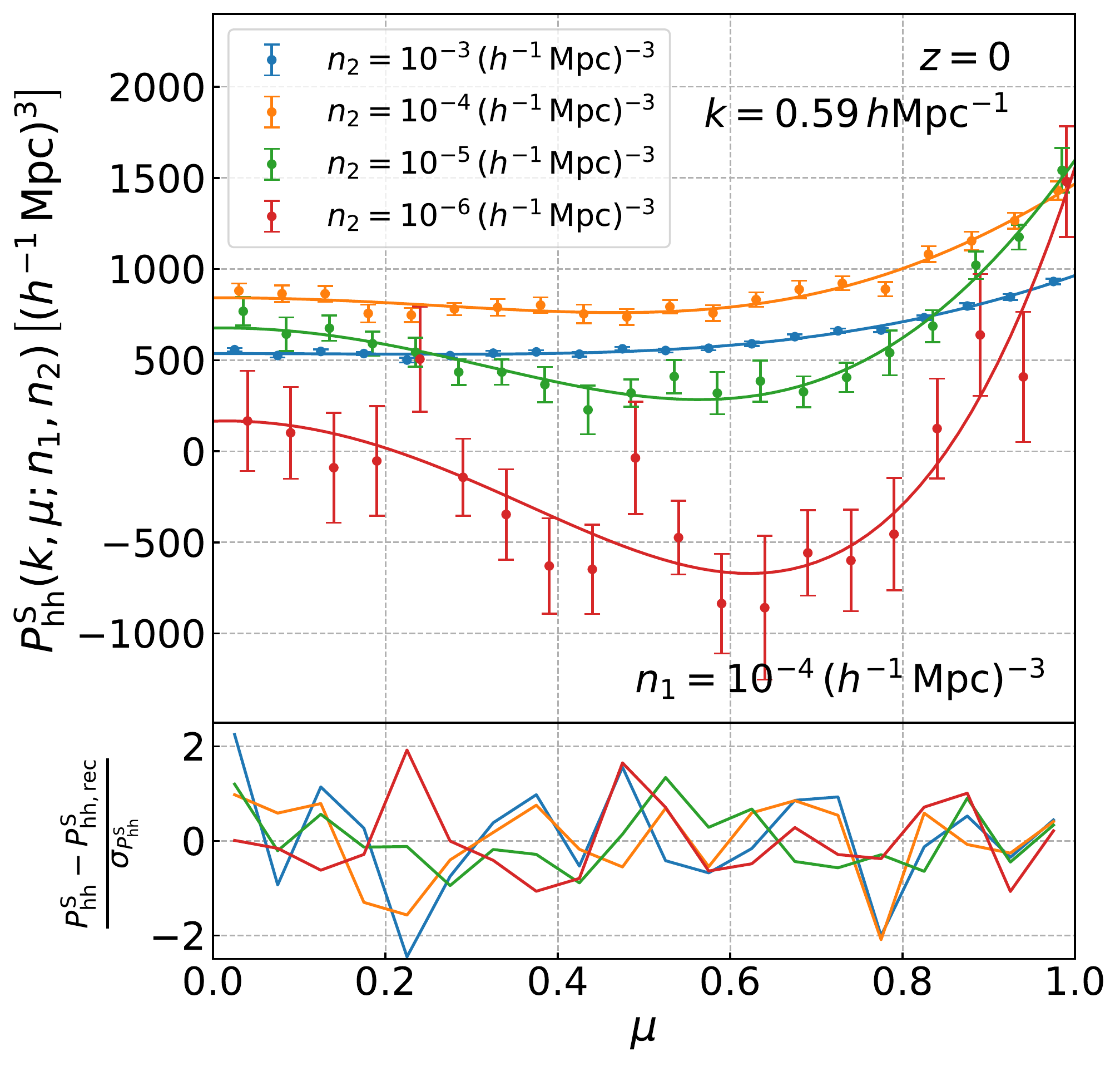}
    \end{minipage}
  \end{tabular}
\caption{
An assessment of the approximation, Eq.~(\ref{eq:pkmu_rec}), to model the redshift-space power spectrum $P^{\rm S}_{\rm hh}(k,\mu)$ in terms of the lowest four multipole moments.
In the upper panel in each plot, the symbols with error bars are the cross power spectrum between the halo samples of number densities ($n_1$ and $n_2$), $P^\mathrm{S}_\mathrm{hh}(k,\mu; n_1, n_2)$, measured from one simulation realization at $z=0$ for the {\it Planck} cosmology, where we consider the fixed number density for one sample, $n_1=10^{-4} \, (h^{-1}\,{\rm Mpc})^{-3}$ and consider the other sample of different number densities, $n_2 = 10^{-3}$ (blue), $10^{-4}$ (orange), $10^{-5}$ (green), or $10^{-6}$ (red) $(h^{-1} \, {\rm Mpc})^{-3}$, respectively.
The error bars are the standard deviation among 15 realizations for the {\it Planck} cosmology, corresponding to the statistical errors in the band power measurement for a volume of $8~(h^{-1}\,{\rm Gpc})^3$.
The solid lines are the results of Eq.~(\ref{eq:pkmu_rec}), i.e., the power spectrum reconstructed from the multipole moments up to degree $\ell = 6$ that are measured from the same halo samples in the same realization.
We show the results at $k = 0.05, 0.19, 0.39, \text{ and } 0.59 \, h \, {\rm Mpc}^{-1}$, from upper left to lower right plots.
We also show, in each lower panel, the differences between the reconstructed spectra and the simulation results in $(k,\mu)$ bins, relative to the scatter in the bin.
}
\label{fig:pkmu_reconstruct}
\end{figure*}

The power spectrum signals measured in $(k,\mu)$ bins from each simulation
are noisy due to the small number of modes averaged within each $(k,\mu)$ bin.
This inaccuracy is particularly problematic for low number density samples.
To overcome this obstacle, we use the lowest four multipole moments to approximate the two-dimensional power spectrum as 
\begin{align}
\label{eq:pkmu_rec}
  P^\mathrm{S}_\mathrm{hh}(k,\mu) \simeq \sum_{\ell = 0,2,4,6} P^\mathrm{S}_{\mathrm{hh},\ell}(k) 
  \, {\cal L}_\ell(\mu),
\end{align}
Here we ignore contributions from the higher-order multipoles of $\ell \ge 8$.
Since the hexadecapole ($\ell = 4$) and tetra-hexadecapole ($\ell = 6$) 
moments are highly noisy and have almost zero amplitudes at low $k$ for most cosmological models, we avoid to directly learn the individual multipoles, and instead choose to feed the reconstructed two-dimensional power spectrum, based on 
Eq.~(\ref{eq:pkmu_rec}), into the neural network.
Using Eq.~(\ref{eq:pkmu_rec}), we reconstruct the approximated power spectrum in $31$ linearly spaced bins of $k$ in the range $[0.01, 0.61]\,h\,{\rm Mpc}^{-1}$ and $20$ linearly spaced bins of $\mu$ in the range $[0.025, 0.975]$.
We use $620$ $(k,\mu)$ bins in total.

A validation of the approximation [Eq.~(\ref{eq:pkmu_rec})] is given by Fig.~\ref{fig:pkmu_reconstruct}.
The symbols show the power spectrum $P_\mathrm{hh}^\mathrm{S}(k,\mu)$ as a function of $\mu$ in some representative bins of $k$, directly measured
from a particular realization of the fiducial {\it Planck} cosmology. 
On the other hand, the solid lines are the results obtained using Eq.~(\ref{eq:pkmu_rec}), where we used the multipole moments of $\ell=0, 2, 4$ and 6 measured from the same realization. 
The solid lines show a good agreement with the direct measurements and do not show any systematic deviation for any value of $(k,\mu)$, confirming that 
the higher-order multipoles do not give a significant contribution to the two-dimensional power spectrum.

In addition to the approximation [Eq.~(\ref{eq:pkmu_rec})], we employ the following linear transformation to reduce the dynamic range of data vector.
For every sampling point of ($k,\mu$), we transform the data vector so that
the mean and variance of data over all the inputs (cosmology, redshift and two distinct number densities) are reduced to zero and unity, respectively:
\begin{align}
    P^\mathrm{S}_\mathrm{hh}(k,\mu) \mapsto \frac{P^\mathrm{S}_\mathrm{hh}(k,\mu) - \bar{P}^\mathrm{S}_\mathrm{hh}(k,\mu)}{\sqrt{{\rm Var}\left[P^\mathrm{S}_\mathrm{hh}(k,\mu) \right]}},
    \label{eq:ps_redefinition}
\end{align}
where $\bar{P}^\mathrm{S}_\mathrm{hh}(k,\mu)$ and ${\rm Var}\left[P^\mathrm{S}_\mathrm{hh}(k,\mu) \right]$ are the mean and variance 
among all the power spectra in each $(k,\mu)$ bin over all the training and validation datasets.
We feed these transformed data into the neural network.

\subsection{Regression using a neural network}
\label{subsec:training}

The machine learning using neural networks has been rapidly developed on the back of the recent progress of machine power and 
the success of the back-propagation method, in addition to the vast increase of available data.
Its applicability to a broad range of learning tasks has already been recognized in the community of cosmology as well as astrophysics.
For regression tasks, feed-forward neural networks perform accurately serving as a
``universal function approximator'' \cite{Cybenko1989}, i.e., it can approximate almost any continuous functions $f(\bx)$ with high precision, provided that it has a sufficiently large number of parameters.
As we mentioned above, another strength of neural networks is their relatively easy handle on multi-output functions.

\begin{figure}
\centering
\includegraphics[width=0.48\textwidth]{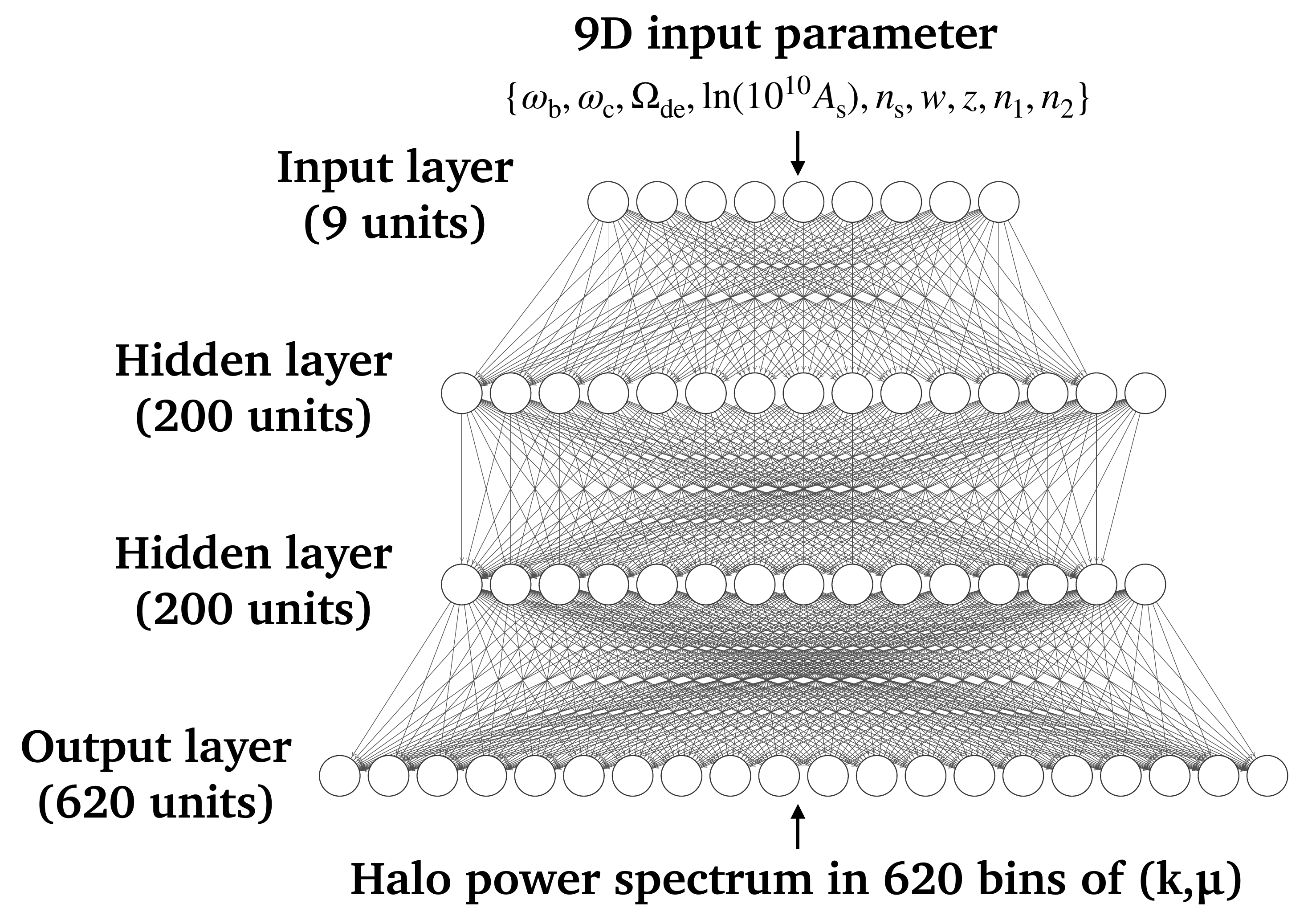}
\caption{
The architecture of the feed-forward neural network which we adopt for the regression of the input power spectrum data (training dataset).
The input layer has nine units corresponding to the nine-dimensional input parameters [Eq.~(\ref{eq:param_in})].
We adopt two hidden layers that contain 200 units to give a large flexibility to the mapping from the input to output vectors.
Finally the output layer has 620 units, which is equal to the number of the $(k,\mu)$ bins we use.
}
\label{fig:neural_net}
\end{figure}

In this work we found that a feed-forward neural network with a simple architecture enables us to perform a multi-dimensional regression of the power spectrum data measured from the simulations.
Figure~\ref{fig:neural_net} shows the network architecture we adopt for the regression.
We adopt the fully-connected network with two hidden layers;
the input layer takes the nine input parameters, i.e., six $w$CDM cosmological parameters, redshift $z$ and two distinct number densities,
$n_1, n_2$:
\begin{align}
\label{eq:param_in}
{\bf p}_{\rm in} = \{\omega_{\rm b},\omega_{\rm c},\Omega_{\rm de}, \ln(10^{10}A_{\rm s}), n_\mathrm{s}, w, z, n_1, n_2 \}.
\end{align}
On the other hand, the network output layer corresponds to the vector of $P^S_\mathrm{hh}(k,\mu)$ values, i.e., the output dimension is equal to the number of $(k,\mu)$ bins, $N_{\rm bin} = 620$.
We set two hidden layers which have a large number (200 for each) of hidden units.
In Appendix~\ref{sec:architecture}, we describe how we chose the optimal number of hidden units.
As the activation function, we impose the Gaussian Error Linear Units (as known as GELUs) \cite{hendrycks2016gaussian}, which is a smooth variant of the Rectified Linear Units (ReLUs) that are typically used in various machine learning tasks, to both two hidden layers.
This is because we expect that the response (i.e., the derivative) of the power spectrum to any of the input variables is smooth without discontinuity from which the standard ReLU function often suffers.
Through these activation functions, the input parameter vectors are nonlinearly transformed to represent the target quantities, i.e., the data of the halo power spectrum.

We implement the neural network and the training procedure by using \textsc{PyTorch} (\url{https://pytorch.org/}) \cite{paszke2019pytorch},
which is an open-source Python framework for the deep learning.
The training of the network is done by using the adaptive stochastic optimization algorithm \textsc{Adam} \cite{kingma2014adam}.
The training performance is highly sensitive to the learning rate for the \textsc{Adam} optimizer.
We set the learning rate to $10^{-3}$, which we found is the best choice among $10^{-4}, 5\times10^{-4}, 10^{-3}, 5\times10^{-3}, \text{ and } 10^{-2}$.

We train the neural network to {\it learn} the correspondence between the input and output variables,
\begin{align}
    {\bf p}_{\rm in} \mapsto P^\mathrm{S}_\mathrm{hh}(k,\mu | {\bf p}_{\rm in}),
\end{align}
where ${\bf p}_{\rm in}$ is a set of nine parameters [Eq.~(\ref{eq:param_in})], and $P^\mathrm{S}_\mathrm{hh}(k,\mu | {\bf p}_{\rm in})$ is the redshift-space halo power spectrum, based on the transformation of Eq.~(\ref{eq:ps_redefinition}).
The training dataset consists of the power spectrum data for combinations of
80 cosmologies, 21 redshifts, and 10 number density bins for each of $n_1$ and $n_2$; hence, the size of the training data amounts to 168,000 instances.
Through the training procedure, we optimize the neural network so that the network output $P^\mathrm{S}_{\rm emu}(k,\mu)$ precisely approximates the simulation data $P^\mathrm{S}_{\rm sim}(k,\mu)$.
In the optimization, we obtain the network parameters by minimizing the loss function that we define as
\begin{align}
\label{eq:loss_func}
  \tilde{\chi}_m^2 \equiv \frac{1}{m} \sum_{i=1}^m \frac{1}{N_{\rm bin}} \sum_{(k,\mu)}^{N_{\rm bin}} \left[ \frac{P^\mathrm{S}_{{\rm sim}}(k,\mu | {\bf p}_i) - P^\mathrm{S}_{\rm emu}(k,\mu | {\bf p}_i)}{\sigma_{P^\mathrm{S}_{{\rm fid}}}(k,\mu | {\bf p}_i)} \right]^2,
\end{align}
where $N_{\rm bin}$ is the number of $(k,\mu)$ bins ($N_{\rm bin}=620$ in our case), $m$ is the number of training data in one ``mini-batch'' 
(see below), and $\sigma_{P^\mathrm{S}_{{\rm fid}}}(k,\mu | {\bf p}_i)$ is the error of the power spectrum in the $(k,\mu)$ bin for the 
$i$th training dataset.
The mini-batch is a subset of the training dataset that is used to train the network parameters.
The training of neural network is done by feeding data into the network in the form of mini-batch, and repeatedly updating the network parameters according to the derivative of the loss function back-propagated to each unit until the loss function is sufficiently minimized.
The use of mini-batches is beneficial because it is not only memory efficient, but also leads to an improved optimization performance compared to feeding all the training data all at once, as it adds a certain degree of stochasticity to the parameter updates and this helps to escape from local minima in the high-dimensional network parameter space.
The training period during which all the mini-batches in the training dataset are fed into the network is called as an ``epoch.''
At the beginning of each epoch, we set up the mini-batches by randomly shuffling the whole training dataset and dividing it into the mini-batches each of which contains 2000 instances, i.e., $m=2000$ in Eq.~(\ref{eq:loss_func}).
For the error $\sigma_{P^\mathrm{S}_{\rm fid}}$, we use the standard deviation of the power spectra, for a given set of redshift, $n_1$ and $n_2$ in the $i$th parameter (${\bf p}_i$), computed from the 15 realizations for the {\it Planck} cosmology.
That is, we ignore the dependence of the power spectrum error on cosmological models, as we have only one realization for each cosmological model in the training sets. Note that the simulations for the {\it Planck} cosmology is not included in the training dataset. 

We use 1000 epochs to train the neural network, and after the training we obtain the optimized network parameters that give a parametrized fitting formula of the redshift-space power spectrum, $P_{\rm emu}^\mathrm{S}(k,\mu)$.
We design the loss function [Eq.~(\ref{eq:loss_func})] to approximately correspond to the $\chi^2$ value between the data and  ``model'' (network output in this case), averaged over all the $(k,\mu)$ bins of all the instances in the mini-batch.
Hence, we expect that the loss function roughly goes to unity when the training successfully converges to the optimal result.
In fact, the trained network we obtain shows the loss function value to be about 2 for {\it both} the training and validation dataset (while the validation dataset is not used for the training itself), and this value is sufficiently saturated by the end of the training.
With the trained network we can compute the redshift-space power spectrum, $P^\mathrm{S}_\mathrm{hh}(k,\mu)$, for an arbitrary input set of the model parameters that are covered within the ranges of nine parameters.
From the neural network output which has $N_\mathrm{bin} = 620$ values, we can obtain the prediction at any point of $(k,\mu)$ within the range that we consider, by using the bivariate cubic spline interpolation.

\subsection{Large-scale limit: Stitching with linear theory prediction}
\label{sec:linear_theory}

The redshift-space power spectrum measured from simulations is considerably noisy on very large scales due to the lack of large-scale Fourier modes or 
the significant sample variance of a finite volume, $8~(h^{-1}\,{\rm Gpc})^3$.
Thus, the power spectrum predicted by the neural network output does not meet our requirements at roughly $k \lesssim 0.02\,h\,{\rm Mpc}^{-1}$.
To overcome this inaccuracy, we stitch the linear theory prediction with the neural network output to obtain the emulator predictions over a wide range of scales.
Specifically, we smoothly stitch the neural network output and the linear theory prediction as
\begin{align}
  P^\mathrm{S}_{\mathrm{hh}}(k,\mu) = P^\mathrm{S}_{\rm hh, lin}(k,\mu) e^{-(k/k_{\rm switch})} + P^\mathrm{S}_{\rm hh, NN}(k,\mu) \left[ 1 - e^{-(k/k_{\rm switch})} \right],
  \label{eq:ps_siwtch}
\end{align}
where $P^\mathrm{S}_{\rm hh, NN}(k,\mu)$ is the neural network output
we have described above, and $P^\mathrm{S}_{\rm hh, lin}(k,\mu)$ is the linear theory prediction.
For the latter, we employ the following model:
\begin{align}
  P^\mathrm{S}_{\rm hh, lin}(k,\mu; n_1, n_2) = \left[b_{\rm h}(n_1) + f\mu^2 \right] \left[b_{\rm h}(n_2) + f\mu^2 \right] P_{\rm lin}(k),
\end{align}
where $f = {\rm d}\ln D_+/{\rm d}\ln a$ is the linear growth rate, and $b_{\rm h}(n)$ is the linear bias for the halo sample of a given number density $n$. 
We use the \textsc{Dark Emulator} developed in \citet{Nishimichi_2019} for real-space halo statistics to compute the halo bias for the halo sample of a given number density.
Throughout this paper, we adopt the switching scale $k_{\rm switch} = 0.03 \, h \, {\rm Mpc}^{-1}$ for all cosmological models.
Including this stitching, our emulator implementation can compute $P^\mathrm{S}_\mathrm{hh}(k,\mu)$ in a few of $10^{-2}$ CPU seconds on a 2.8 GHz quad-core Intel Core i7 processor, for given input parameters.

\subsection{Emulator performance}

\begin{figure}
\centering
\includegraphics[width=0.48\textwidth]{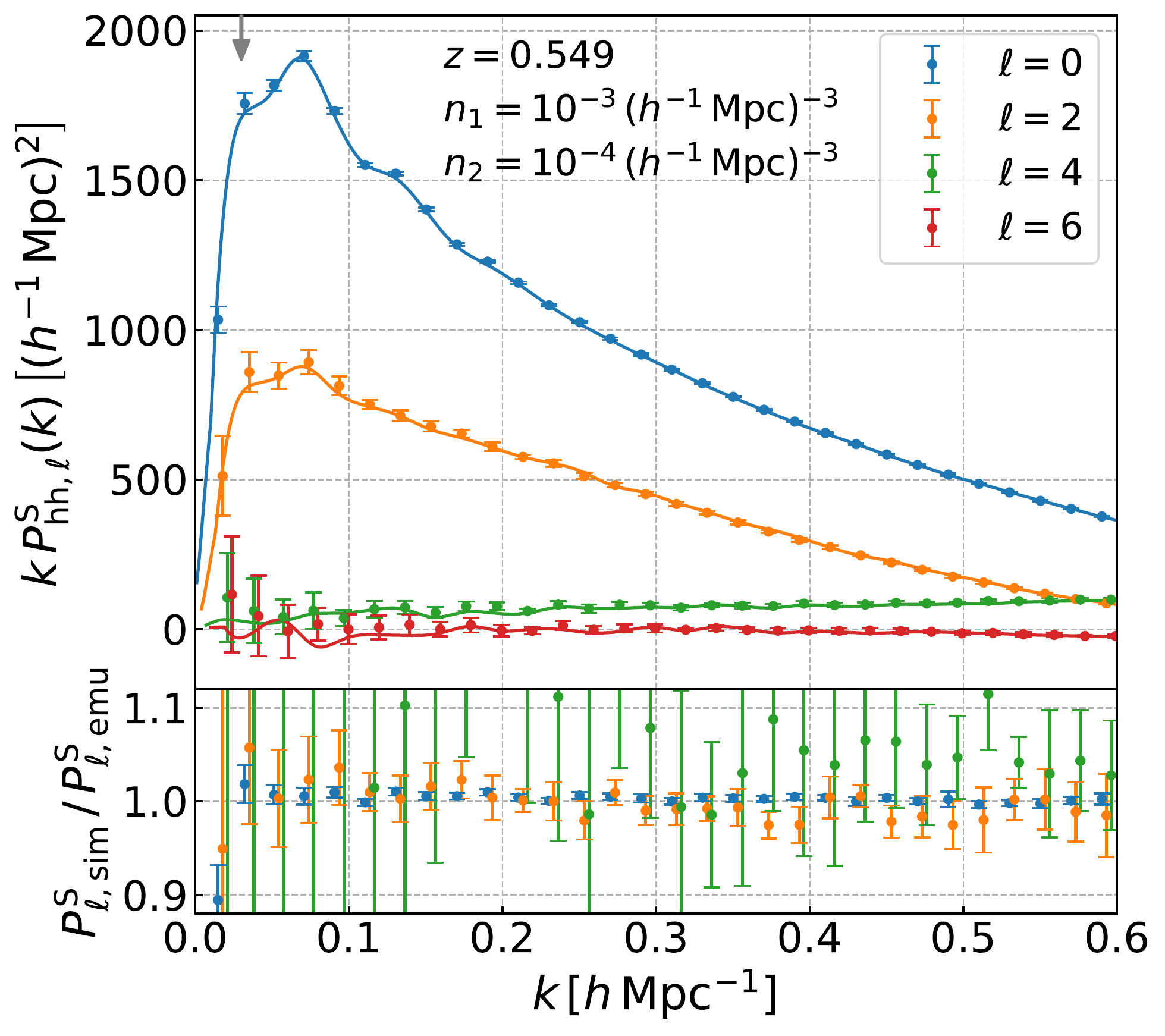}
\caption{
An example of the emulator prediction for the multipole moments of the redshift-space cross power spectrum between the halo samples 
with number densities, $n_1=10^{-3}$ and $n_2=10^{-4}~(h^{-1}{\rm Mpc})^3$ at $z=0.549$, for the {\it Planck} cosmology that is not used 
in the training dataset.
The blue, orange, green and red lines show the predictions for the monopole ($\ell=0)$, quadrupole ($\ell=2)$, hexadecapole ($\ell=4$), and tetra-hexadecapole ($\ell=6$) moments, respectively.
The symbols with error bars denote the moments measured from one realization for the {\it Planck} cosmology, 
where the errors are the statistical errors for a volume of $8~(h^{-1}\,{\rm Gpc})^3$, as in the previous figure.
The gray down arrow on the upper horizontal axis denotes the scale $k_{\rm switch} = 0.03\,h\,{\rm Mpc}^{-1}$ that is the switching scale between the linear theory prediction and the emulator output [see around Eq.~(\ref{eq:ps_siwtch}) for details].
}
\label{fig:pl_emu_demo}
\end{figure}
\begin{figure}
\centering
\includegraphics[width=0.48\textwidth]{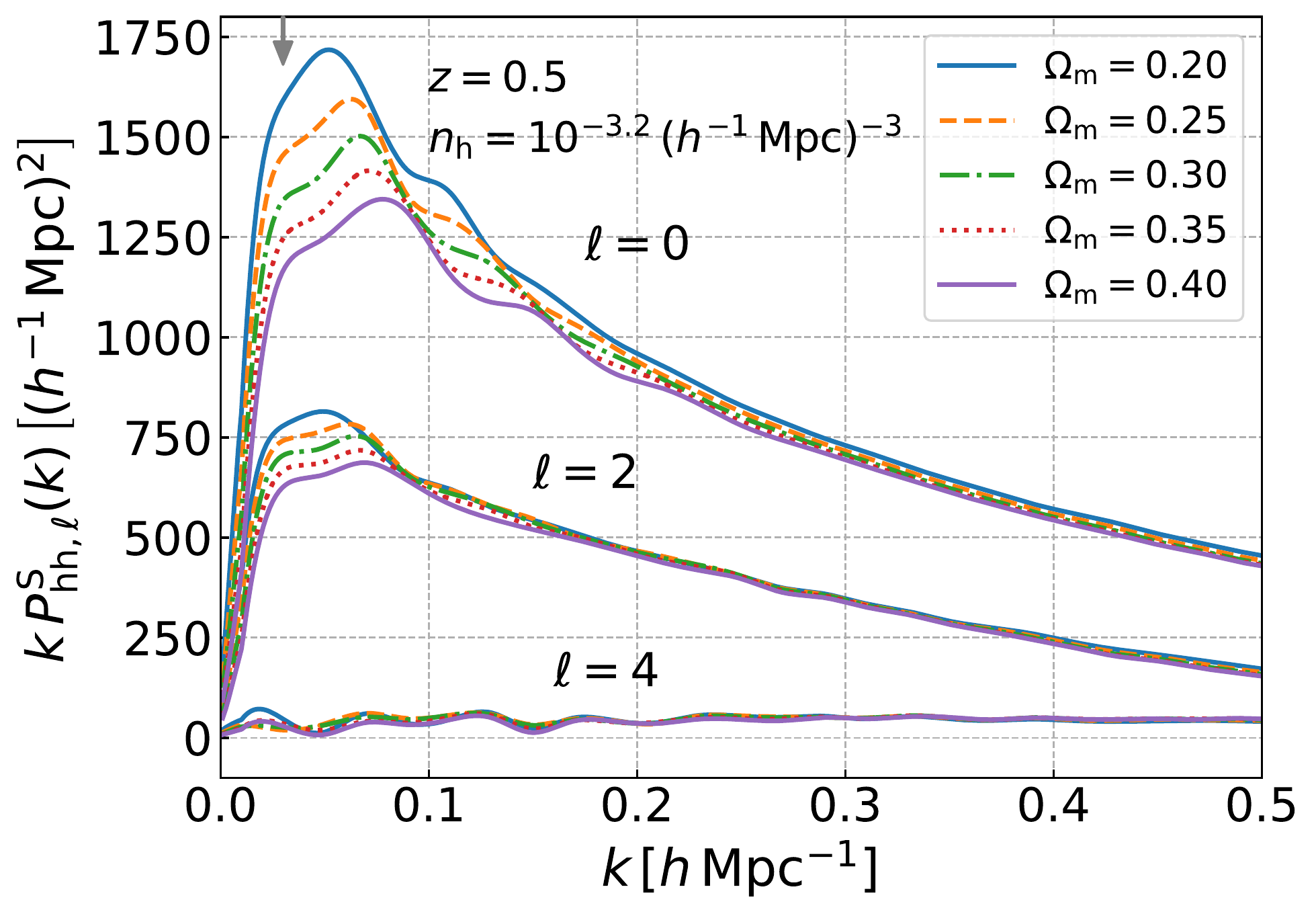}
\caption{
A demonstration of the emulator predictions in the cosmological parameter space.
We vary $\Omega_{\rm m}$ in the range $[0.2,0.4]$ while other cosmological parameters are fixed to the values for the {\it Planck} cosmology.
We show the predictions for the halo sample with $n_{\rm h} = 10^{-3.2}\,(h^{-1} \, {\rm Mpc})^{-3}$ at $z=0.5$, which are not at the sampling points of redshift and number density in the training dataset.
}
\label{fig:p0p2p4_emu_demo2}
\end{figure}
\begin{figure}
\centering
\includegraphics[width=0.48\textwidth]{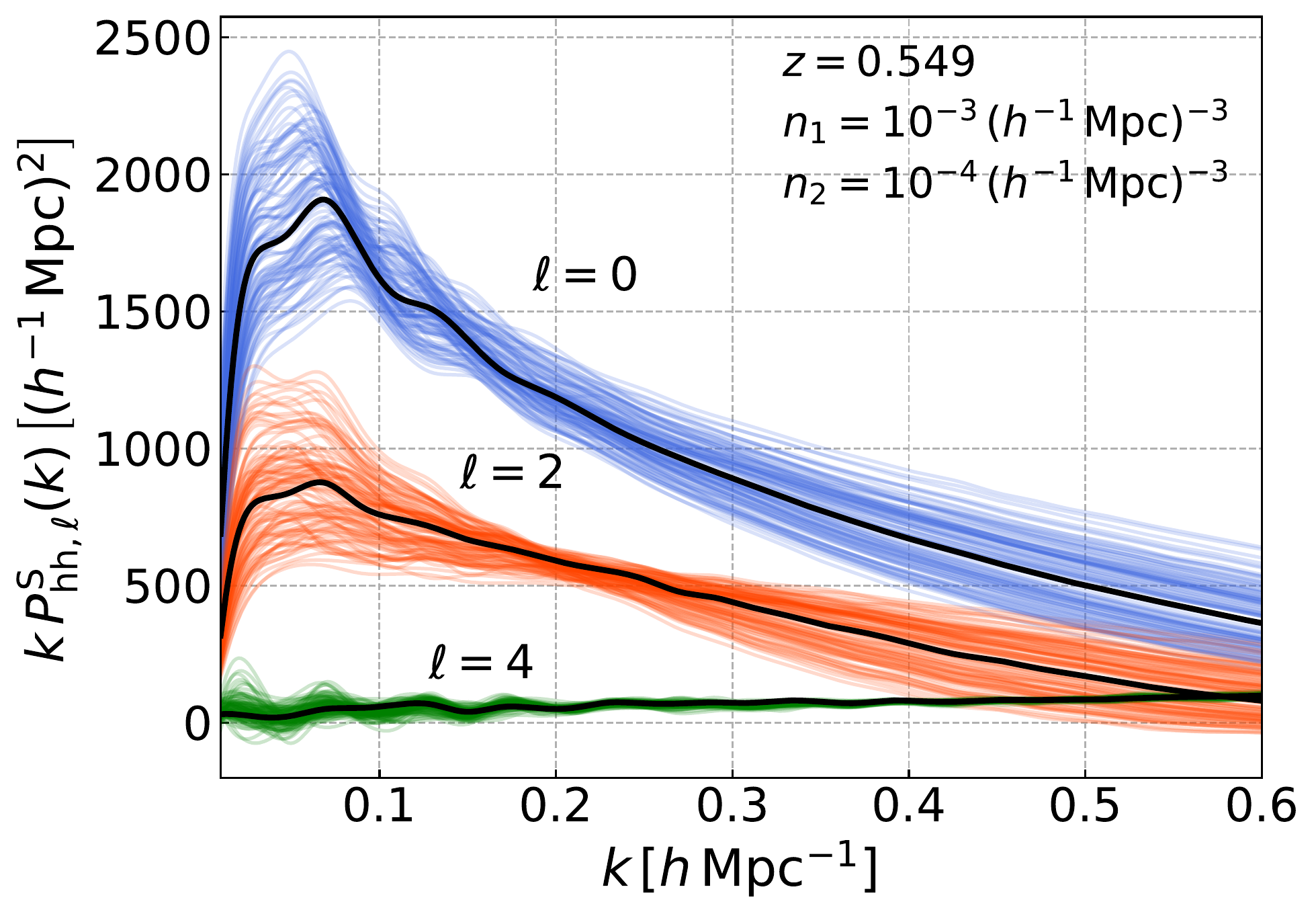}
\caption{
The emulator predictions for all the 101 cosmologies covered by the \textsc{Dark Quest} simulation suite.
We show the moments for the halo samples with $(n_1, n_2) = (10^{-3}, 10^{-4})\,(h^{-1}\,{\rm Mpc})^{-3}$ at $z=0.549$.
For comparison the black thick lines show the predictions for the {\it Planck} cosmology.
}
\label{fig:p0p2p4_emu_all_cosmo}
\end{figure}

We below discuss the validation and performance of the emulator that we have explained in the preceding section.

Figure~\ref{fig:pl_emu_demo} shows an example of the emulator predictions, for the {\it Planck} cosmology that is not used in the training set.
Here we consider the multipole moments of the redshift-space cross power spectrum between the halo samples of number densities,
$n_1=10^{-3}~(h^{-1}\,{\rm Mpc})^{-3}$ and $n_2=10^{-4}~(h^{-1}\,{\rm Mpc})^{-3}$,  which can be obtained by numerically integrating 
the emulator output, $P^\mathrm{S}_\mathrm{hh}(k,\mu)$, over $\mu$, weighted by the Legendre polynomials corresponding to the multipole order.
The figure shows that the monopole and quadrupole from the simulations are reproduced well by our network, by better than 5\% in the fractional difference (even over the range of $k$ scales where the quadrupole moment has small amplitudes).
The $\ell=4$ and $6$ moments have smaller amplitudes and noisy, but the emulator still explains the overall trend with wave number $k$ fairly well, especially in large $k$ bins compared to the statistical errors for a volume of $8~(h^{-1}\,{\rm Gpc})^3$.
The gray arrow on the upper axis denotes the switching scale between the linear theory and the direct neural network prediction as discussed in Sec.~\ref{sec:linear_theory}.

In Fig.~\ref{fig:p0p2p4_emu_demo2}, we show how we can use the emulator to study variations in the monopole, quadrupole and hexadecapole moments for cosmological models with different $\Omega_{\rm m} (= 1-\Omega_{\rm de})$.
Note that we assumed the flat-geometry universe, and the other five parameters, i.e., $\{\omega_{\rm b},\omega_{\rm c},\ln (10^{10}A_{\rm s}),n_{\rm s},w\}$, are kept to their values for the {\it Planck} cosmology.
The figure shows a clear dependence of the moments on $\Omega_{\rm m}$, including variations in the feature originating from the baryonic acoustic oscillations (BAO).
The changes in the quadrupole and hexadecapole are not so prominent compared to the monopole because of their smaller amplitudes.

Figure~\ref{fig:p0p2p4_emu_all_cosmo} gives another demonstration of the emulator. 
We here show variations in the multipole moments of the redshift-space power spectrum for all the 101 cosmological models that are sampled in the \textsc{Dark Quest} simulation suite.
Here we consider the halo samples with different number densities of $n_1=10^{-3}$ and $10^{-4}~(h^{-1}\,{\rm Mpc})^{-3}$ at $z=0.549$, as an example of general cases. 
The figure shows that the emulator covers a wide dynamic range for each multipole moment, and describes the cosmological dependence of the BAO features.

\begin{figure*}
\centering
  \begin{tabular}{c}
    \begin{minipage}{0.33\hsize}
    \centering
    \includegraphics[width=0.99\textwidth]{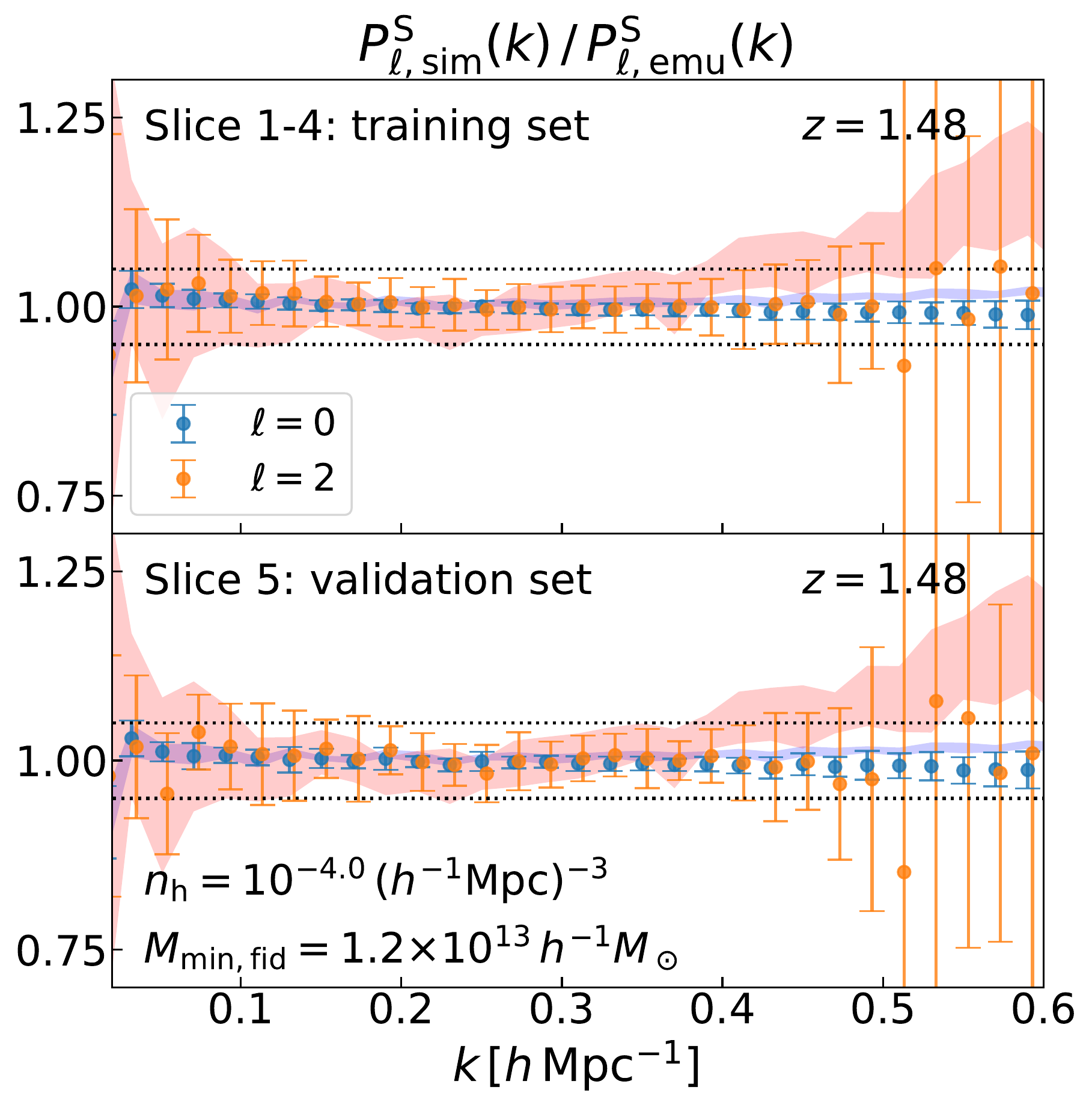}
    \end{minipage}
    
    \begin{minipage}{0.33\hsize}
    \centering
    \includegraphics[width=0.99\textwidth]{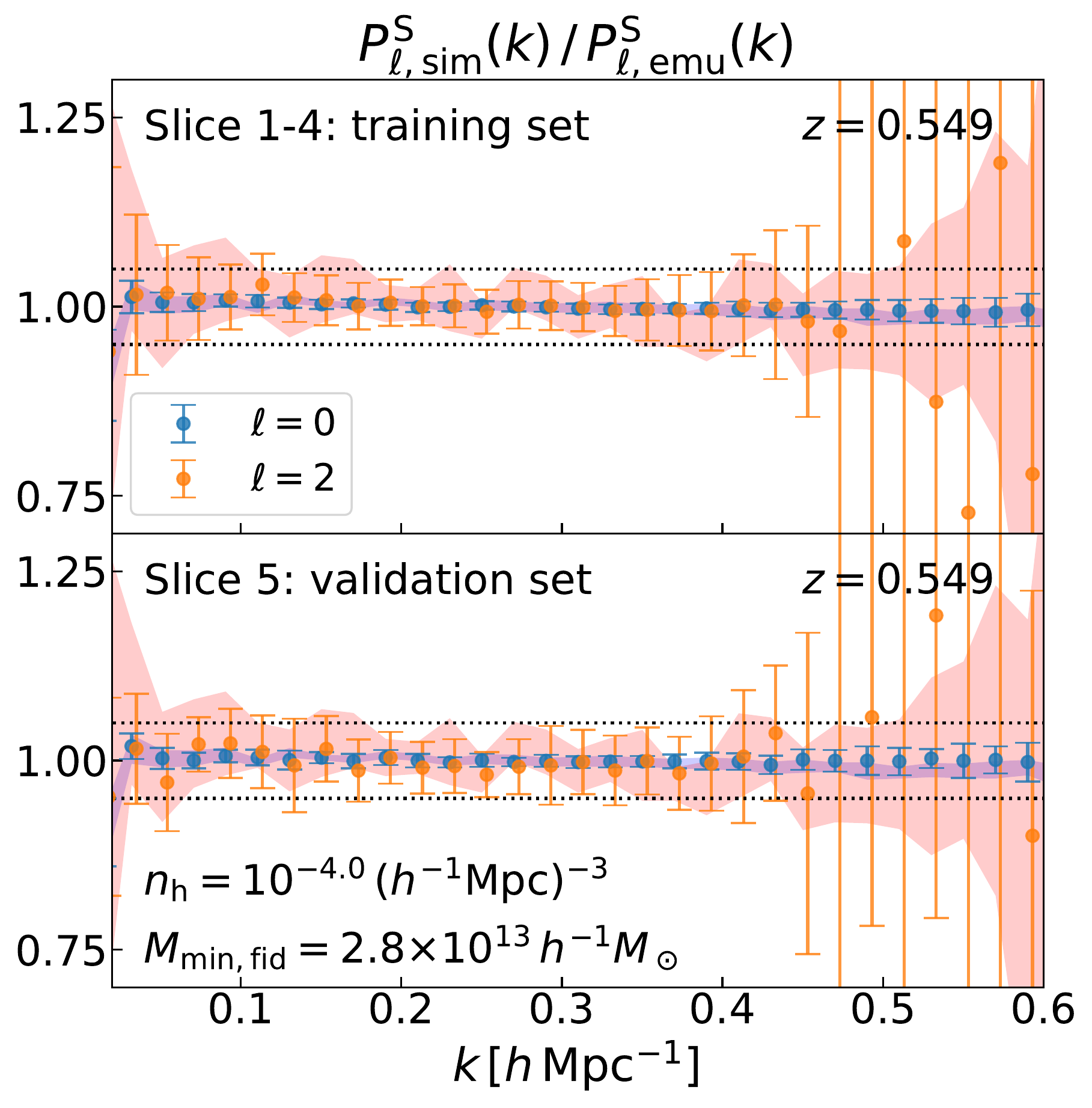}
    \end{minipage}

    \begin{minipage}{0.33\hsize}
    \centering
    \includegraphics[width=0.99\textwidth]{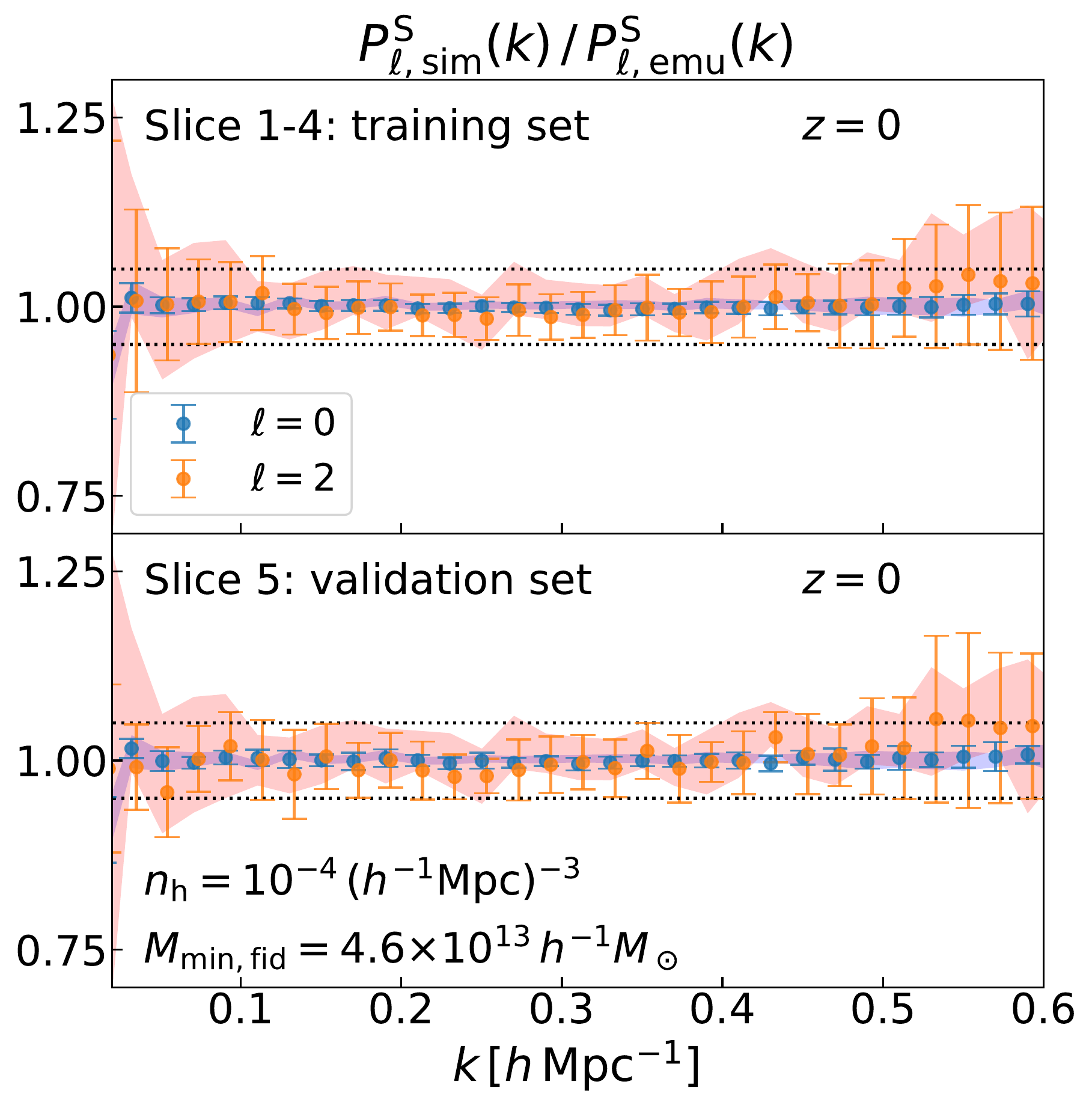}
    \end{minipage}
  \end{tabular}
\caption{
A validation of the emulator predictions. 
We compare, by the ratio, the emulator predictions with the simulation results for the monopole and quadrupole moments, for the halo sample with $n_{\rm h} = 10^{-4} \, (h^{-1}\,{\rm Mpc})^{-3}$ at three representative redshifts $z=1.48$, 0.549, and $0.0$, respectively.
The upper panel in each plot shows the comparison for 80 cosmological models in the training dataset, while the lower panel shows the comparison for 20 cosmological models in the validation dataset.
The symbols and error bars are the mean and standard deviation among 80 or 20 results, respectively.
For comparison the blue and red shaded regions are the statistical errors around the ratio for the fiducial {\it Planck} cosmology, where the errors are for $V=8~(h^{-1}\,{\rm Gpc})^3$.
$M_{\rm min,fid}$ in the legend denotes the halo mass threshold corresponding to the number density $n_{\rm h} = 10^{-4} \, (h^{-1}\,{\rm Mpc})^{-3}$, for the {\it Planck} cosmology at each redshift.
The black dotted lines indicate $\pm 5\%$ fractional errors.
}
\label{fig:p0p2_emu_validation}
\end{figure*}
\begin{figure*}
\centering
  \begin{tabular}{c}
    \begin{minipage}{0.33\hsize}
    \centering
    \includegraphics[width=0.99\textwidth]{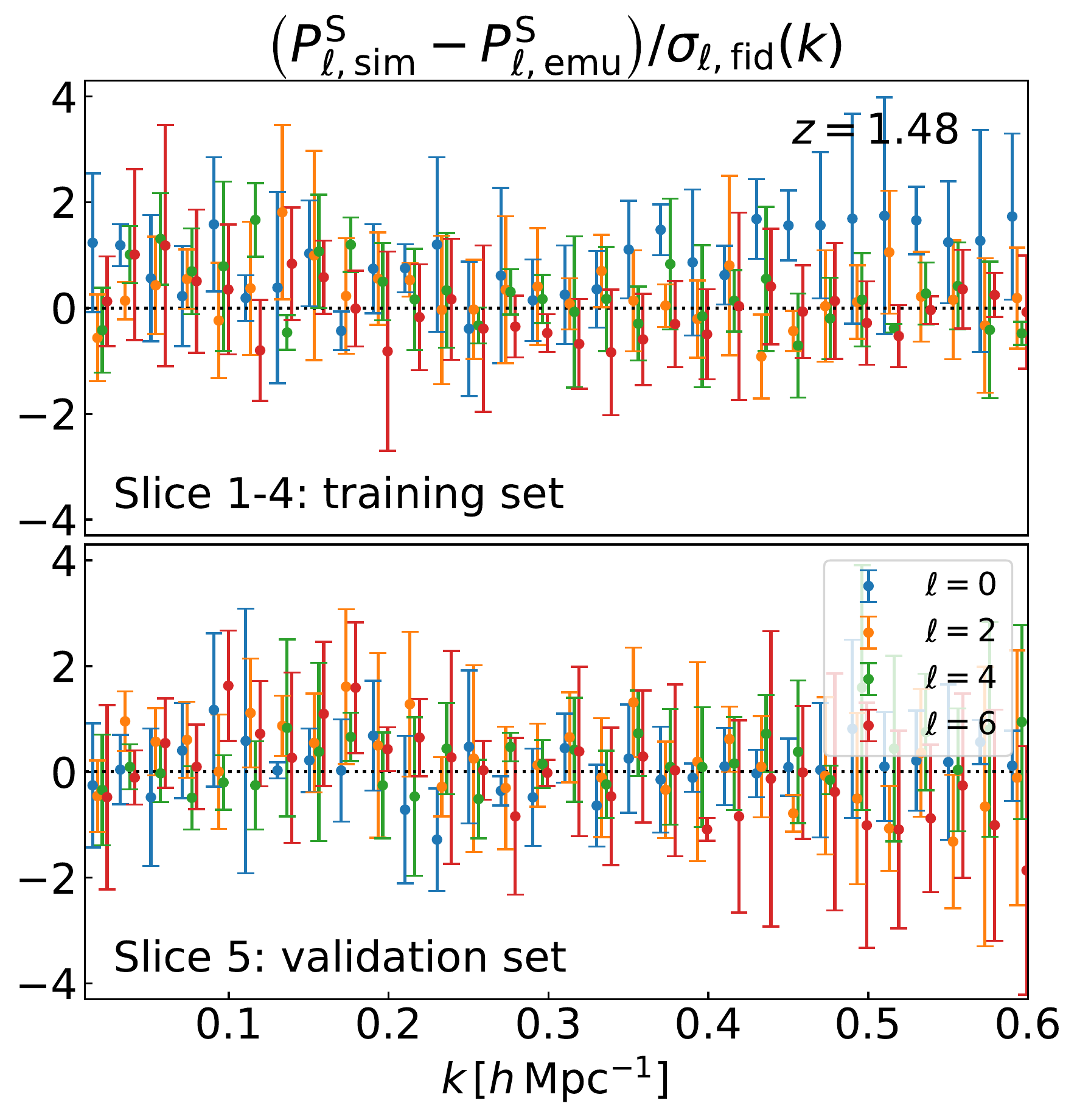}
    \end{minipage}
    
    \begin{minipage}{0.33\hsize}
    \centering
    \includegraphics[width=0.99\textwidth]{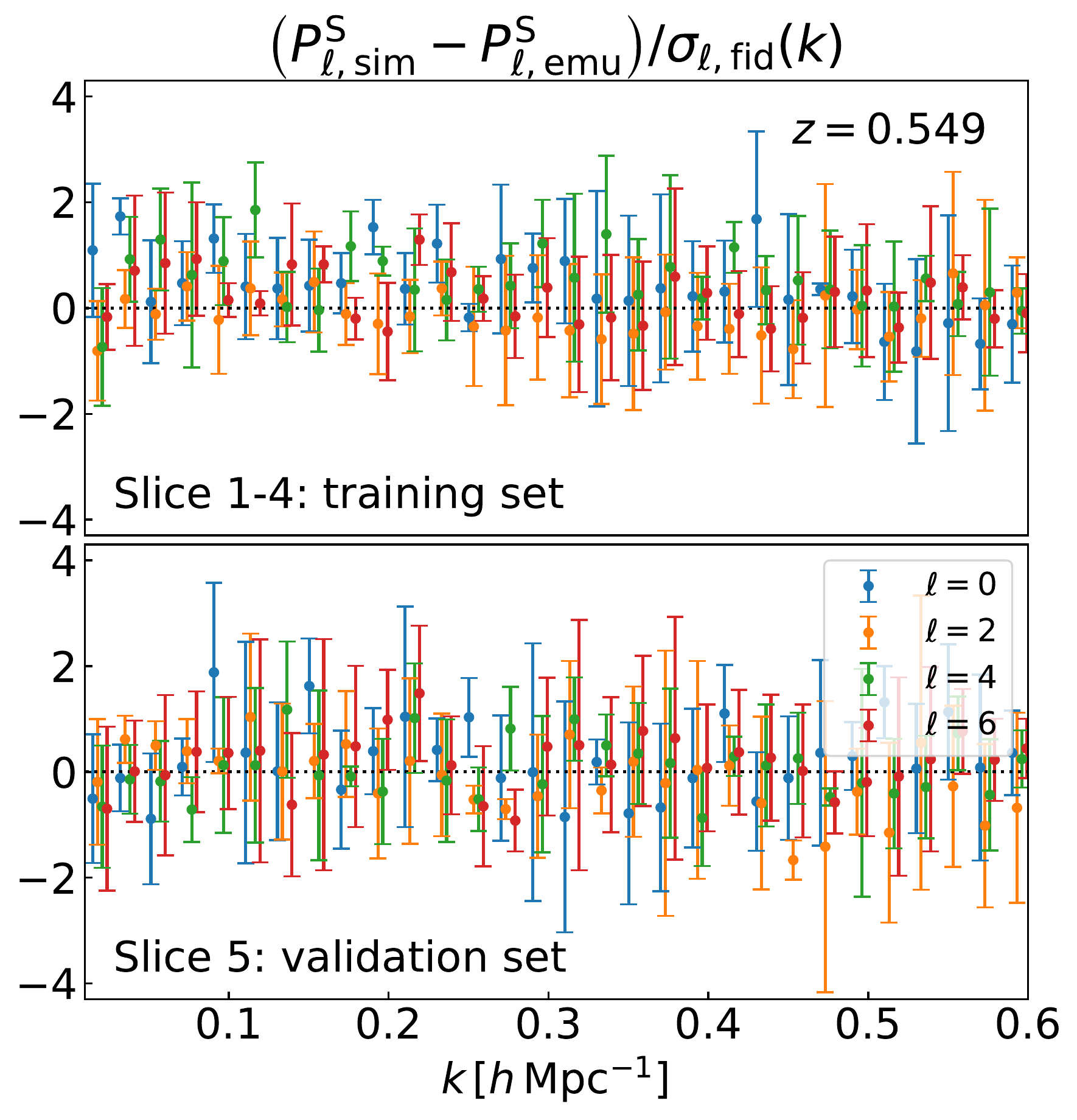}
    \end{minipage}

    \begin{minipage}{0.33\hsize}
    \centering
    \includegraphics[width=0.99\textwidth]{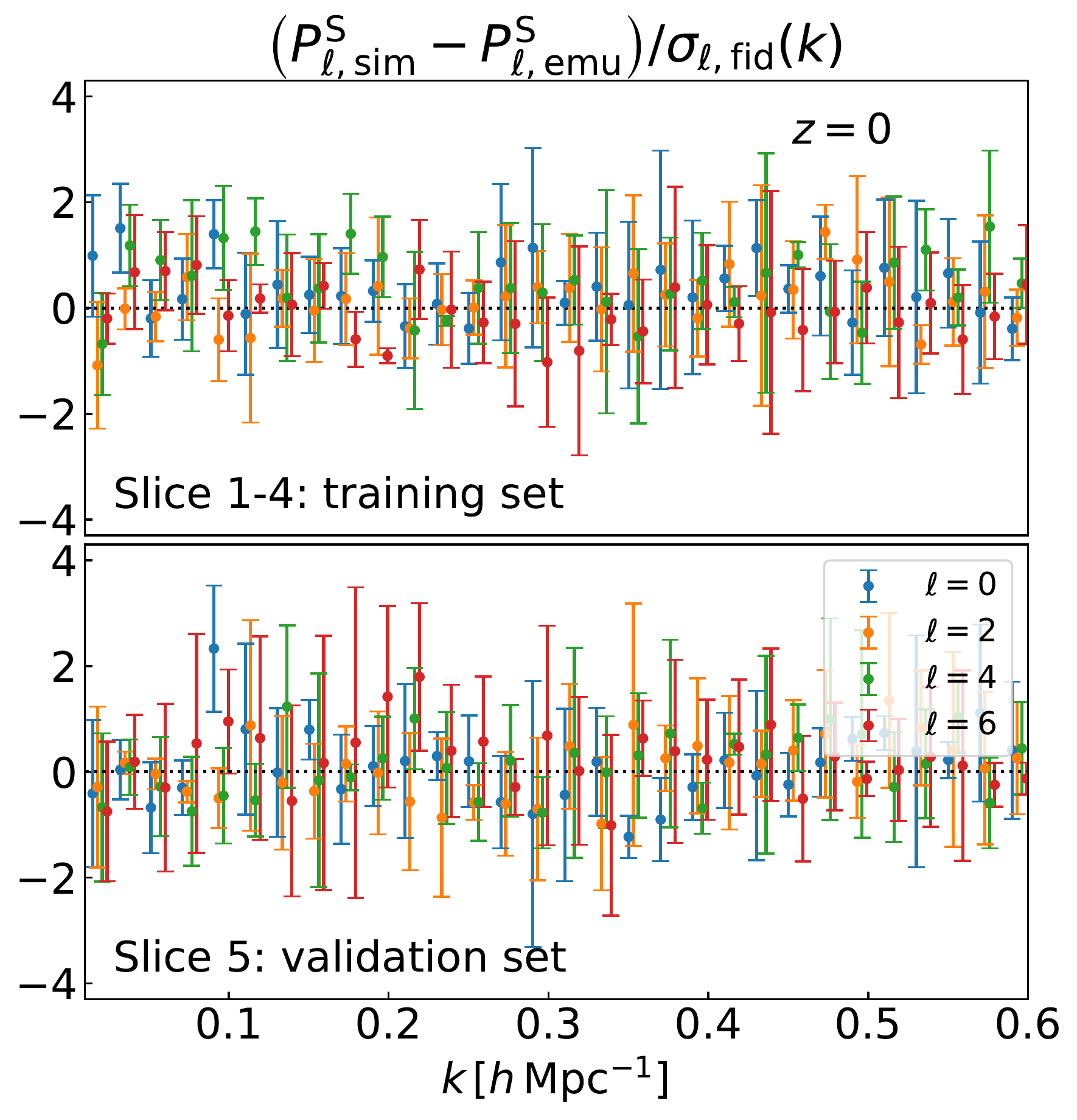}
    \end{minipage}
  \end{tabular}
\caption{
Similar to the previous figure, but another validation of the emulator predictions using the quantify to evaluate the accuracy of the emulator prediction (Eq.~\ref{eq:chi_criterion}),
which is defined by the difference between the emulator predictions and the simulation results relative to the statistical errors in each multipole moment for a volume of $8~(h^{-1}\,{\rm Gpc})^3$.
We here show the results up to the multipole moments of $\ell=6$.
The symbols and error bars are the mean and standard deviation among 80 and 20 realizations for the training and validation sets, respectively.
}
\label{fig:pl_emulation_chi}
\end{figure*}
Figure~\ref{fig:p0p2_emu_validation} gives an assessment of the emulator outputs.
We consider the multipole moments of redshift-space power spectra for the halo sample with number density $n_{\rm h} = 10^{-4}\,(h^{-1}\,{\rm Mpc})^{-3}$ at redshifts $z=1.48$, 0.549, and 0.0 from left to right panels, respectively.
The sample at $z=0.549$ roughly corresponds to host halos of the SDSS BOSS galaxies.
The upper panel of each plot shows that the emulator well recovers the input power spectrum data, meaning that the neural network does not degrade the accuracy after the regression. 
The lower panel gives a validation of the emulator, which shows the comparison of the emulator predictions with the multipole moments directly measured 
from simulations for each of 20 validation cosmological models in slice~5, which are not used in the training. 
The neural network reproduces equally well the simulation results for each of the validation models.
The accuracy of the emulator predictions is comparable between the training and validation datasets.
It implies that the neural network successfully avoids the overfitting.
We also emphasize that our emulator can predict the monopole and quadrupole moments of halo power spectrum with the number density $n_{\rm h} = 10^{-4}\,(h^{-1}\,{\rm Mpc})^{-3}$ with about 1 and 5\% accuracies, respectively, in the fractional errors. 
% Thus the emulator meets our requirements.
For comparison the blue and red shaded regions denote the statistical errors expected for measurements of the monopole and quadrupole moments, respectively,  
for a volume of $8~(h^{-1}\,{\rm Gpc})^3$, which are estimated from the standard deviations among 15 realizations for the {\it Planck} cosmology, where the simulation volume is larger than that of the SDSS BOSS survey,
$V_s\simeq 5.7~(h^{-1}\,{\rm Gpc})^3$.
The quadrupole moment for the \textit{Planck} cosmology (red shaded region) at $z=1.48$ shows a deviation at high $k$. It should be by chance due to the inaccuracy of learning, since we do not see the same behavior at other redshifts or number densities.
One might notice relatively larger variances in the quadrupole moment (i.e., orange error bars) at $z=1.48 \text{ and }0.549$ for both the training and validation sets at $k \gtrsim 0.4\,h\,{\rm Mpc}^{-1}$.
This is due to the fact that the quadrupole moments happen to have a transition from positive to negative values around these scales for some of the cosmological models, for the halo sample at these redshifts and number density.
In Appendix~\ref{sec:accuracy_number_density}, we show the prediction accuracy for the monopole and quadrupole moments for the halo samples of different number densities.

For the hexadecapole ($\ell = 4$) and tetra-hexadecapole ($\ell = 6$) moments, it is tricky to make a similar fractional comparison of the emulator predictions with the simulation results, especially at large scales, because the higher-order moments are noisy in the simulation measurements, and have small (almost zero-consistent) amplitudes.
Instead, we perform another comparison as shown in Fig.~\ref{fig:pl_emulation_chi}. 
In this figure we compare the differences between the emulator predictions and the simulation data (both in the training and validation sets), relative to
the standard deviation among 15 realizations for the fiducial {\it Planck} cosmology; we use the following quantity to evaluate the accuracy 
of the emulator predictions:
\begin{align}
\label{eq:chi_criterion}
  \frac{ P^\mathrm{S}_{\ell, {\rm sim}}(k) - P^\mathrm{S}_{\ell, {\rm emu}}(k) }{\sigma_{P^\mathrm{S}_{\ell, {\rm fid}}} (k)},
\end{align}
where $P^\mathrm{S}_{\ell, {\rm sim}}(k)$ and $P^\mathrm{S}_{\ell, {\rm emu}}(k)$ are the power spectrum multipole of degree $\ell$ measured from the simulation halo catalogs and predicted by our emulator, respectively, and $\sigma_{P^\mathrm{S}_{\ell, {\rm fid}}}(k)$ is the standard deviation among 15 realizations for the {\it Planck} cosmology.
We show the results for the same halo samples of three redshifts and number density as in Fig.~\ref{fig:p0p2_emu_validation}.
The four color (blue, orange, green, and red) symbols with error bars are their mean and standard deviation over 80 or 20 cosmologies in the training (upper) or validation (lower) sets, respectively.
The figure shows that the accuracies of the emulator predictions for the higher-order moments are roughly comparable between the training and validation sets as well as among different multipole moments, over all the $k$ range that our emulator covers.
This means that the training of neural network has been successfully done so that the all the terms in the loss function [Eq.~(\ref{eq:loss_func})] are on average equally minimized, and the training procedure did not cause a serious overfitting.

\subsection{Derivatives of the the power spectrum with respect to cosmological parameters}
\label{subsec:derivative}

\begin{figure*}
\centering
  \begin{tabular}{c}
    \begin{minipage}{0.5\hsize}
    \centering
    \includegraphics[width=0.99\textwidth]{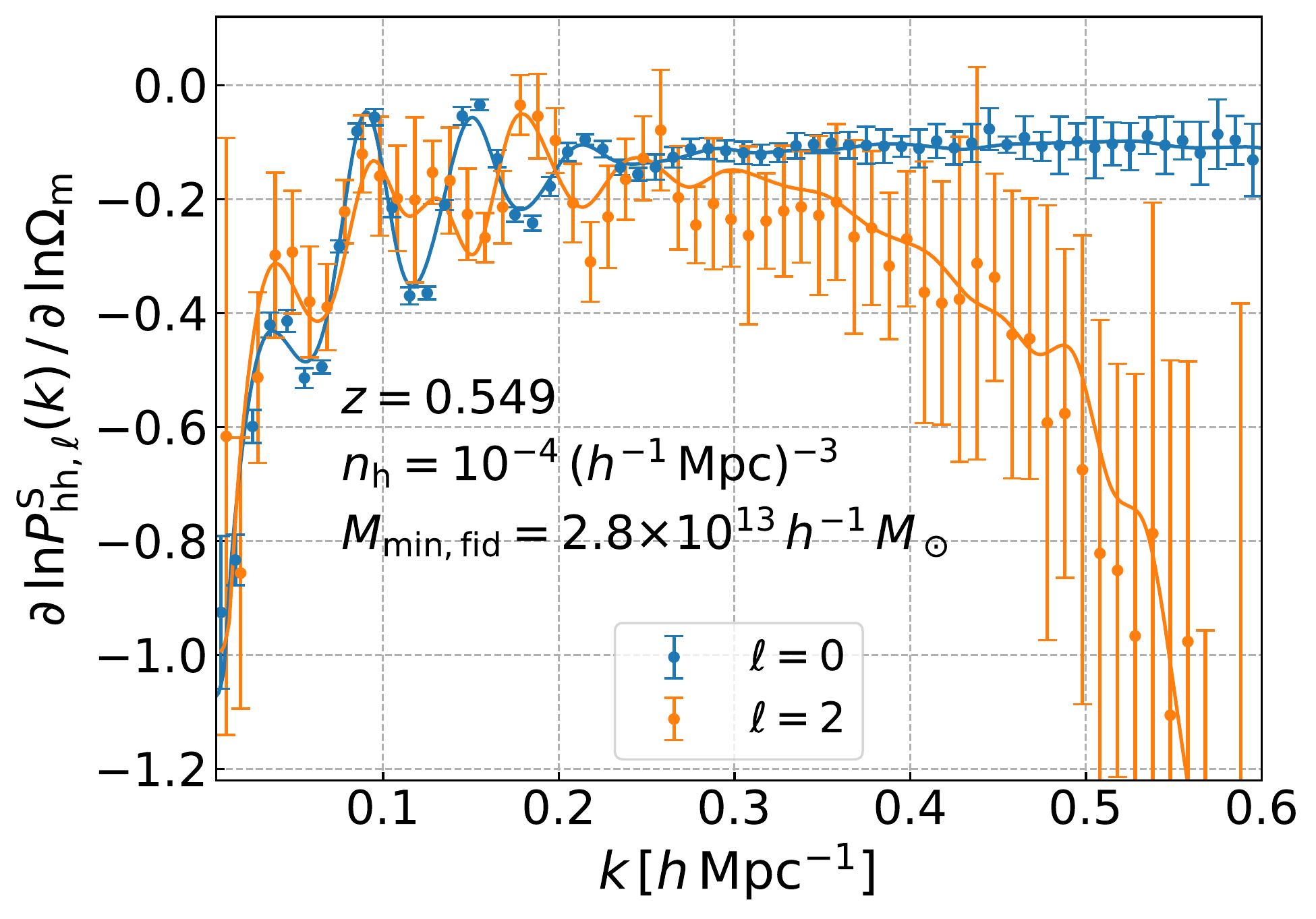}
    \end{minipage}
    
    \begin{minipage}{0.5\hsize}
    \centering
    \includegraphics[width=0.99\textwidth]{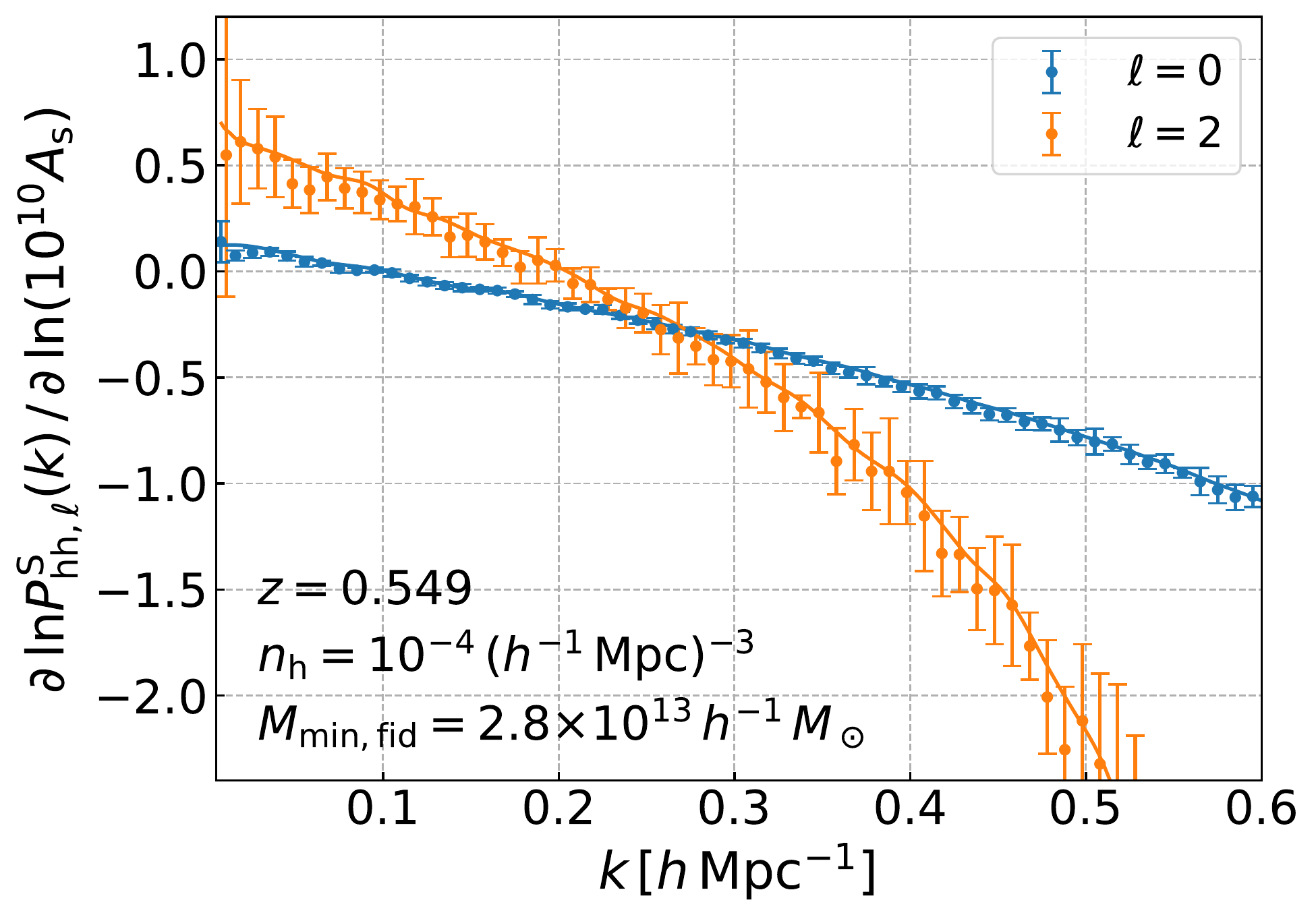}
    \end{minipage}
  \end{tabular}
\caption{
The logarithmic derivatives of the halo power spectrum with respect to $\Omega_{\rm m}$ (left) and $A_{\rm s}$ (right), around the fiducial \textit{Planck} cosmology.
We show the derivatives of the monopole (blue) and quadrupole (orange) moments.
The solid lines are the emulator predictions, while the symbols are the signals measured from the $N$-body simulations. For the latter, 
we used the additional $N$-body simulations, with varying $\Omega_{\rm m}$ or $A_{\rm s}$ from its {\it Planck} cosmology value, to numerically 
evaluate the derivatives from the measured power spectra in different simulations (see text for details).
The error bars are the standard deviation among the 10 realizations each of which has a volume of $8~(h^{-1}\,{\rm Gpc})^3$.
}
\label{fig:derivative}
\end{figure*}

Figure~\ref{fig:derivative} shows the derivatives of the halo power spectrum with respect to two cosmological parameters, $\Omega_{\rm m}$ (left) and $A_{\rm s}$ (right).
% \tnrv{[TN: Once again, $A_{\rm s}$ is a derived parameter from the $6$ input parameters, and thus we how to specify how exactly we change $\omega_c$, $\omega_b$ and $\Omega_{\rm de}$.]}
% \ykrv{[YK: We take $\ln (10^{10} A_{\rm s})$ as one of the primary input parameters, and the variation of $A_{\rm s}$ simply follows that of $\ln (10^{10} A_{\rm s})$. I added a description to avoid the ambiguity below in cyan.]}
While our training procedure of the neural network is such that it minimizes the differences between the training data and the network outputs, the \textit{rate} of change in the data in response to changes in the input parameters is not explicitly taken into account.
Hence there is no guarantee that the emulator gives accurate predictions on the derivatives.
In this figure, we focus on the derivatives with respect to the cosmological parameters around the fiducial \textit{Planck} cosmology.
We show the case of the halo sample with number density $n_{\rm h} = 10^{-4}\,(h^{-1}\,{\rm Mpc})^{-3}$ at $z=0.549$, which corresponds to the halo mass threshold $M_{\rm min} = 2.8 \times 10^{13}~h^{-1}\,M_\odot$ for the {\it Planck} cosmology.
For the monopole (blue) and quadrupole (orange) moments, the emulator predictions (solid lines) and the measured derivative signals (symbols) show a good agreement with each other.
For the measured signals, we used the additional $N$-body 
simulations in Ref.~\cite{PhysRevD.101.023510} to numerically evaluate the derivatives, where we used a shifted value of 
$\Omega_{\rm de} = 1-\Omega_{\rm m}$ or $\ln (10^{10} A_{\rm s})$ by $\pm 5\%$ from the fiducial value of {\it Planck} cosmology to run the 
simulations (the other five parameters in Eq.~(\ref{eq:cosmo_param_range}) are kept to their fiducial values).
The simulation settings such as the box side length and the number of particles are identical to the \textsc{Dark Quest} LR simulations.
The error bars are the standard deviation among 10 realizations (we have two different simulations of $+5\%$ and $-5\%$ to take the two-sided derivative with respect to each of $\Omega_{\rm m}$ and $A_{\rm s}$, and thus we used in total 40 realizations for this study).
We compute the emulator predictions by the two-sided numerical derivatives in which we shift $\Omega_{\rm m}$ or $\ln (10^{10} A_{\rm s})$ by $\pm 1\%$, while changing this rate in the range of 1\%--5\% gives almost no change in the predictions.

\section{Utility of emulator: galaxy power spectrum}
\label{sec:discussion}

\begin{table*}
\begin{center}
\begin{tabular}{l|l|l}\hline\hline
Class & Function & Description \\ \hline
Primary & $P_\mathrm{hh}^\mathrm{S}(\bk;z, M_1, M_2, \bp)$ & Redshift-space power spectrum for halos of mass thresholds $M_1$ and $M_2$, given 
as a function of \\
&& redshift ($z$), wave vector ($\bk$) and a set of cosmological parameters $\bp=\{\omega_{\rm b},\omega_{\rm c}, \Omega_{\rm de}, \ln(10^{10}A_{\rm s}),
n_\mathrm{s},w\}$  \\ \hline
Nuisance & $\langle N_{\rm c}\rangle(z,M)$ & HOD for central galaxies\\
& $\langle N_\mathrm{s}\rangle(z, M)$ & HOD for satellite galaxies \\
& $\tilde{\cal H}(\bk;z, M, ...)$ & Real-space position distribution
of galaxies in host halos of mass $M$ 
\\
& $\tilde{\cal F}(k_\parallel;z, M, ...)$ & Velocity distribution of galaxies inside halos of mass $M$ \\ 
\hline\hline
\end{tabular}
\caption{
A summary of the functions that we use in this paper. ``Primary'' function is the redshift-space power spectrum for halos, 
$P^\mathrm{S}_\mathrm{hh}(\bk)$. 
It is the primary output of the emulator we develop in this paper.
``Nuisance'' functions are needed to model the relation between halos and target galaxies in redshift space. 
These functions need to be flexible enough to model the range of effects of galaxy physics in the redshift-space galaxy power spectrum. 
\label{tab:functions}}
\end{center}
\end{table*}

We have so far described the construction and validation of the emulator for the halo power spectrum.
Our primary aim is to have accurate model predictions for the redshift-space galaxy power spectrum, a direct observable in galaxy redshift surveys.
In this section we describe how we can use the emulator output to make model predictions for the galaxy power spectrum, and demonstrate that it has a sufficient functionality to give the theoretical templates for a cosmological analysis of actual galaxy redshift surveys.

\subsection{Galaxy power spectrum based on the halo model formalism}
\label{subsec:halo_model_formalism}

The redshift-space galaxy power spectrum is among the most important observables in galaxy redshift surveys.
However, the {\it ab initio} modeling of galaxy formation and evolution is still quite challenging due to complexity in the physical processes.
We instead adopt an empirical prescription to model the relation between galaxies and halos, the halo occupation distribution (hereafter HOD)
\cite{1998ApJ...494....1J,peacock:2000qy,seljak:2000uq,scoccimarro:2001fj,2005ApJ...633..791Z}
(also see \cite{Cooray02} for a review).
We then combine the HOD prescription with the emulator output analytically at the level of equations to compute model predictions for the redshift-space 
power spectrum of galaxies (also see \cite{Nishimichi_2019} for the similar method for the real-space galaxy clustering statistics).
In doing so, we keep a large flexibility by providing a dedicated functionality module so that a user can adopt a desired HOD prescription to model the halo-galaxy connection for a sample of target galaxies. 
This implementation is done based on the halo model approach \cite{seljak:2000uq,peacock:2000qy,scoccimarro:2001fj}.
The requirement for an application of our emulator to the halo model approach is that galaxies in a sample of consideration reside in host halos with masses $M\ge 10^{12}~h^{-1}M_\odot$ at a redshift for a cosmological model within the ranges covered by our training set.

We here summarize representative model ingredients of HOD that we already implemented in the emulator modules. 
Once again, a user can extend the model to include other effects, so the following items should be considered as a working example: 
\begin{itemize}
\item $\avrg{N_{\rm c}}(M)$ --- The HOD for central galaxies that model the average number of ``central'' galaxies in the host halos of mass $M$. 
\item $\avrg{N_\mathrm{s}}(M)$ --- The HOD for satellite galaxies. 
\item ${\cal H}(r; M)$ --- The normalized radial profile of satellite galaxies in the host halo of mass $M$. 
One can employ the spherically-symmetric profile in the average sense, and the profile needs to be defined so as to satisfy the normalization condition of $\int_0^{R_{200}}\!4\pi r^2\mathrm{d}r~{\cal H}(r;M)=1$.

\item ${\cal F}(\Delta r_\parallel; r, M)$ --- The distribution function of the relative line-of-sight displacement due to the RSD effect caused by the internal (virial) velocities of satellite galaxies in the host halos of mass $M$ \citep{2001MNRAS.321....1W,2001MNRAS.325.1359S,hikage12a}.
This leads to the FoG effect \cite{jackson72}.
We can assume a spherically symmetric profile with a dependence only on the radial distance from the halo center, in the average sense, but the radial and tangential velocity dispersions with respect to the halo center can be different. 
The velocity function satisfies the normalization condition $\int_{-\infty}^\infty\!\mathrm{d}\Delta r_\parallel~{\cal F}(\Delta r_\parallel)=1$.

\item ${\cal P}(r_{\rm off}; M)$ --- Some of central galaxies can have an off-centering effect with respect to the halo center (the density maximum) as a consequence of merger or accretion in the hierarchical structure formation, as indeed indicated by the actual data \cite{hikage:2013kx} or 
by the simulation study \cite{masaki13}. We can assume the spherically symmetric distribution for ${\cal P}$ in the average sense, and the profile satisfies the normalization condition, $\int_0^\infty\!\!4\pi r_{\rm off}^2\mathrm{d}r_{\rm off}~ {\cal P}(r_{\rm off})=1$.
The off-centered galaxies would have internal motions with respect to the halo center, so the velocity distributions of the off-centered galaxies need to be given if one wants to include the RSD effect, as we do for satellite galaxies using ${\cal F}$.
\end{itemize}
Note that the above functions depend on redshift $z$, but we omit $z$ in the argument for notational simplicity.
One can employ parametrized functions to model these ingredients or inject a numerical table into the emulator modules to implement the halo-galaxy connection.
In Table~\ref{tab:functions}, we summarize the functions that we use in this paper.

In the following, as a demonstration for the application of our emulator, we employ the same halo-galaxy connection model in Ref.~\cite{PhysRevD.101.023510} that resembles the SDSS BOSS galaxies. 
The details of the model parameters are given in Appendix~\ref{sec:hod}. 
In the halo model approach, the redshift-space power spectrum of galaxies is given by the sum of one- and two-halo terms [Eq.~(\ref{eq:psgg_general})], which are expressed in terms of the above functions as
\begin{align}
\label{eq:PS1h}
    P^{\rm S, 1h}_\mathrm{gg}(\bk) &= \frac{1}{\bar{n}_{\rm g}^2}
\int\!\!{\rm d}M~{\color{blue}{\frac{{\rm d}n}{{\rm d}M}}(M)}
\avrg{N_{\rm c}}\!(M) \left[
2\lambda_\mathrm{s}(M) \,\tilde{{\cal H}}^\mathrm{S}(\bk; M)
\right. \nonumber \\
&\hspace{2em}\left.+ \lambda_\mathrm{s}(M)^2 \,\tilde{{\cal H}}^\mathrm{S}(\bk; M)^2
\right],
\end{align}
and
\begin{align}
\label{eq:PS2h}
P^{\rm S,2h}_\mathrm{gg}(\bk)&=
\frac{1}{\bar{n}_{\rm g}^2}
\int\!{\rm d}M_1 {\color{blue}{\frac{{\rm d}n}{{\rm d}M}}(M_1)}\left[\avrg{N_{\rm c}}\!(M_1)+\avrg{N_\mathrm{s}}\!(M_1)\,
\tilde{{\cal H}}^\mathrm{S}(\bk; M_1)\right]\nonumber\\
&\times \int\!{\rm d}M_2{\color{blue}{\frac{{\rm d}n}{{\rm d}M}}(M_2)}\left[\avrg{N_{\rm c}}\!(M_2)
+\avrg{N_\mathrm{s}}\!(M_2)\,\tilde{{\cal H}}^\mathrm{S}(\bk; M_2)\right] \nonumber \\
&\times {\color{blue}{P^\mathrm{S}_\mathrm{hh}(\bk; M_1, M_2)}},
\end{align}
with the mean number density of galaxies, defined as
\begin{align}
\bar{n}_\mathrm{g}=\int\!\!\mathrm{d}M~{\color{blue}{\frac{\mathrm{d}n}{\mathrm{d}M}}(M)}\left[\avrg{N_\mathrm{c}}(M)+\avrg{N_\mathrm{s}}(M)\right].
\label{eq:ng_hod}
\end{align}
The model ingredients in blue color fonts (in electronic version) denote the quantities obtained from the emulator, and other ingredients in black are the functions needed for the halo-galaxy connection. For this particular example, 
we assumed that satellite galaxies reside only in a halo that already hosts a central galaxy. Furthermore, we assume that the number distribution of satellite galaxies in a given host halo of mass $M$ follows the Poisson distribution with mean $\lambda_\mathrm{s}(M)$.
The function $\tilde{{\cal H}}^\mathrm{S}(\bk; M)$ is the Fourier transform of the normalized distribution function of galaxies in redshift space including the RSD effect due to virial motions of satellite galaxies.
Hence, $\tilde{{\cal H}}^\mathrm{S}(\bk)$ is a two-dimensional function depending on $k$ and the angle between $\bk$ 
and the line-of-sight direction. 
We give expressions of these functions for our default HOD in Appendix~\ref{sec:hod}.
Note that the above equations give detailed forms of Eq.~(\ref{eq:psgg_general}).

When we further include the off-centering effects of central galaxies, we need to replace 
$\avrg{N_{\rm c}}$ in Eqs.~(\ref{eq:PS1h})
and (\ref{eq:PS2h}) with
\begin{align}
\label{eq:off-center}
\left[(1-p_{\rm off})+p_{\rm off}\exp\left\{-\frac{1}{2}{k^2{R}_{\rm off}^2}\right\} \tilde{{\cal F}}(k_\parallel;M)\right]\avrg{N_{\rm c}}\!(M),
\end{align}
where $p_{\rm off}$ is a parameter to characterize the probability that each central galaxy is off-centered from the true center of its host halo, 
$R_{\rm off}$ is a parameter to characterize the typical off-centering radius, and $\tilde{{\cal F}}(k_\parallel;M)$ is the Fourier transform of the velocity distribution function of satellite galaxies ($k_\parallel$ is the line-of-sight component of $\bk$).
We here assumed that the off-centered central galaxies follow the same velocity distribution as that of satellite galaxies.

We should emphasize that the form of $P^\mathrm{S}_\mathrm{hh}(k, \mu; M_1,M_2)$ in the emulator output makes it straightforward to include the FoG effect due to the virial motions of galaxies in the host halo and the AP geometrical distortion effect (see below) to obtain the redshift-space power spectrum of galaxies. 
This is not the case if the emulator output is in the form of the multipole moments.
This is one of the requirements to which we stick when building the emulator in this paper.

We note that, strictly speaking, our standard implementation of the one-halo term in Eq.~(\ref{eq:PS1h}) behaves as a shot-noise-like term of $k^0$ at the limit of $k\to 0$ and this violates the mass and momentum conservation at this limit \cite{peebles1980}. 
Nevertheless, for the following tests for the SDSS BOSS-like galaxies, we did not find any signature of failure caused by our implementation of the one-halo term over the range of $k$ we consider, so we ignore this limitation for now. 
For further improvement, one can introduce an empirical function to give a cutoff of the one-halo term at very small $k$, e.g., following the method in Ref.~\cite{2014MNRAS.445.3382M} (see also \cite{Valageas11a}). 

Our default implementation implicitly assumes that the halo-galaxy connection is determined solely by the host halo mass. 
This would be violated if more complicated conditions apply to the
target galaxies, which is often referred to as the assembly bias, i.e., the existence of additional parameter, beyond halo mass, in the halo-galaxy connection such as the halo mass concentration, the halo ellipticity, and environments in more general terms.
The previous studies discussed that the assembly bias hardly affects the RSD effect due to bulk motions of host halos, partly because the RSD effect is a gravitational effect \citep{2019MNRAS.487.2424M,2019MNRAS.486..582P,PhysRevD.101.023510}.
The assembly bias effect should be carefully taken into account when one performs the cosmological parameter estimation, but we do not discuss it further in this paper.

\subsection{Implementation of galaxy power spectrum}
\label{subsec:galaxy_power_spectrum}

\begin{figure*}
\centering
  \begin{tabular}{c}
    \begin{minipage}{0.5\hsize}
    \centering
    \includegraphics[width=0.99\textwidth]{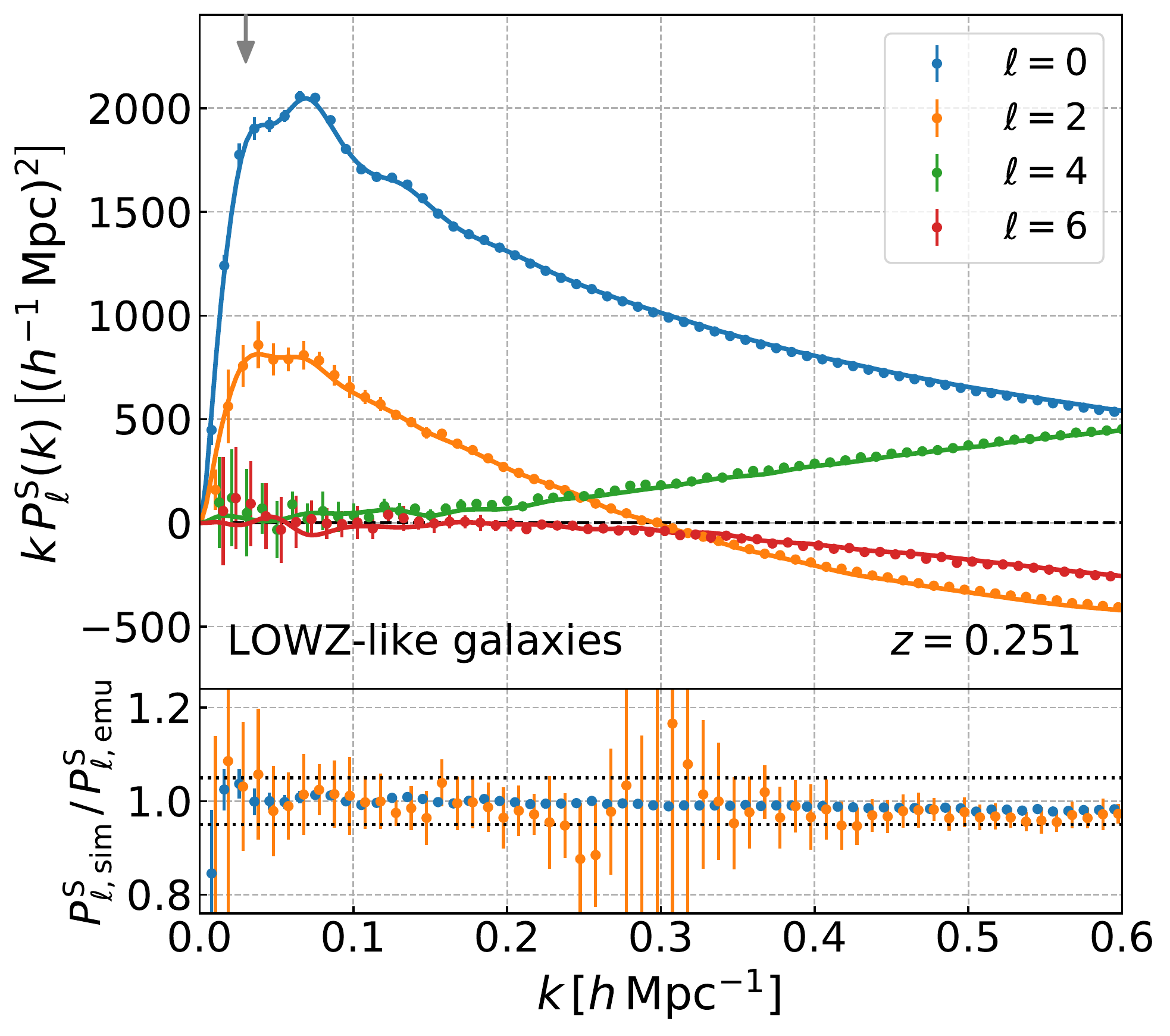}
    \end{minipage}
    
    \begin{minipage}{0.5\hsize}
    \centering
    \includegraphics[width=0.99\textwidth]{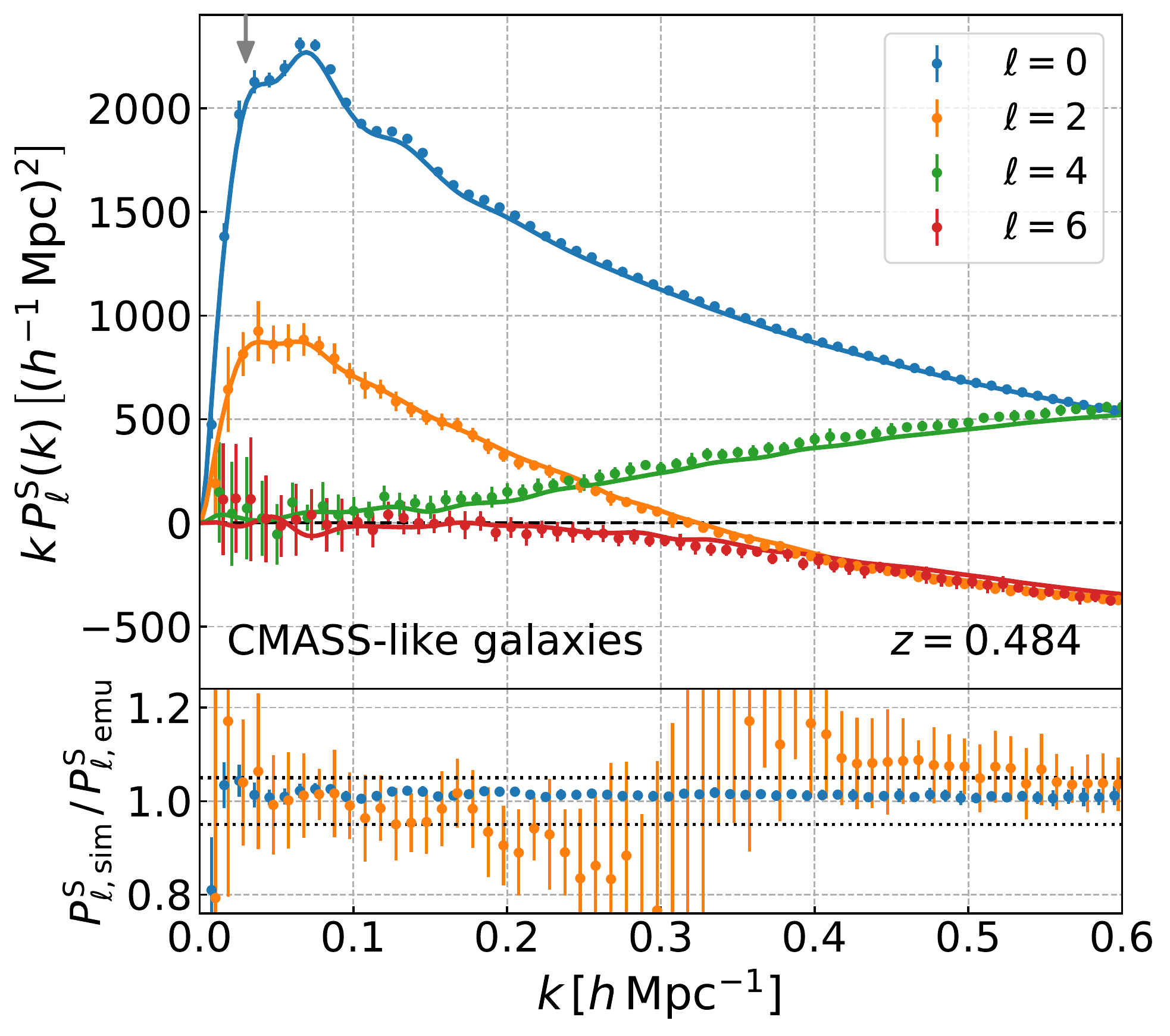}
    \end{minipage}
  \end{tabular}
\caption{
The solid lines show the emulator predictions for the multipole moments of redshift-space power spectrum for galaxies that mimic the SDSS BOSS LOWZ- and CMASS-like galaxies at $z=0.251$ and $z=0.484$, respectively, for the {\it Planck} cosmology. 
Here we adopt the HOD method to combine with the emulator outputs to compute the redshift-space galaxy power spectra for the BOSS-like galaxies.
The symbols with error bars are the spectra measured from the mock catalogs, where we employ the same HOD to populate galaxies into halos in each simulation realization, include the RSD effect, and then measure the multipole moments from the mocks. 
The mock results are for one particular realization, and the errors are for a volume of $8~(h^{-1}\,{\rm Gpc})^3$. 
The lower panels show the ratio for the monopole and quadrupole moments.
}
\label{fig:power_emu_lowz}
\end{figure*}
\begin{figure}
\centering
    \includegraphics[width=0.49\textwidth]{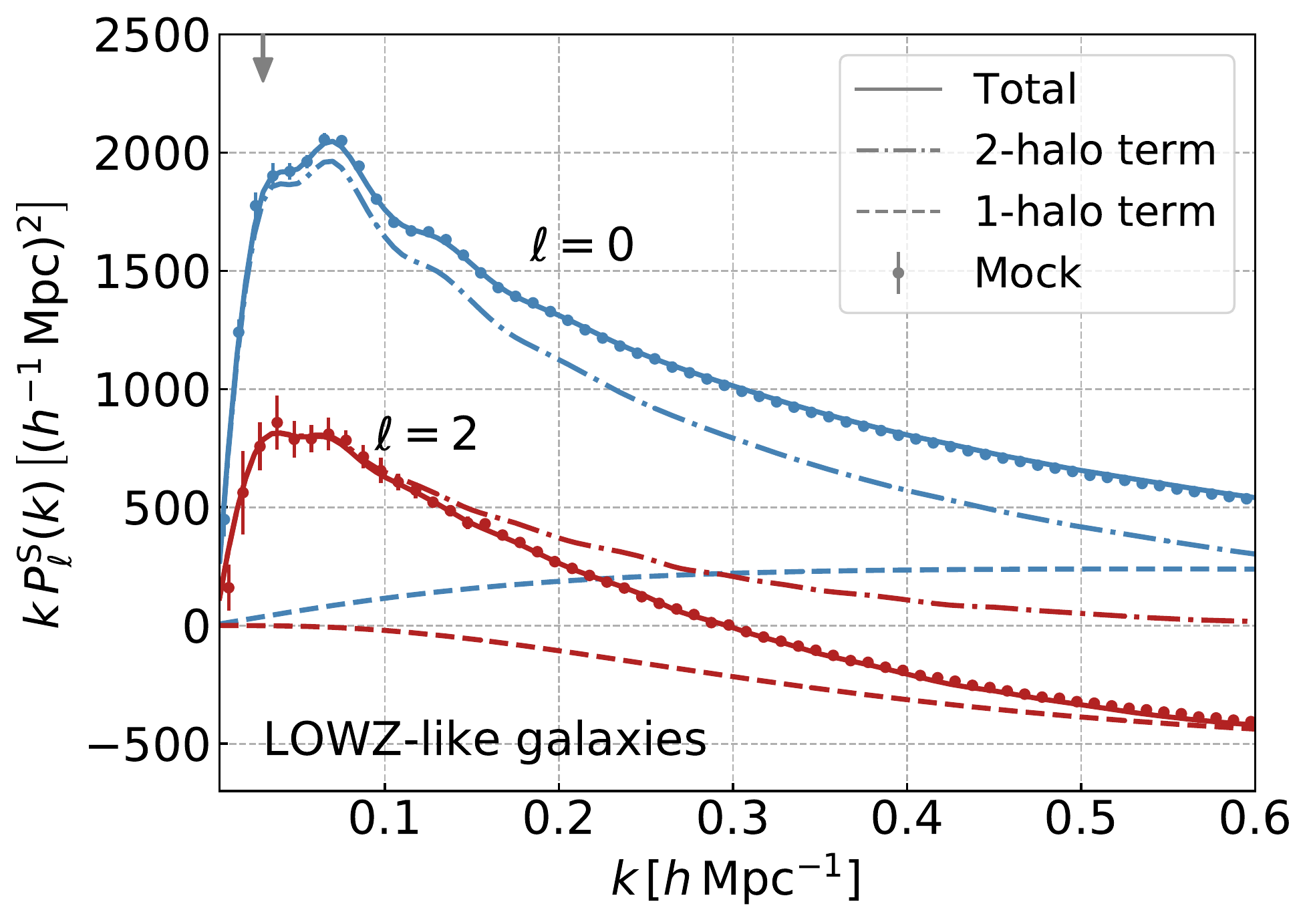}
\caption{
The data points with error bars and the solid curves 
are the same as those for the LOWZ-like sample in Fig.~\ref{fig:power_emu_lowz}. The dashed and dot-dashed lines are the one- and two-halo term contributions to the total power for the monopole and quadrupole moments, respectively. 
}
\label{fig:power_emu_lowz_each_term}
\end{figure}
\begin{figure*}
\centering
  \begin{tabular}{c}
    \begin{minipage}{0.5\hsize}
    \centering
    \includegraphics[width=0.99\textwidth]{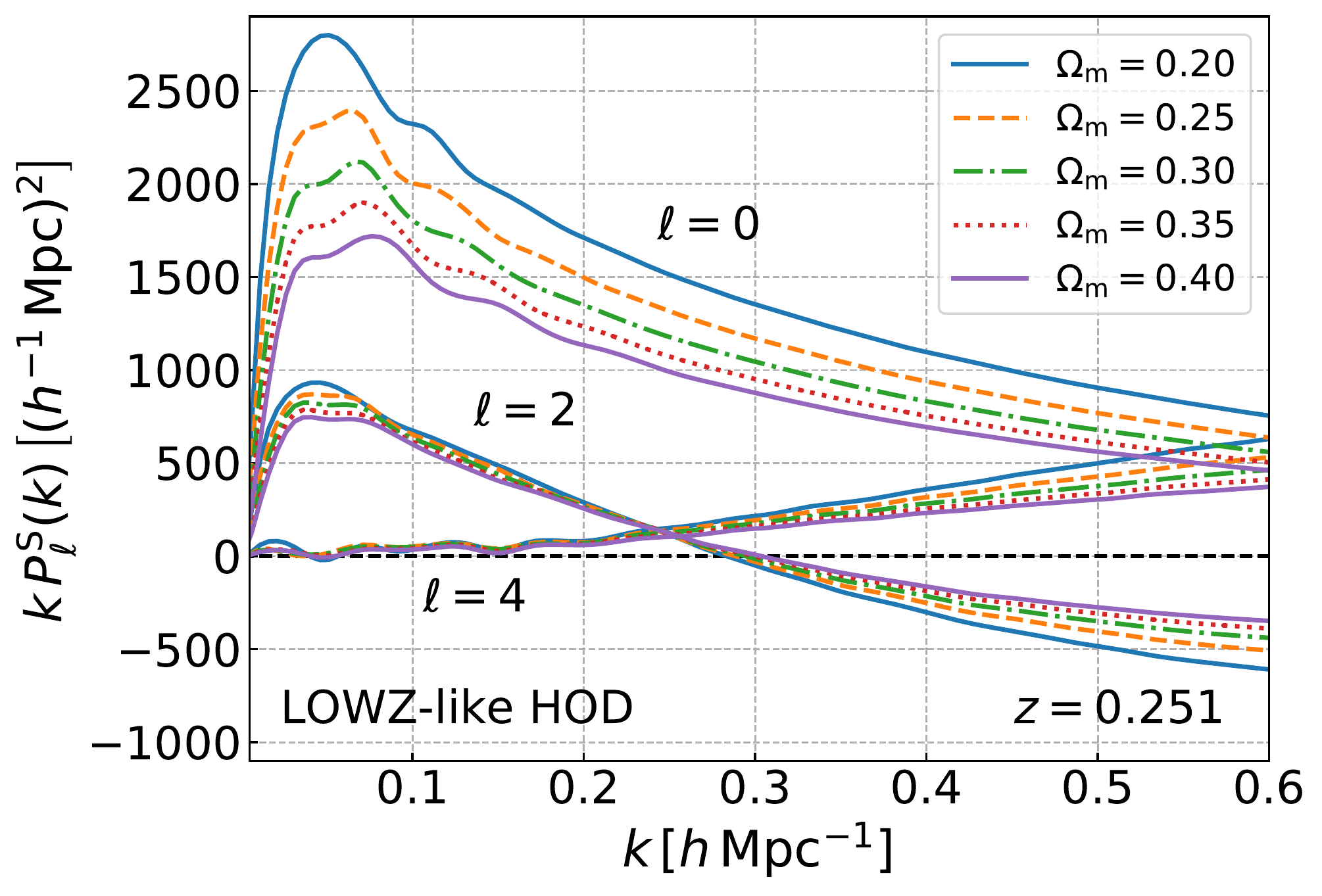}
    \end{minipage}
    \begin{minipage}{0.5\hsize}
    \centering
    \includegraphics[width=0.99\textwidth]{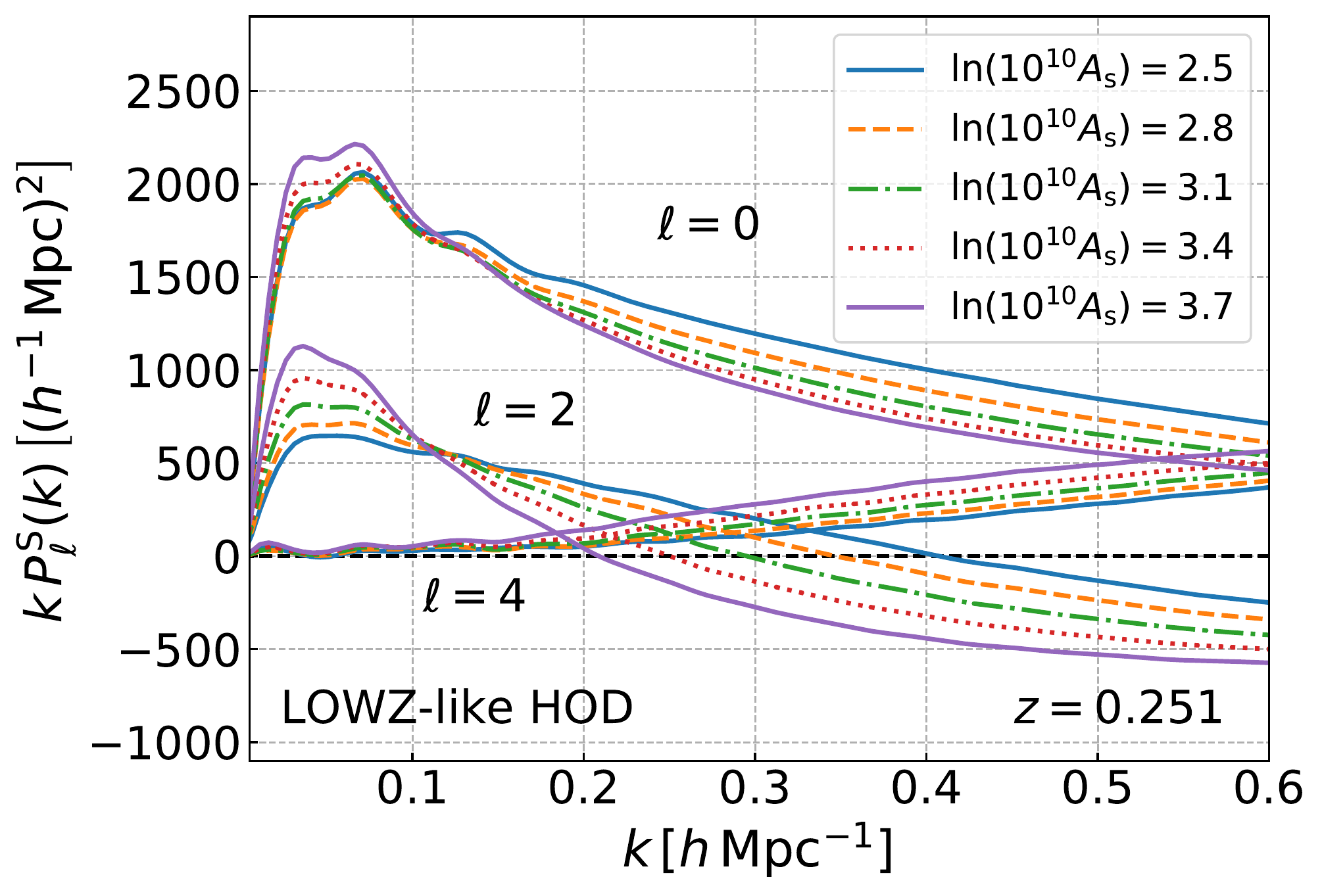}
    \end{minipage}
  \end{tabular}
\caption{
A demonstration for the use of the emulator. 
Here we use the emulator to study variations in the multipole moments of redshift-space galaxy power spectrum for cosmological models with varying $\Omega_{\rm m}$ or $A_{\rm s}$ for LOWZ-like galaxies as in Fig.~\ref{fig:power_emu_lowz}.
Other model parameters, besides a varied parameter ($\Omega_{\rm m}$ or $A_{\rm s}$), are kept to their fiducial values.
}
\label{fig:power_emu_lowz_cosmo_varied}
\end{figure*}
We now present a demonstration of the application of our emulator to predicting the redshift-space power spectrum of galaxies.
To do this, we consider galaxy samples from the SDSS-III Baryon Oscillation Spectroscopic Survey (BOSS) as a working example.
More specifically, we use the same mock catalogs as used in Ref.~\citep{PhysRevD.101.023510} that mimic the LOWZ sample at $z=0.251$ and the CMASS sample at $z=0.484$ \citep{2013AJ....145...10D}.
The survey volumes for these samples $V_s \simeq 1.0~(h^{-1}\,{\rm Gpc})^3$ for the fiducial {\it Planck} cosmology.
Here we populate galaxies into halos in each simulation realization using the specific halo model ingredients, as also given in Appendix~\ref{sec:hod}, and will compare the emulator predictions for the redshift-space power spectra using the same halo model with the spectra measured from the mock catalogs.
Although details of the model ingredients are not essential for the demonstration, we use up to 18 model parameters for each galaxy sample: 
six cosmological parameters, two parameters to model the AP effect, 
and 10 parameters to model the halo-galaxy connection (see Table~II in \citet{PhysRevD.101.023510}).
Among these, 12 parameters (the parameters besides the cosmological parameters)
are different for the LOWZ and CMASS galaxy samples.
Even for this fairly complex model, our emulator enables one to compute the redshift-space power spectrum of galaxies in about 0.35 seconds on a 2.8 GHz quad-core Intel Core i7 processor including a two-dimensional integral for the two mass variables.

In Fig.~\ref{fig:power_emu_lowz}, we compare the emulator predictions for the multipole moments of the redshift-space galaxy spectrum with those measured from the mock catalogs for SDSS LOWZ- and CMASS-like samples.
Note that, for this result, we did not include the off-centering effect, and will discuss it below separately. 
The figure clearly shows that the emulator fairly well reproduces the mock measurements over all scales up to $k=0.6~h\,{\rm Mpc}^{-1}$. 
To be more quantitative the lower panels show the ratio, compared to the statistical errors for a volume of $8~(h^{-1}\,{\rm Gpc})^3$. 
The agreement is well within the errors, definitely within the expected errors for an actual survey volume of $V_s\simeq 1~(h^{-1}\,{\rm Gpc})^3$, which have a factor of 3 larger errors than those plotted in the figure. 
The lower panel displays a relatively large discrepancy (bias) around $k\simeq 0.3~h\,{\rm Mpc}^{-1}$ for the quadrupole moments due to the zero-crossing in the amplitude. 
The mock measurements are quite computationally expensive; run high-resolution simulations (a few days for each with multiple processors), identify halos, populate galaxies into halos, and then measure the redshift-space power spectrum and the moments. The emulator enables a computation of these galaxy spectra in ${\cal O}(0.1)$ CPU second, and allows for huge improvements in the computation time, more than 6 orders of magnitudes (at least days time scale with multiple CPUs vs. 0.1 seconds with a single CPU).

In Fig.~\ref{fig:power_emu_lowz_each_term}, we show respective contributions of the one- and two-halo terms [Eqs.~(\ref{eq:PS1h}) and (\ref{eq:PS2h})] to the total power of the multipole moments for the LOWZ sample in Fig.~\ref{fig:power_emu_lowz}. 
The one-halo term gives a non-negligible or even significant contribution to each of the moments, starting from quite small-$k$ scales, around $k\simeq 0.05~h\,{\rm Mpc}^{-1}$ for the monopole, and from $k\simeq 0.1~h\,{\rm Mpc}^{-1}$ for the quadrupole, respectively.
The nice agreements between the emulator predictions and the mock measurements 
cannot be realized unless we include the one-halo term contributions even on such large scales. 
Hence, this means that we have to marginalize over the halo-galaxy connection parameters, which preferentially affect the one-halo term, to obtain robust constraints on cosmological parameters.

We can easily use the emulator to study the dependence of the galaxy power spectrum on cosmological parameters.
Figure~\ref{fig:power_emu_lowz_cosmo_varied} shows how the multipole moments of the galaxy power spectrum vary with changes in either $\Omega_{\rm m}$ or $A_{\rm s}$.
Here again we vary $\Omega_{\rm m}$ through $\Omega_{\rm de}$ using the spatial flatness ($\Omega_{\rm m} = 1 - \Omega_{\rm de}$), and the other cosmological parameters in Eq.~(\ref{eq:cosmo_param_range}), besides a varied parameter ($\Omega_{\rm de}$ or $A_{\rm s}$), and the halo-galaxy connection parameters are kept fixed to their fiducial values.
Thus our emulator quite easily enables us to evaluate the sensitivity of the galaxy power spectrum to cosmological parameters, which would be useful to explore an optimal survey design for a galaxy survey.

\begin{figure*}
\centering
  \begin{tabular}{c}
    \begin{minipage}{0.33\hsize}
    \centering
    \includegraphics[width=0.99\textwidth]{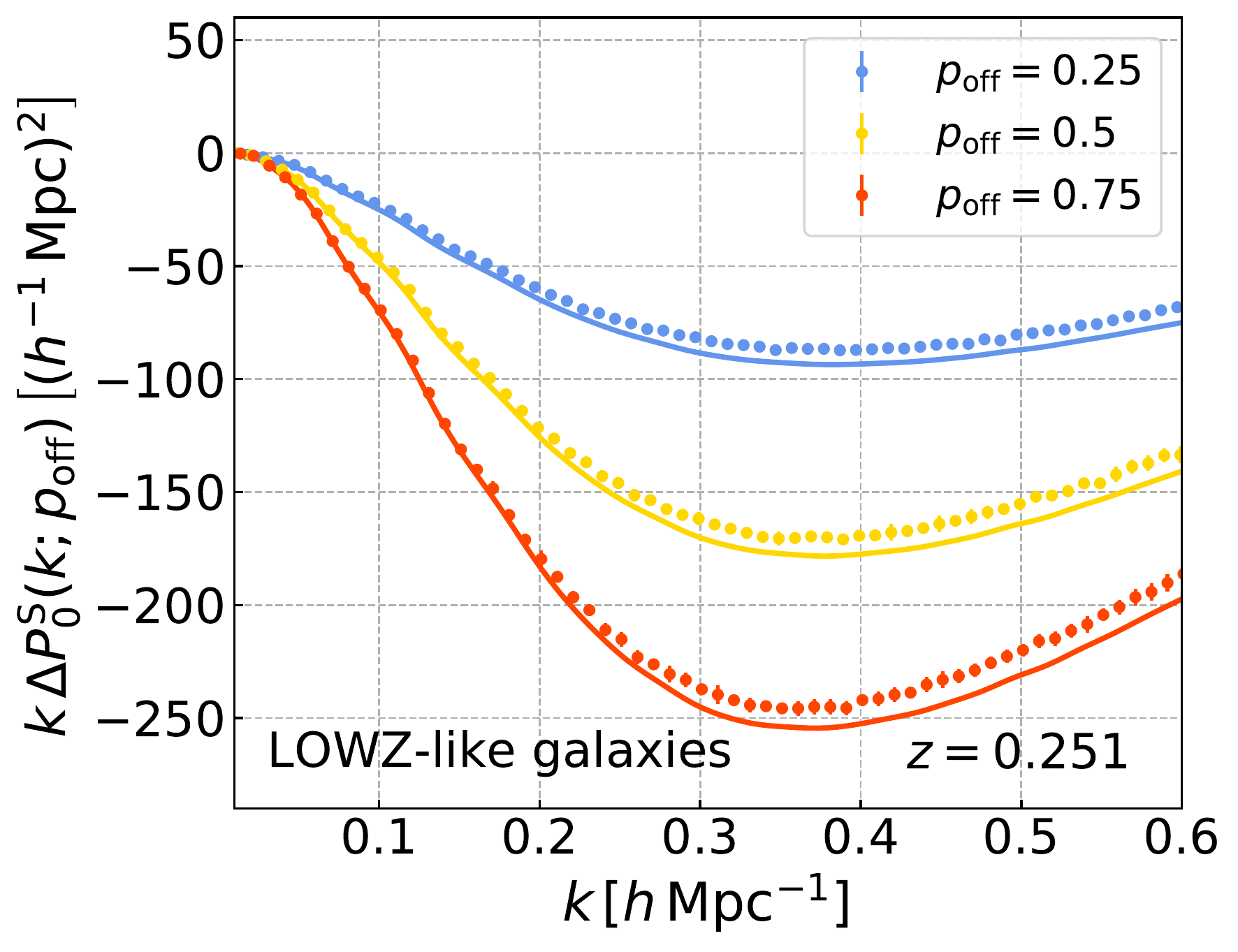}
    \end{minipage}
    
    \begin{minipage}{0.33\hsize}
    \centering
    \includegraphics[width=0.99\textwidth]{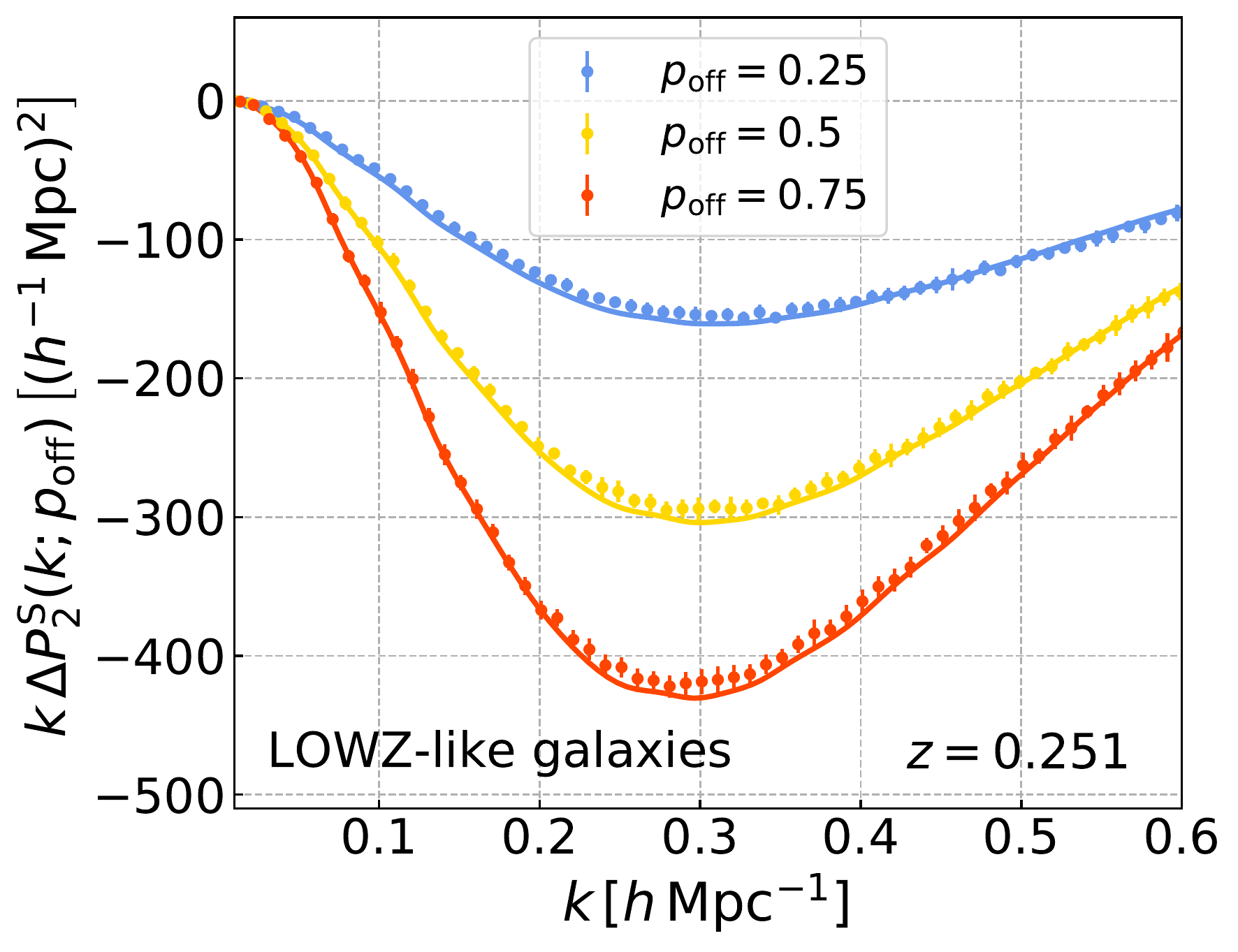}
    \end{minipage}

    \begin{minipage}{0.33\hsize}
    \centering
    \includegraphics[width=0.99\textwidth]{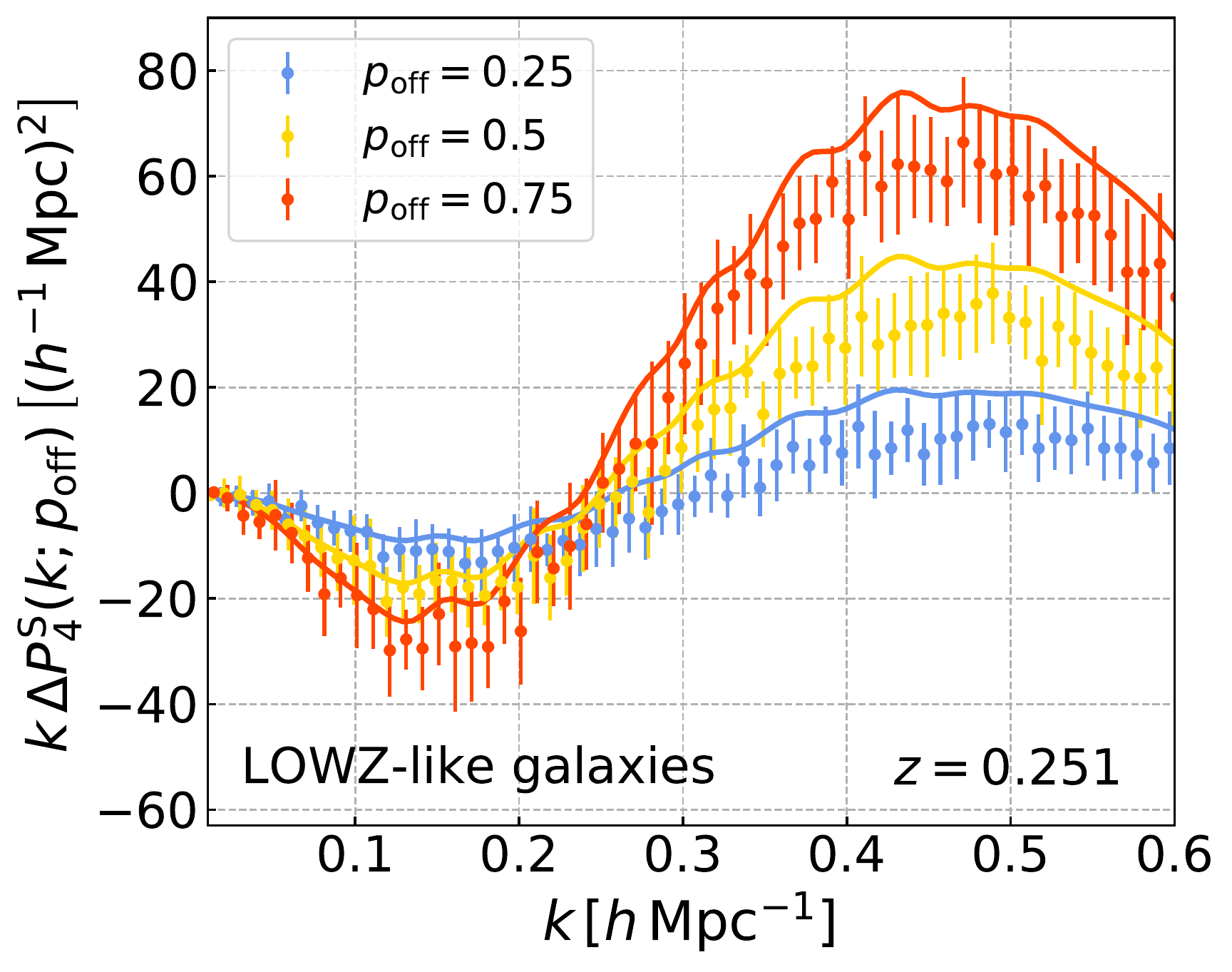}
    \end{minipage} \\
    
    \begin{minipage}{0.33\hsize}
    \centering
    \includegraphics[width=0.99\textwidth]{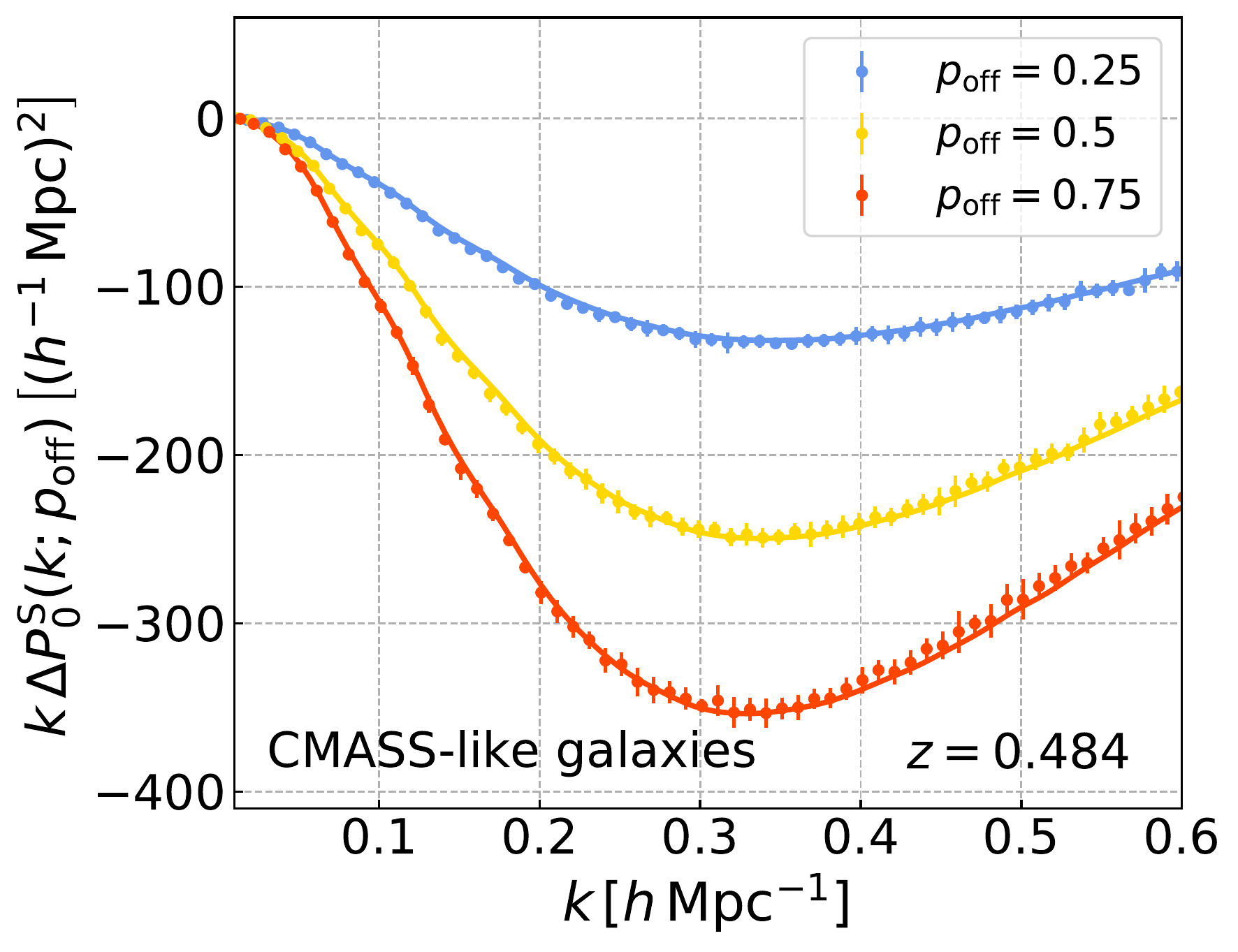}
    \end{minipage}
    
    \begin{minipage}{0.33\hsize}
    \centering
    \includegraphics[width=0.99\textwidth]{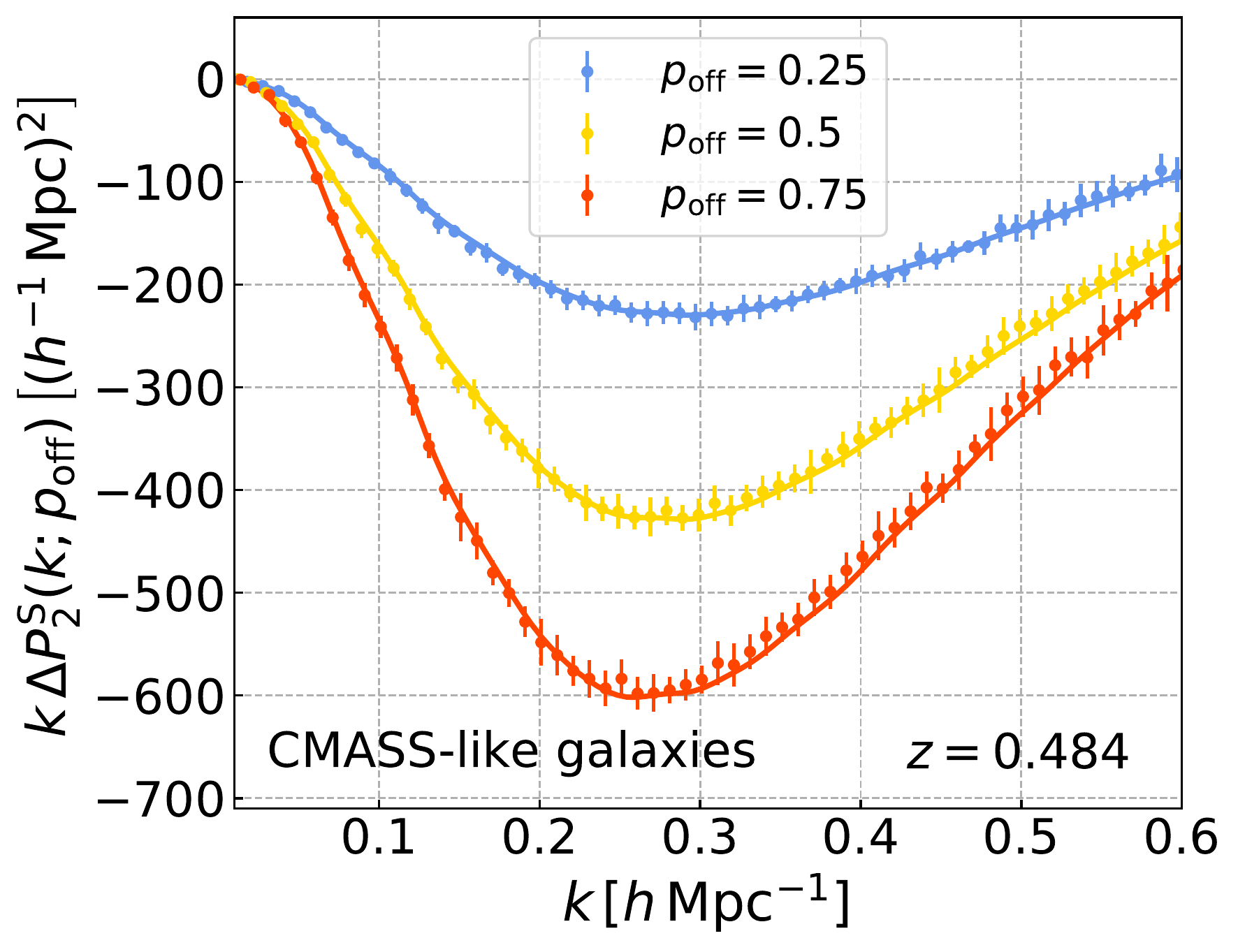}
    \end{minipage}

    \begin{minipage}{0.33\hsize}
    \centering
    \includegraphics[width=0.99\textwidth]{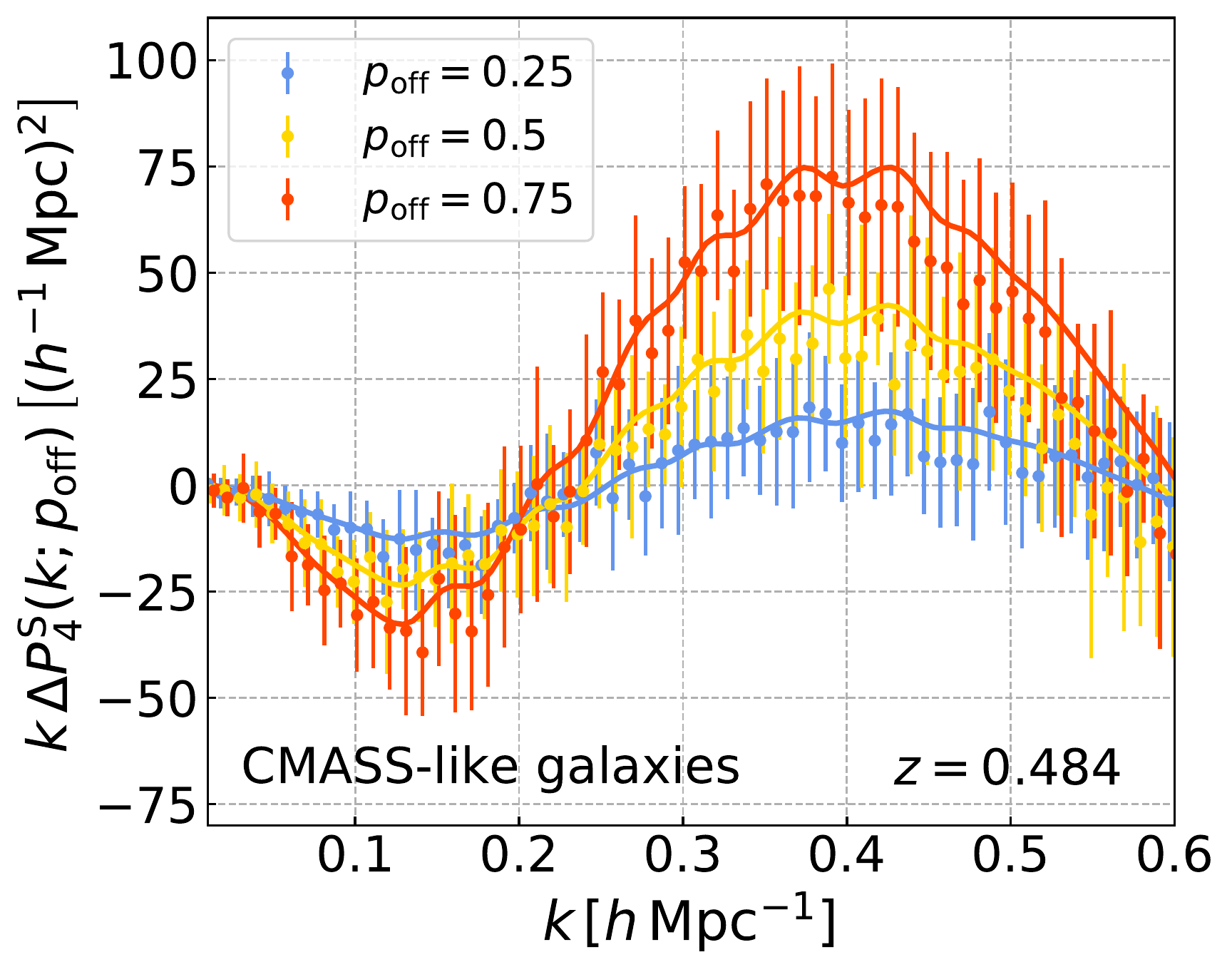}
    \end{minipage}
  \end{tabular}
\caption{
Shown is how the off-centering effects of central galaxies affect the monopole, quadrupole and 
hexadecapole moments for the LOWZ- and CMASS-like galaxies as in Fig.~\ref{fig:power_emu_lowz}. Here we quantify the effects by the differences between 
the power spectra with and without the off-centering effects [see Eq.~(\ref{eq:diff_offcen})].
The solid lines are the emulator predictions, while the symbols with the error bars are the results measured from the mock catalogs.
We model the off-centering effects by the two parameters, $p_{\rm off}$ and ${\cal R}_{\rm off}$, where $p_{\rm off}$ is a fraction of 
central galaxies that are off-centered in host halos of a given mass (here we assume the constant fraction across all halo masses), and 
${\cal R}_{\rm off}$ is a characteristic radius relative to the scale radius of NFW profile for the host halo (see text for details).
We consider three cases for $p_{\rm off}$, as indicated in the legend, and consider ${\cal R}_{\rm off}=2$.
The case without the off-centering effect corresponds to $p_{\rm off}=0$.
}
\label{fig:power_emu_lowz_pl_offcenter}
\end{figure*}

Since dark matter halos are not relaxed nor in dynamical equilibrium and have no clear boundary, there is no unique definition of the halo center. 
Common choices include the potential minimum, the mass density peak, the center of mass of member particles, or the position of massive subhalos. 
Throughout this paper we employ the mass density maximum traced by the center-of-mass position of a certain fraction of innermost particles as a proxy of halo center, as provided as the \textsc{Rockstar} output. 
In addition, central galaxies might have an offset from the halo center (any of the above centers) as a consequence of merger and accretion (e.g., see Fig.~11 in \cite{masaki13}). 
If a galaxy-galaxy weak lensing measurement is available for spectroscopic galaxies used in the redshift-space power spectrum measurements, it might be possible to observationally constrain the off-centering effects \cite{hikage12a,hikage:2013kx}.
In any case, the off-centering effect is uncertain or difficult to accurately model, so a conservative approach would be to include the possible contamination in the model template, which should be marginalized over. 
We here use the emulator to study the impact of off-centering effects on the redshift-space galaxy power spectrum. 
To quantify the impact, we study the differences between the multipole moments of redshift-space power spectrum with and without the 
off-centering effects, defined as
\begin{align}
\label{eq:diff_offcen}
    \Delta P^\mathrm{S}_\ell(k; p_{\rm off}) \equiv P^\mathrm{S}_\ell(k; p_{\rm off}) - P^\mathrm{S}_\ell(k; p_{\rm off}=0),
\end{align}
where $p_{\rm off}$ is a parameter to specify the fraction of central galaxies in halos of a given mass $M$ that are off-centered.
Here we do not consider the halo mass dependence of $p_{\rm off}$; that is, we assumed the same fraction of off-centered galaxies across different host-halo masses. 
For a characteristic off-centering radius, we adopt ${\cal R}_{\rm off}=2$, where ${\cal R}_{\rm off}$ is the parameter to specify the characteristic off-centering radius relative to the scale radius $r_{\rm s}$ of the Navarro-Frenk-White (NFW) profile \cite{NFW}, i.e., we use $R_{\rm off} = {\cal R}_{\rm off} r_{\rm s}$ in Eq.~(\ref{eq:off-center}).

Figure~\ref{fig:power_emu_lowz_pl_offcenter} shows variations in the multipole moments with the different $p_{\rm off}$ values, for the LOWZ- and CMASS-like galaxies. 
We also implemented the same off-centering effects into the mock galaxy catalogs, and then measured the multipole moments from the varied mocks.
The figure shows that the off-centering effects affect the multipole moments at $k \gtrsim 0.1\,h\,{\rm Mpc}^{-1}$, by more than the statistical errors of $8~(h^{-1}\,{\rm Gpc})^3$ volume. 
Our emulator nicely captures the variations in the monopole, quadrupole and hexadecapole moments due to the off-centering effect, although there still remain subtle differences between the predictions and the mock measurements.
We checked that the subtle difference can be resolved by slightly changing the off-centering parameters.
Hence, we would like to suggest that the off-centering parameters need to be included, and then be marginalized over the uncertainties in cosmological analyses, rather than precisely fixing them from theoretical considerations.

\subsection{Alcock-Paczy\'{n}ski effect}

The galaxy power spectrum obtained from the galaxy redshift surveys also contains the geometric information through the AP effect \cite{alcock79,matsubara:1996qy}.
The AP geometrical test offers a unique, powerful probe of cosmological distances at a redshift(s) of a given galaxy survey.
The AP effect is caused by the discrepancy between the true cosmology and the ``reference'' cosmological model assumed in order to convert the measured redshifts and angular positions of galaxies to the three-dimensional comoving coordinates when measuring the clustering statistics.
From this discrepancy the wave vector in the reference frame, $\bk_{\rm ref}$, is related to that in the true frame, $\bk$, by
\begin{align}
  k_{{\rm ref},\perp} = \frac{D_{\rm A}(z)}{D_{\rm A, ref}(z)} k_\perp, \hspace{1em} k_{{\rm ref},\parallel} = \frac{H_{\rm ref}(z)}{H(z)} k_\parallel,
\end{align}
where the subscripts $\perp$ and $\parallel$ mean the perpendicular and parallel components to the line-of-sight direction, $D_{\rm A}(z)$ and $H(z)$ are the angular diameter distance and Hubble parameter at redshift $z$ where the target galaxies reside, respectively.
Quantities with subscript ``ref'' denote those for the reference cosmology.
This geometric distortion induces an additional apparent anisotropy in the redshift-space galaxy clustering, and the redshift-space galaxy power spectrum including the AP effect can be given as
\begin{align}
P^\mathrm{S}_{\rm gg,ref}(k_{{\rm ref},\parallel},k_{{\rm ref},\perp})&=
\frac{1}{\alpha_\perp^2\alpha_\parallel}
P^\mathrm{S}_\mathrm{gg}(k_{\parallel},k_\perp),
\label{eq:PS_gg_AP}
\end{align}
where $\alpha_\perp$ and $\alpha_\parallel$ are the distortion parameters defined as
\begin{align}
\label{AP_params}
\alpha_\perp \equiv \frac{D_{\rm A}(z)}{D_{\rm A, ref}(z)}, \hspace{1em} \alpha_\parallel \equiv \frac{H_{\rm ref}(z)}{H(z)}.
\end{align}
Equivalently, this effect can also be represented in terms of $(k,\mu)$ as
\begin{align}
\label{eq:PS_gg_AP_kmu}
  P^\mathrm{S}_{{\rm gg,ref}}(k_{\rm ref},\mu_{\rm ref}) = \frac{1}{\alpha^2_\perp\alpha_\parallel} P^\mathrm{S}_\mathrm{gg}\left [k(k_{\rm ref}, \mu_{\rm ref}), \mu(\mu_{\rm ref}) \right ],
\end{align}
where
\begin{align}
&k(k_{\rm ref}, \mu_{\rm ref}) \equiv \sqrt{k_\parallel^2+k_\perp^2}= k_{\rm ref} \frac{1}{\alpha_\perp} \left [1+\mu_{\rm ref}^2 \left (\frac{\alpha_\perp^2}{\alpha_\parallel^2}-1 \right ) \right ]^{1/2}, \\
&\mu(\mu_{\rm ref}) \equiv \frac{k_\parallel}{k}=  \mu_{\rm ref} \frac{\alpha_\perp}{\alpha_\parallel}  \left [1+\mu_{\rm ref}^2 \left (\frac{\alpha_\perp^2}{\alpha_\parallel^2}-1 \right ) \right ]^{-1/2}.
\label{eq:kmu_AP}
\end{align}
The parameter combination, $\alpha_\perp^2 \alpha_\parallel$, is often referred to as the isotropically averaged shift that is related 
to the spherically averaged distance $D_V(z)$ \cite{2017MNRAS.470.2617A}; the parameter handles the isotropic dilation of the BAO feature and the overall amplitude of the power spectrum.
On the other hand, $\alpha_\perp/\alpha_\parallel$ is called the AP parameter.
While different multipole moments of the power spectrum are entangled through the AP effect, our emulator enables to easily include the AP effect, because it is designed to predict the two-dimensional power spectrum
 $P_{\rm hh}^{\rm S}(k,\mu)$.
After computing Eq.~(\ref{eq:PS_gg_AP_kmu}), one can also obtain the moments of power spectra by numerically integrating Eq.~(\ref{eq:PS_gg_AP_kmu}) over $\mu_{\rm ref}$ weighted by the Legendre polynomials ${\cal L}_\ell(\mu_{\rm ref})$:
\begin{align}
P^\mathrm{S}_{{\rm gg, ref},\ell}(k_{\rm ref}) = \frac{2\ell+1}{2\alpha^2_\perp\alpha_\parallel } \int_{-1}^1 {\rm d}\mu_{\rm ref}~P^\mathrm{S}_\mathrm{gg}\left [k(k_{\rm ref}, \mu_{\rm ref}), \mu(\mu_{\rm ref}) \right ]
{\cal L}_\ell(\mu_{\rm ref}).
\label{eq:PS_l_AP}
\end{align}

\begin{figure*}
\centering
  \begin{tabular}{c}
    \begin{minipage}{0.33\hsize}
    \centering
    \includegraphics[width=0.99\textwidth]{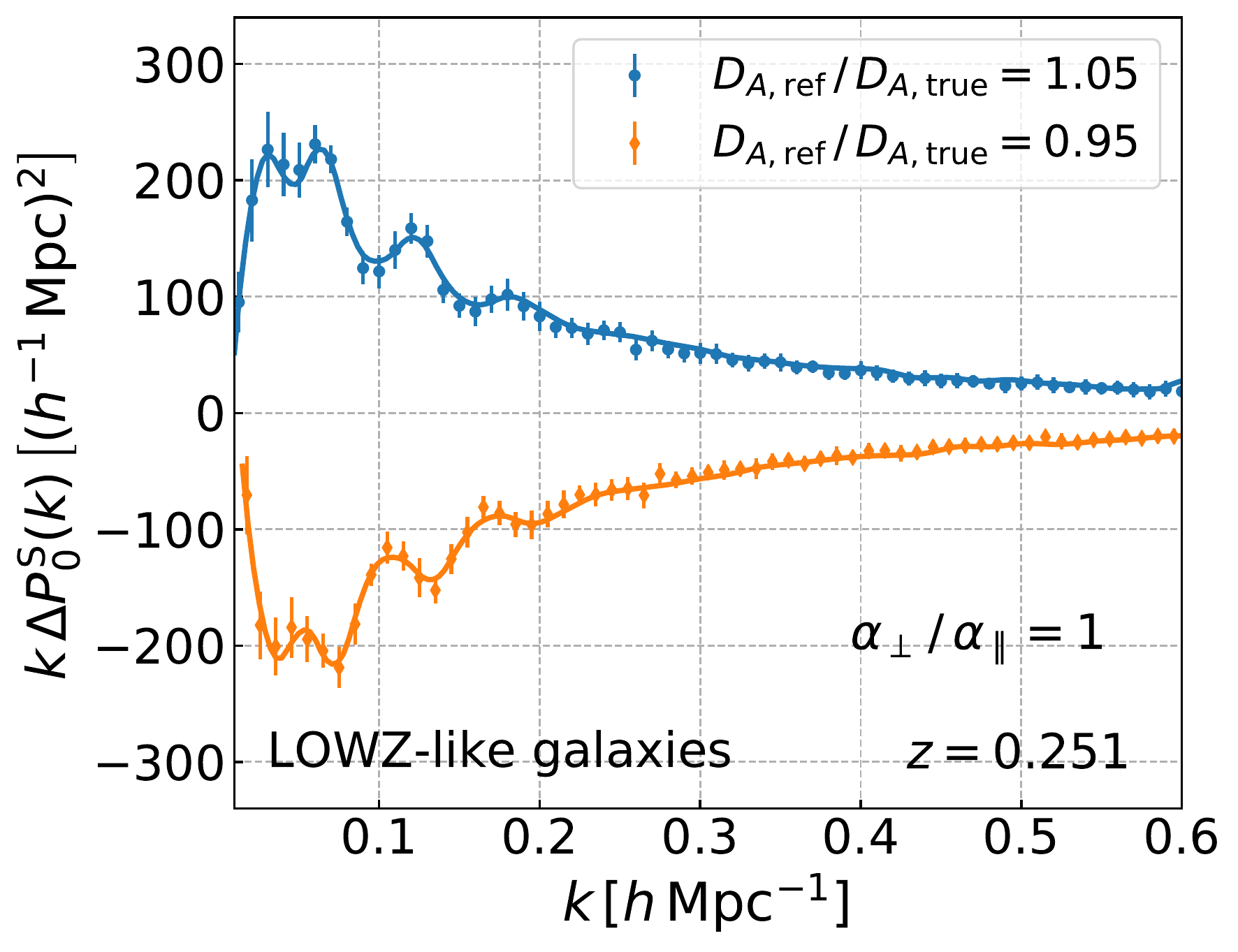}
    \end{minipage}
    
    \begin{minipage}{0.33\hsize}
    \centering
    \includegraphics[width=0.99\textwidth]{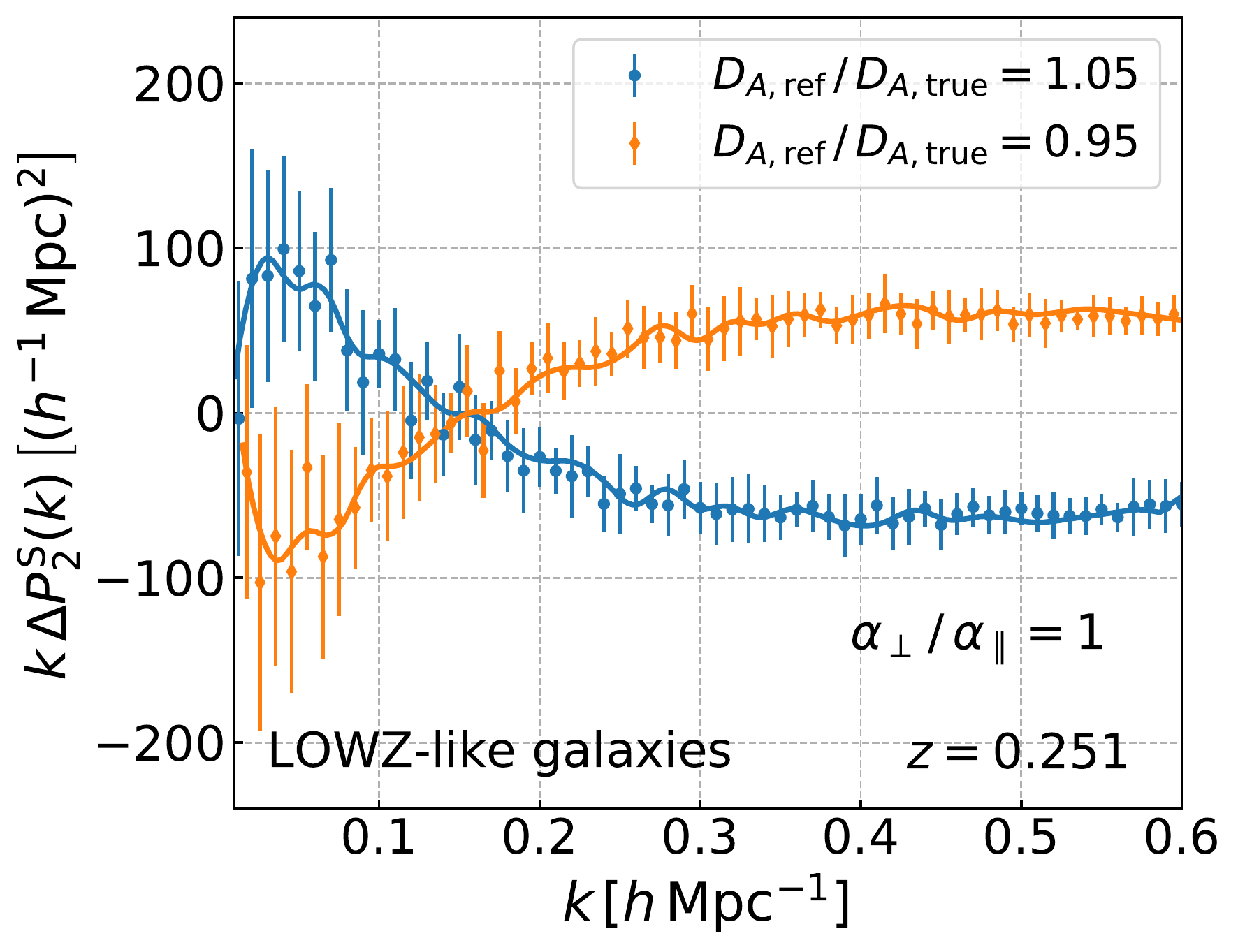}
    \end{minipage}

    \begin{minipage}{0.33\hsize}
    \centering
    \includegraphics[width=0.99\textwidth]{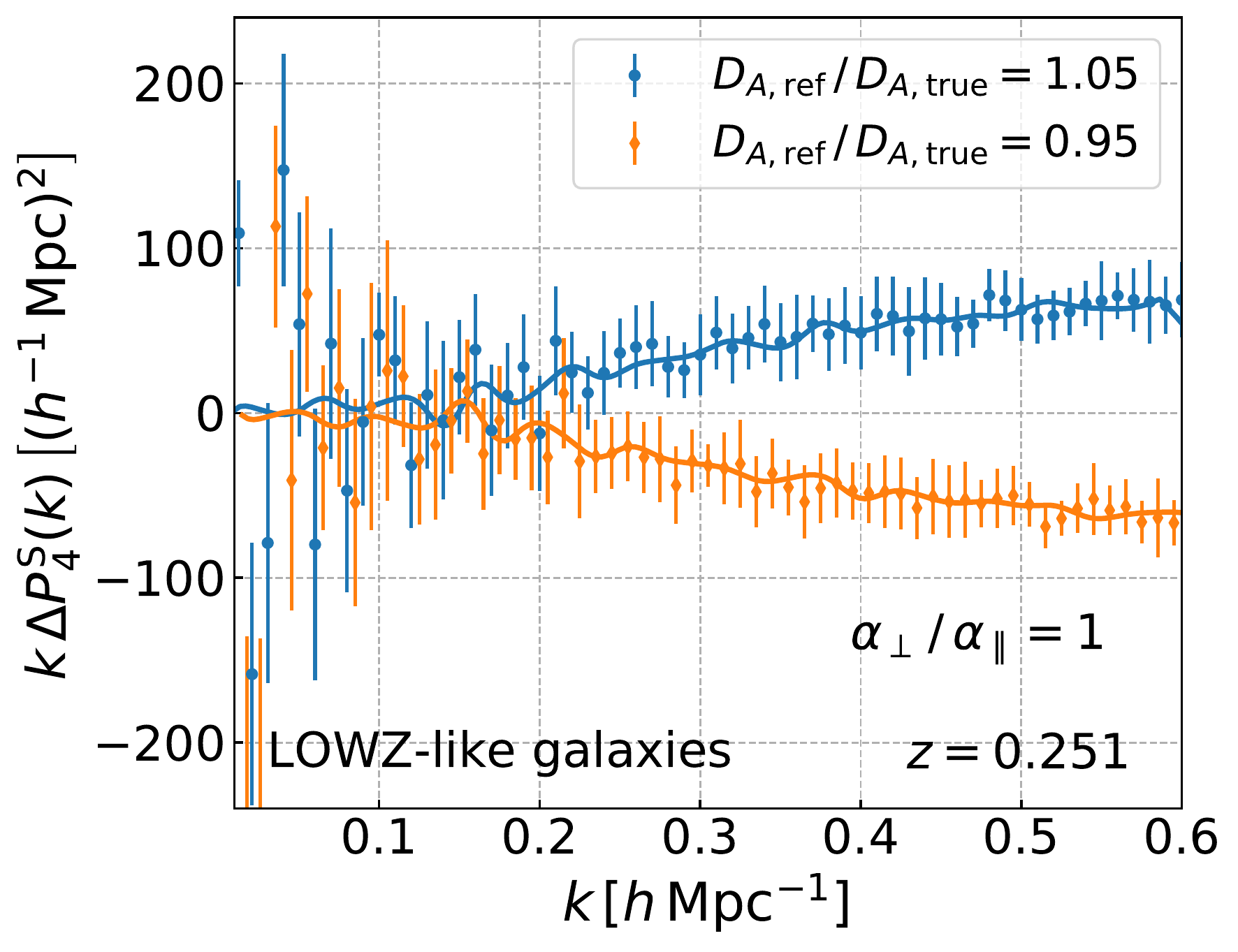}
    \end{minipage} \\

    \begin{minipage}{0.33\hsize}
    \centering
    \includegraphics[width=0.99\textwidth]{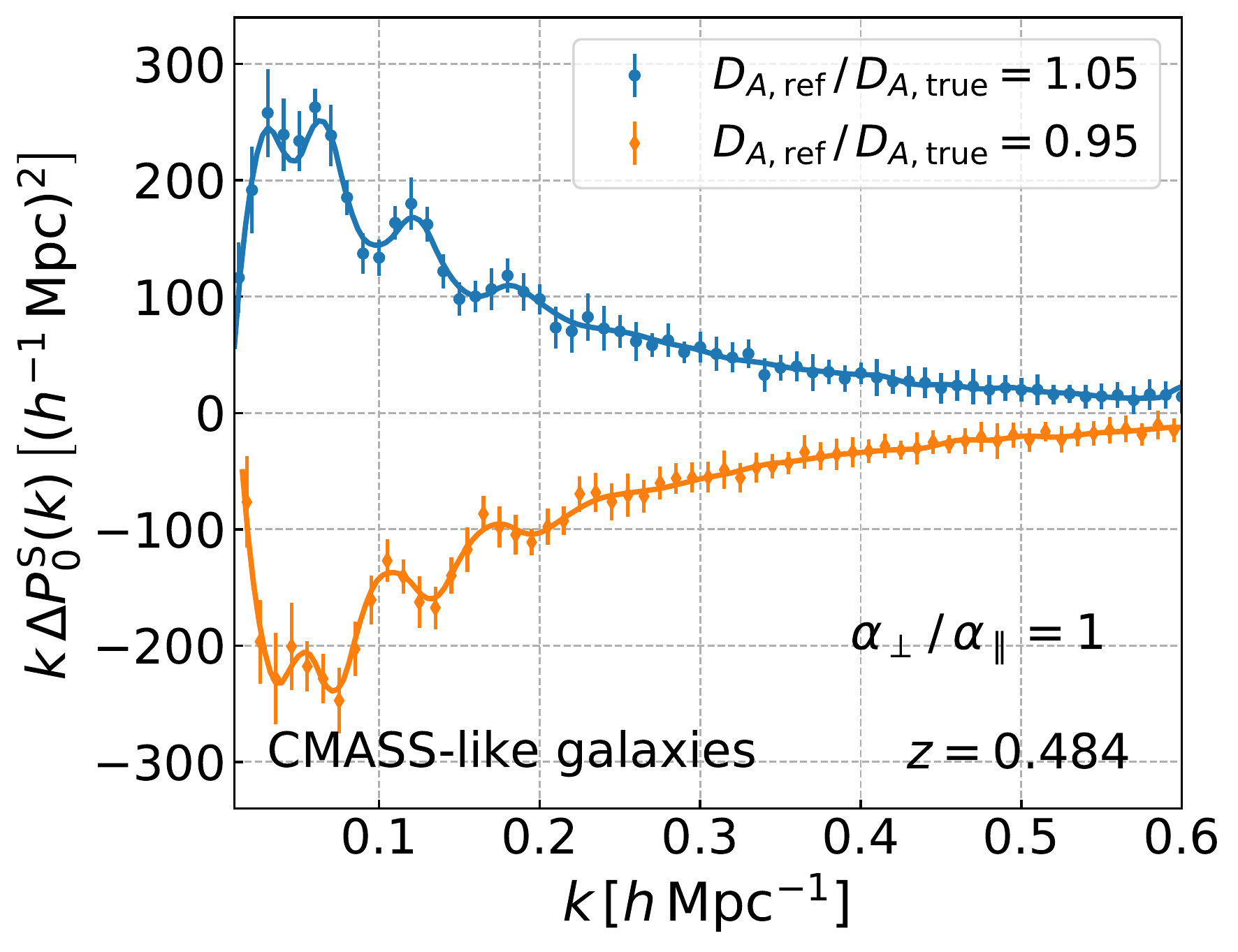}
    \end{minipage}
    
    \begin{minipage}{0.33\hsize}
    \centering
    \includegraphics[width=0.99\textwidth]{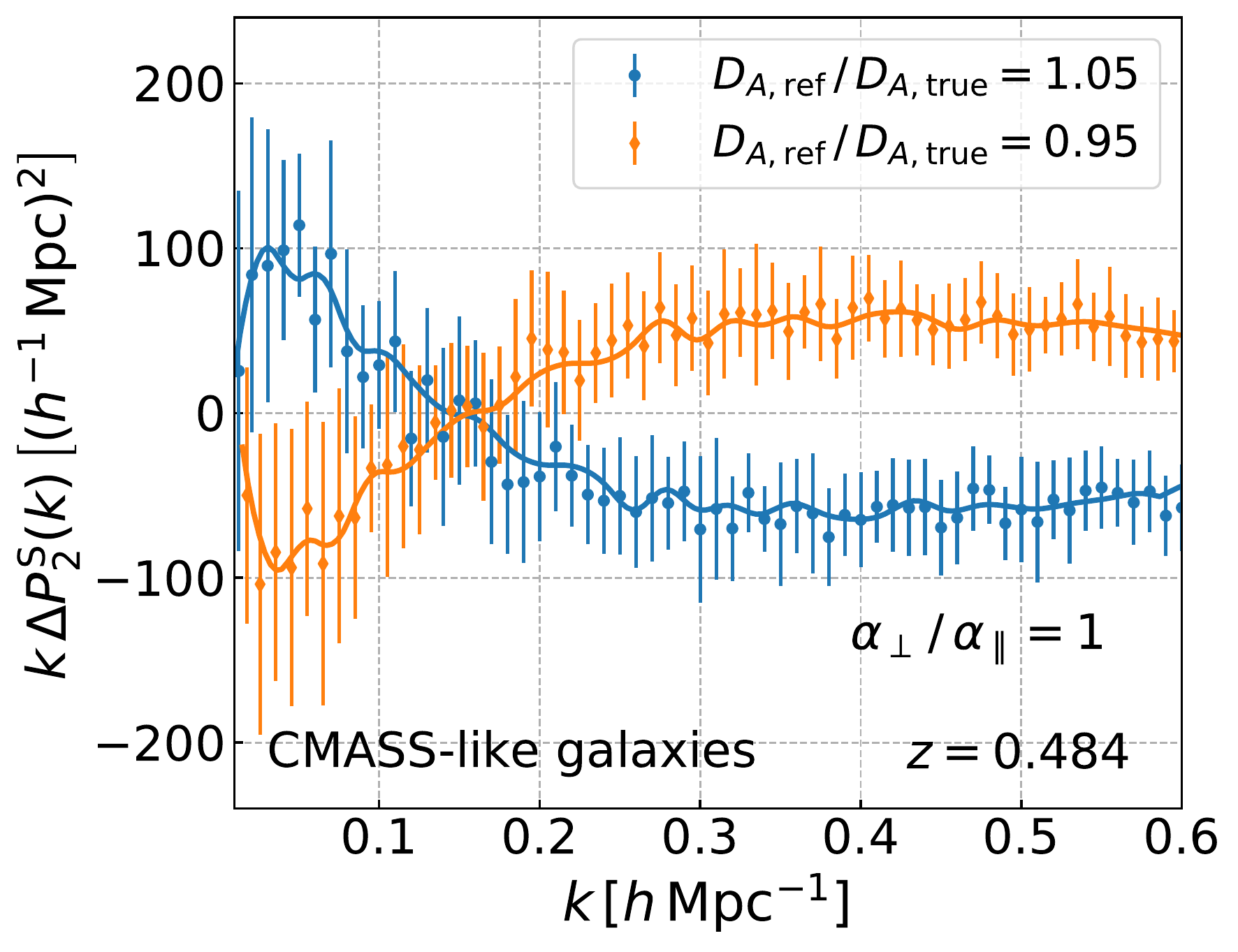}
    \end{minipage}

    \begin{minipage}{0.33\hsize}
    \centering
    \includegraphics[width=0.99\textwidth]{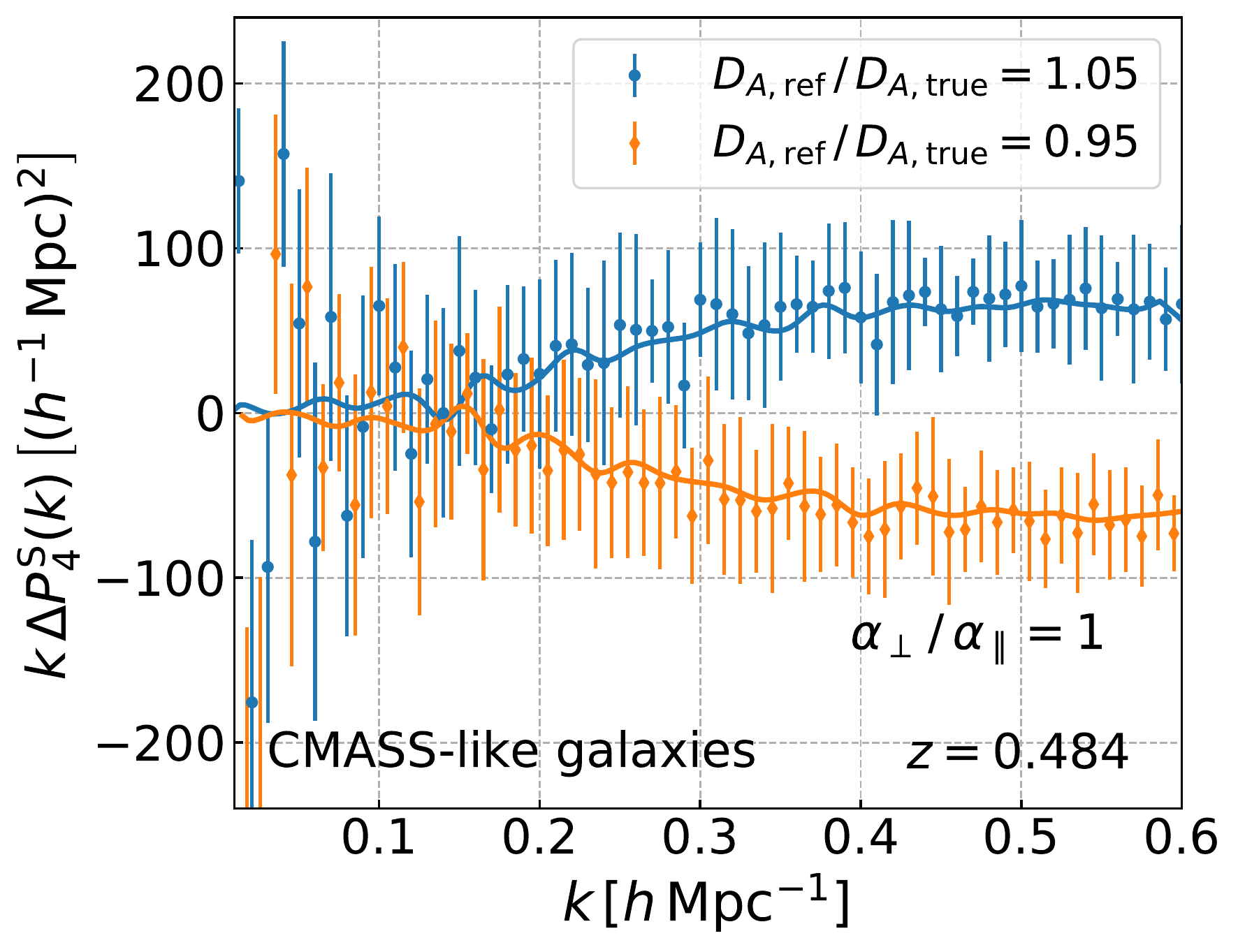}
    \end{minipage}
  \end{tabular}
\caption{
The AP distortion effect on the monopole and quadrupole moments in the galaxy power spectrum for the LOWZ- and CMASS-like galaxies. 
Here we vary either the angular diameter distance $D_{A,{\rm ref}}(z)$ or the Hubble expansion rate $H_{\rm ref}(z)$ at the redshift of galaxy sample
by $\pm 5\%$, keeping the AP distortion parameter $\alpha_\perp/\alpha_\parallel \propto D_A(z)H(z)$ to their true value (i.e. $\alpha_\perp/\alpha_\parallel=1$), 
where $D_{A,{\rm ref}}$ and $H_{\rm ref}$ are the quantities for the reference cosmology that is adopted in the clustering analysis.
The blue and orange solid lines are the emulator predictions, while the symbols with error bars are the results measured from the simulations including the AP distortion. 
The errors are the statistical errors expected for measurements of each moment 
 for a volume of $8~(h^{-1}\,{\rm Gpc})^3$.
}
\label{fig:power_emu_lowz_pl_AP}
\end{figure*}

Figure~\ref{fig:power_emu_lowz_pl_AP} shows how the AP distortion affects the monopole, quadrupole, and hexadecapole moments of the galaxy power spectrum, for the LOWZ- and CMASS-like galaxies. 
We compare the emulator predictions with the measurements from the mock catalogs. We used the same method in Ref.~\cite{PhysRevD.101.023510} to include the AP effects in the mock catalogs.
For illustrative purpose, we here study the differences between the emulator predictions and the mock measurements, defined in a similar way to 
Eq.~(\ref{eq:diff_offcen}).
The AP effect has 2 degrees of freedom in the dependences of the galaxy power spectrum, e.g., the angular diameter distance $D_{\rm A}$ and the Hubble parameter $H$. 
Here we focus on variations in the moments with varying either of $D_{\rm A}$ or $H$ with keeping the AP parameter ($\alpha_\perp/\alpha_\parallel$)
fixed to the fiducial value (its true value).
The figure clearly shows that our emulator well describes the AP distortions in the monopole, quadrupole, and hexadecapole moments at equal accuracies to the mock measurements. Our emulator allows for the computation of these moments in ${\cal O}(0.1)$ CPU second.

All the evaluations of our emulator for the galaxy power spectrum in comparison with the mock measurements are quite encouraging. We conclude that our emulator is ready to apply to actual measurements from galaxy redshift surveys such as the BOSS surveys.

\section{Conclusion}
\label{sec:conclusion}

We have developed an emulator of the redshift-space power spectrum of halos, based on the $N$-body simulations and a machine learning-based technique. 
The fast and accurate prediction of the power spectrum in a multi-dimensional parameter space requires an efficient way for the regression of the simulation data, and for this we adopted a feed-forward neural network with a simple structure.
In the six-dimensional parameter space of flat-geometry $w$CDM cosmology and the redshift range $[0,1.48]$ investigated in the \textsc{Dark Quest} simulation suite, the trained network can provide the halo power spectrum $P^\mathrm{S}_\mathrm{hh}(k,\mu)$ for a given halo mass threshold above $10^{12}\,h^{-1}\,M_\odot$, in several CPU milliseconds.
The prediction accuracy of the emulator is shown to be about 1\%--5\% for the monopole and quadrupole moments of the power spectrum for halos with number density $n_{\rm h} = 10^{-4} \,(h^{-1}\,{\rm Mpc})^{-3}$, which is roughly comparable to the number density of galaxies targeted in the SDSS-III spectroscopic survey.

We demonstrated that we can combine the emulator outputs with the HOD prescription to obtain model predictions for the redshift-space galaxy power spectrum for a galaxy sample of interest.
Since our emulator outputs the redshift-space power spectrum in the form of $P^{\rm S}(k,\mu)$, instead of the multipole moments, it allows one to easily incorporate the Finger-of-God effects due to the random motions of galaxies inside host halos, 
the off-centering effects, and the AP distortions, where these effects generally mix contributions of different multipole moments to a given multipole.
As a working example, we used the HOD models for the SDSS-III LOWZ- and CMASS-like galaxies to obtain the redshift-space galaxy power spectra using the emulator outputs.
We showed that the emulator predictions well match the power spectra measured from the simulation-based mock catalogs that are generated using the same HOD and the same spatial and velocity distributions of galaxies inside halos.
Our emulator can compute the galaxy power spectrum in $\mathcal{O}(0.1)$ CPU seconds, which corresponds to a huge reduction in the computation time compared to 
the brute-force method (constructing the galaxy power spectrum from the simulation-based mock catalogs).

We are planning to perform the cosmological parameter inference by comparing the emulator-based theoretical template with the measurements from the BOSS galaxy data. 
We want to show how the parameter constraints can be improved with the increase of the maximum wave number $k_{\rm max}$ used in the parameter inference. 
To do this, it is important to assess whether the derived parameters are not biased compared to the true values, and blinded cosmology challenges, as done in Ref.~\cite{2020arXiv200308277N}, would be needed to fully validate the performance.
We are also planning to make our emulator code public after completing the cosmology challenges.
These are our future works, and will be presented elsewhere.

\smallskip{\em Acknowledgments}
Y.K. thanks to the Yukawa Institute for Theoretical Physics, Kyoto University for the warm hospitality where this work was partly done.
This work was supported in part by World Premier International Research Center Initiative, MEXT, Japan, and JSPS KAKENHI Grant No.~JP15H03654,
No.~JP15H05887, No.~JP15H05893, No.~JP15K21733, No.~JP17H01131, No.~JP17K14273, No.~JP19H00677, and No.~JP20H04723, by Japan Science and Technology Agency (JST) CREST JPMHCR1414, and by JST AIP Acceleration Research Grant No.~JP20317829, Japan.
Y.K. is also supported by the Advanced Leading Graduate Course for Photon Science at the University of Tokyo.
K.O. is supported by JSPS Overseas Research Fellowships.
The $N$-body simulations and subsequent halo-catalog creation in the \textsc{Dark Quest} simulation suite used in this work were carried out on Cray XC50 at Center for Computational Astrophysics, National Astronomical Observatory of Japan.

\appendix

\section{A RESOLUTION STUDY ON THE POWER SPECTRUM MEASUREMENT}
\label{sec:resolution_study}

In Sec.~\ref{subsec:dataset}, we described our settings used in measurements of 
the redshift-space power spectrum of halos that were in turn used as the datasets of emulator building. 
In this appendix, we present a resolution study in the power spectrum measurement.

\begin{figure*}
\centering
  \begin{tabular}{c}
    \begin{minipage}{0.5\hsize}
    \centering
    \includegraphics[width=0.99\textwidth]{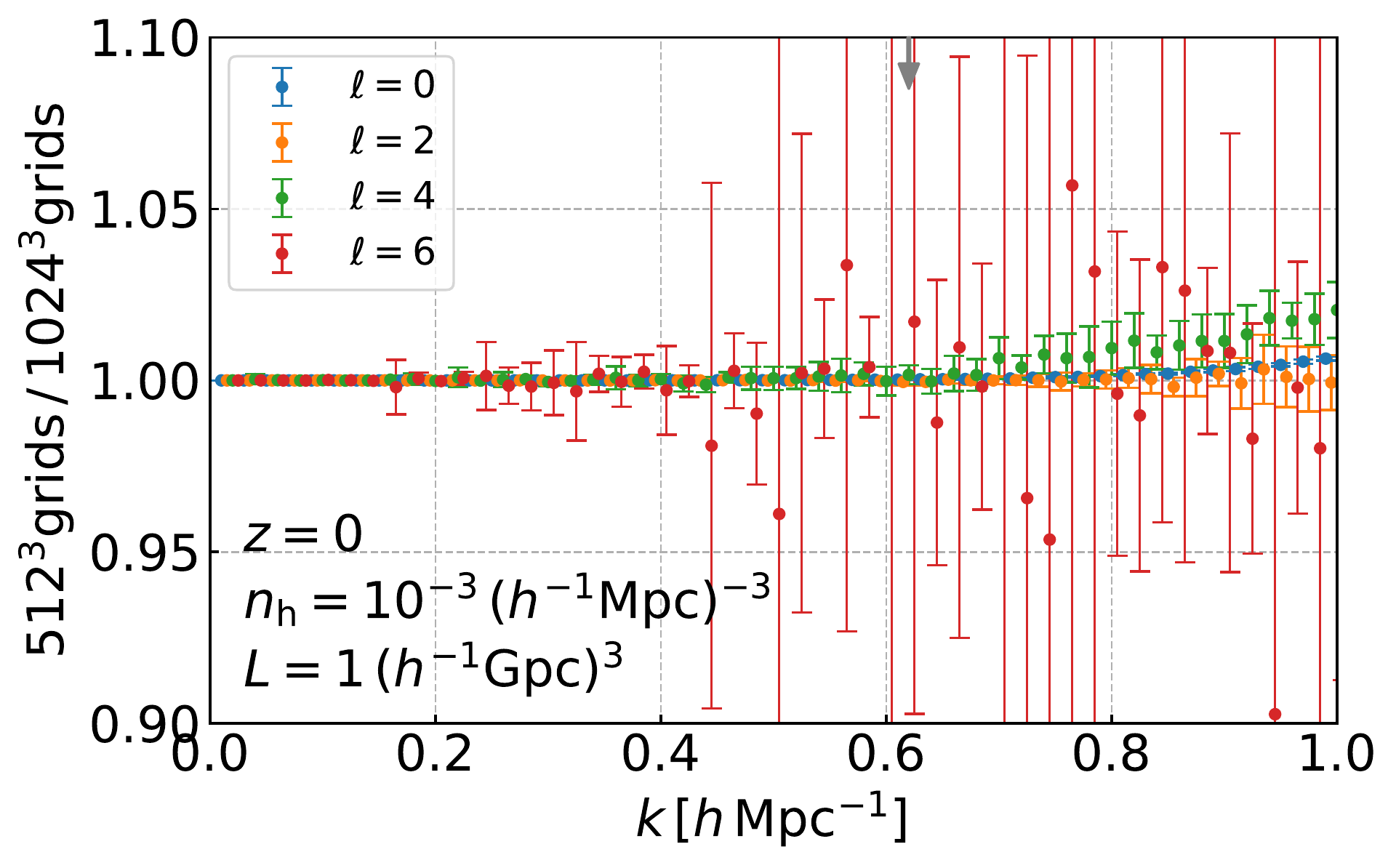}
    \end{minipage}
    
    \begin{minipage}{0.5\hsize}
    \centering
    \includegraphics[width=0.99\textwidth]{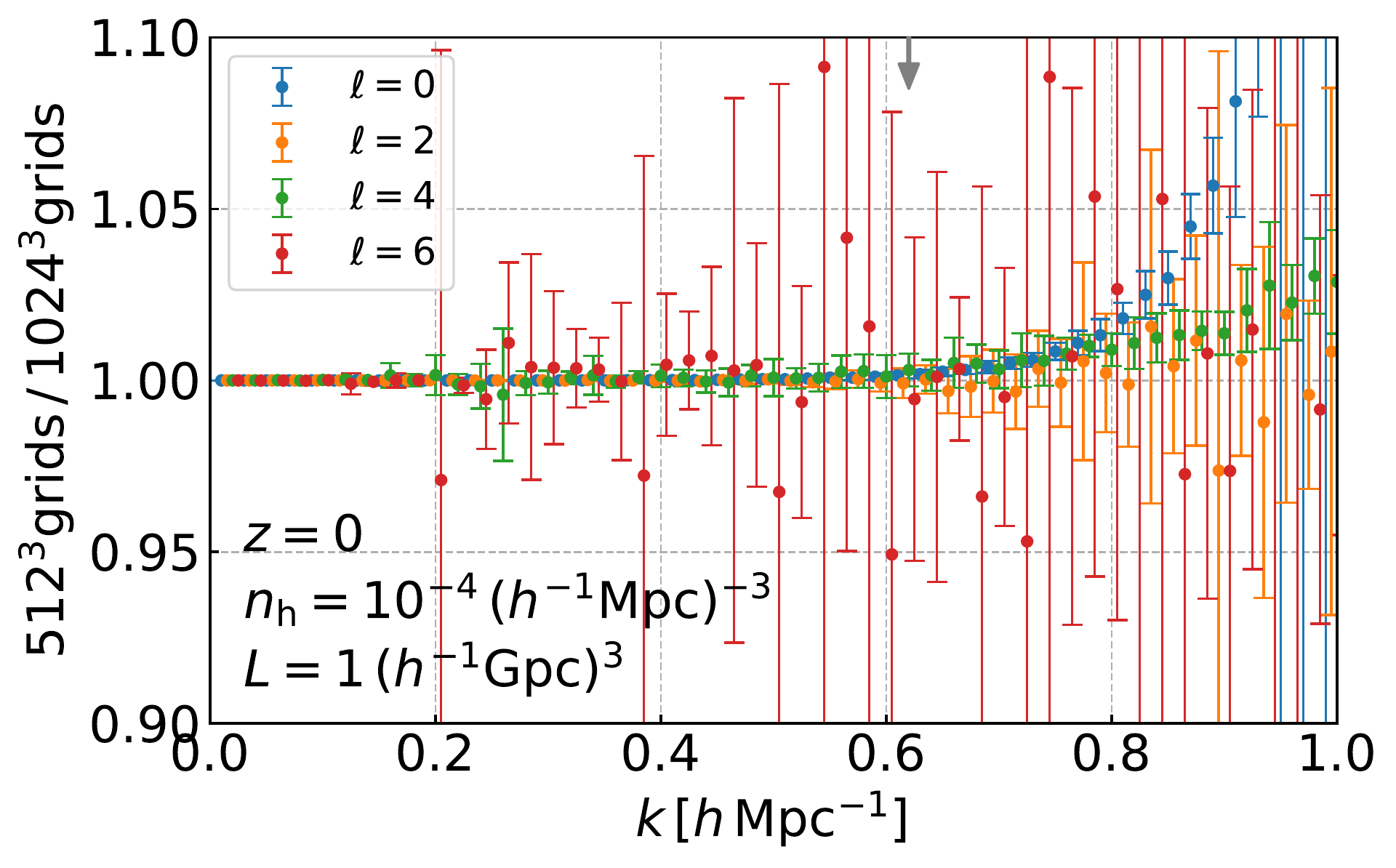}
    \end{minipage}
  \end{tabular}
\caption{
The effect of the FFT grid assignment on the multipole moments of halo power spectrum measured from the simulations.
We compare four multipole moments of degrees $\ell = 0$, 2, 4, and 6 for the {\it Planck} cosmology, between the CIC assignments on $512^3$ grids and on $1024^3$ grids, respectively.
Since we measure from the \textsc{Dark Quest} HR simulations with box size of $1\,h^{-1}\,{\rm Gpc}$, the Nyquist wave number for each setting is $1.61$ and $3.22\,h\,{\rm Mpc}^{-1}$, respectively.
The former FFT resolution is equivalent to our fiducial setting we used for the main results of this paper.
The gray arrow in the upper horizontal axis indicates the maximum wave number of data, $k_{\rm max} = 0.61 \, h \, {\rm Mpc}^{-1}$, which we adopt to construct the emulator.
}
\label{fig:resolution}
\end{figure*}

First, we study how the grid assignment we used affects the power spectrum measurement.
For this purpose we use the \textsc{Dark Quest} HR simulations, which have a box size of $1~h^{-1}{\rm Gpc}$, a halved size of our 
default simulations of $2~h^{-1}\,{\rm Gpc}$.
Figure~\ref{fig:resolution} compares the multipole moments of halo power spectrum using the CIC assignment with different number of grids: $512^3$ grids and $1024^3$, respectively. 
Here the former has the same Nyquist frequency in the FFT computation as that of 
our default setting ($2~h^{-1}\,{\rm Gpc}$ plus $1024^3$).
For both cases we use the interlacing scheme for the aliasing mitigation.
Here we consider the auto power spectrum for two samples of halos with number densities, $n_{\rm h} = 10^{-3}$ and $10^{-4} \, (h^{-1}\,{\rm Mpc})^{-3}$, respectively, at $z=0$ for the {\it Planck} cosmology.
The error bars represent standard deviation among the 15 realizations.
Although the systematics due to the grid assignment is below the errors for the quadrupole and tetra-hexadecapole ($\ell=6$) moments, it significantly affects the monopole and hexadecapole moments at $k \gtrsim 0.7$ or $0.8 \, h \, {\rm Mpc}^{-1}$.
In this paper we use the power spectrum data at $k < 0.61 \, h \, {\rm Mpc}^{-1}$ for the emulator construction, and in this range our default setting accurately 
estimates the multipole moments with the precision better than the statistical errors. 
Hence we conclude that the FFT resolution does not affect our emulator construction.

\begin{figure}
\centering
\includegraphics[width=0.49\textwidth]{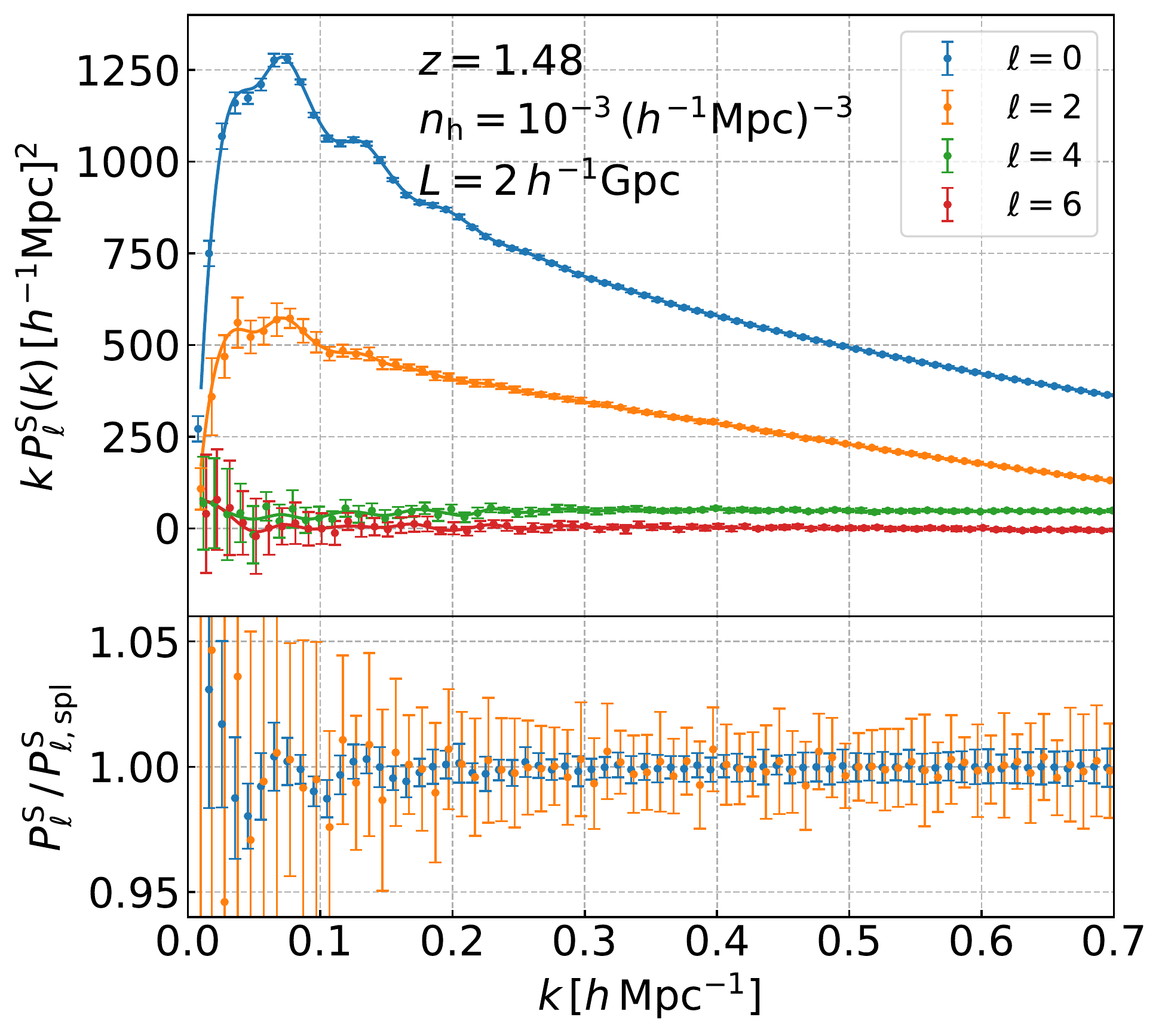}
\caption{
Effects of the $k$-bin width on the multipole moments (up to $\ell=6$)
of halo power spectrum.
The symbols with error bars are the multipole moments measured using the bin width $\Delta k = 0.01 \, h \, {\rm Mpc}^{-1}$.
The solid lines are the results obtained by the spline interpolations of the moments using the bin width $\Delta k = 0.02 \, h \, {\rm Mpc}^{-1}$.
The error bars are estimated from the standard deviation among the 15 realizations of \textsc{Dark Quest} LR simulations ($2~h^{-1}{\rm Gpc}$ on a side) for the {\it Planck} cosmology.
}
\label{fig:binning_effect}
\end{figure}

Second, we study the impact of a finite $k$ binning in the power spectrum measurements. 
There is a trade-off in a choice of the bin width. 
For a finner bin width, the band power measurement in each $k$ bin becomes noisier due to a smaller number of the Fourier modes, but it can well capture features in the power spectrum. 
For a wider bin width, the measurement becomes less noisy but might erase or smooth features in the power spectrum.
For our emulator construction, we need less noisy datasets to avoid any failure of the machine learning due to too large sample variance.
For this reason, we adopt the $k$-bin width $\Delta k = 0.02 \, h \, {\rm Mpc}^{-1}$, although the analyses using the current-generation galaxy redshift surveys usually employ $\Delta k \simeq 0.01 \, h\,{\rm Mpc}^{-1}$
for the $k$-bin width \cite{Beutler:2016arn}.
Hence it is important to check the effect of our binning on the power spectrum measurement.
In Fig.~\ref{fig:binning_effect}, we compare the power spectrum measured with two different bin widths $\Delta k = 0.01$ and $0.02 \, h \, {\rm Mpc}^{-1}$, respectively.
In the upper panel, the solid line represents the spline interpolation of the power spectrum measured at $\Delta k = 0.02 \, h \, {\rm Mpc}^{-1}$, while the symbols are that measured at $\Delta k = 0.01 \, h \, {\rm Mpc}^{-1}$ for the multipole moment of each order.
The error bars show the variances among 15 realizations of the {\it Planck} cosmology.
Over all the $k$ range we are interested in, the power spectrum measured with finner bins shows almost no significant discrepancy from the case of the wider bins.
A caveat is that the wider binning slightly smears out the BAO features. 
However, the primary purpose of this work is to accurately model the nonlinear clustering effects and the RSD effect in the redshift-space power spectrum, so we most care about an unbiased measurement of the power spectrum amplitudes. 
Thus, Fig.~\ref{fig:binning_effect} shows that our binning scheme well captures the amplitudes of the multipole moments, with the precision better than the statistical errors, up to $k\simeq 0.7~h\,{\rm Mpc}^{-1}$.
Hence, we conclude that our choice $\Delta k = 0.02 \, h \, {\rm Mpc}^{-1}$ meets the requirements.

\section{MULTIPLICATION OF THE CROSS POWER SPECTRUM OF HALOS?}
\label{sec:cross_power_spectrum}

In our emulator, we choose to work on the power spectrum of halos with different number densities (masses): 
$P^{\rm S}_{\rm hh}(\bk; n_1,n_2)$.
In linear theory, the redshift-space power spectrum between halos with mass $M_1$ and $M_2$ can be 
expressed in the multiplicative form as
\begin{align}
  P^\mathrm{S}_{\rm hh, lin}(k,\mu; M_1, M_2) = \left[b_{\rm h}(M_1) + f\mu^2 \right] \left[b_{\rm h}(M_2) + f\mu^2 \right] P_{\rm lin}(k),
\end{align}
where $b_{\rm h}(M)$ is the linear bias of halos with mass $M$.
The standard halo model also assumes that the two-halo term of the halo power spectrum is given by such a multiplicative form as $P_{\rm hh}^{\rm 2h}(k; M_1,M_2)=b(M_1)b(M_2)P^{\rm L}_{\rm hh}(k)$ \citep{2012ApJ...745...16T,More15}.
From this consideration, one might ask whether 
the power spectrum of halos in different mass bins can be approximated by the multiplicative form as
\begin{align}
\label{eq:powerM1M2_lin}
  P^\mathrm{S}_{\rm hh, lin}(k,\mu; M_1, M_2) = \sqrt{P^\mathrm{S}_{\rm hh, lin}(k,\mu; M_1) P^\mathrm{S}_{\rm hh, lin}(k,\mu; M_2)}.
\end{align}
If the above approximation or ansatz was valid at nonlinear scales for all the halo mass range, it would be sufficient to study the auto power spectrum of halos in a single mass bin, which reduces the efforts and difficulty of the emulator development. 
Here we study whether the above ansatz is valid using the simulations.

\begin{figure}
\centering
\includegraphics[width=0.49\textwidth]{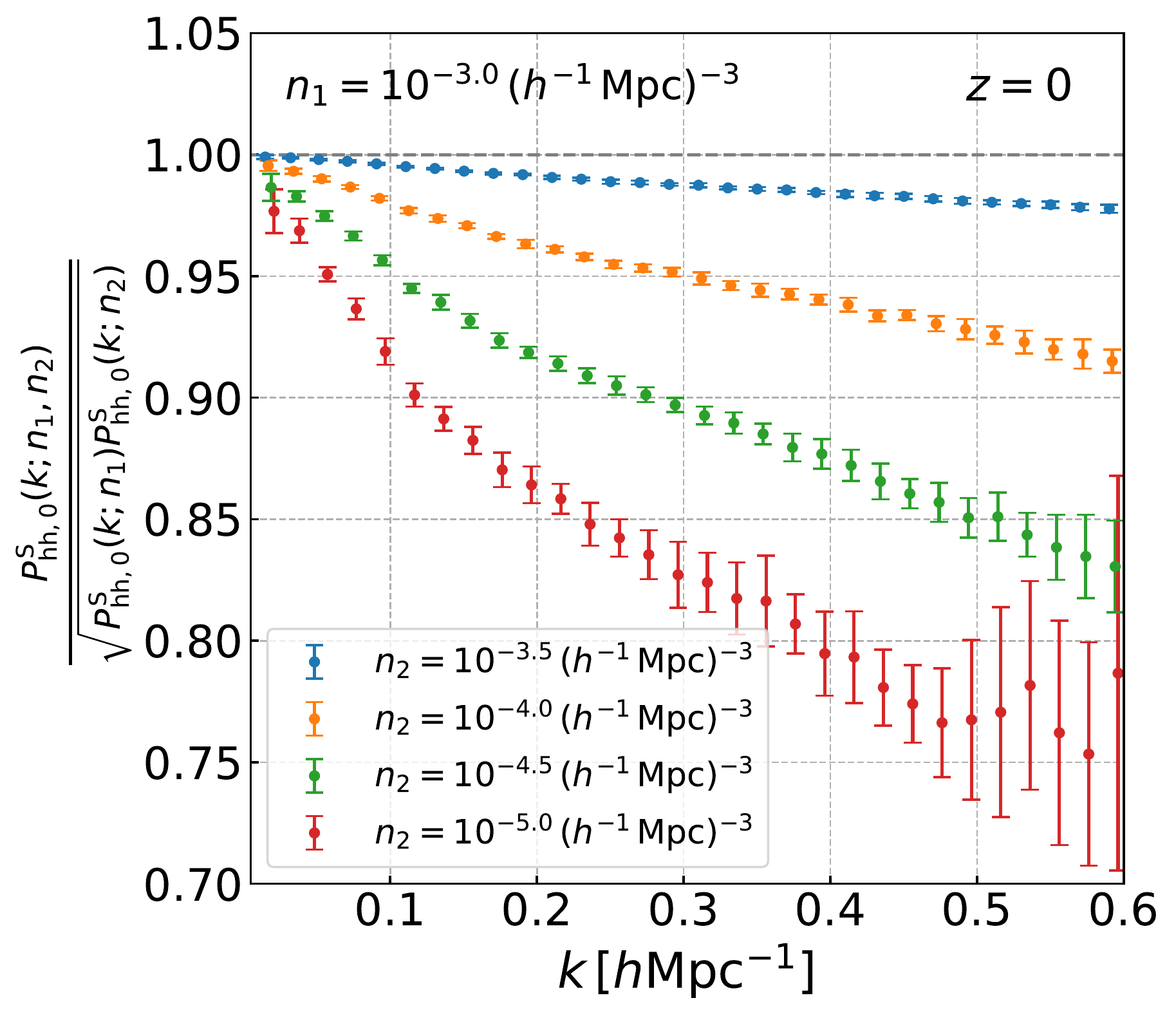}
\caption{
A test of multiplication of the redshift-space halo power spectrum; whether does the multiplication, 
$P^{\rm S}_{\rm hh}(k,\mu;n_1,n_2)=[P_{\rm hh}^{\rm S}(k,\mu; n_1,n_1)P_{\rm hh}^{\rm S}(k,\mu;n_2,n_2)]^{1/2}$, hold? 
This identify holds for the linear theory prediction with the Kaiser RSD effect. If the above identity holds, the cross-correlation coefficients in the $y$ axis should be unity.
Here we consider the case that one halo sample has a fixed number density of $n_1=10^{-3}\,(h^{-1}\,{\rm Mpc})^{-3}$, and the other sample has varying number densities, $n_2$ to be $10^{-3.5}$ (blue), $10^{-4}$ (orange), $10^{-4.5}$ (green) and $10^{-5}$ (red) $(h^{-1}\,{\rm Mpc})^{-3}$, respectively.
The error bars are the standard deviation among the 15 realizations for the {\it Planck} cosmology.
}
\label{fig:p0_corr_coeff}
\end{figure}

In Fig.~\ref{fig:p0_corr_coeff}, we investigate a validity of the ansatz, Eq.~(\ref{eq:powerM1M2_lin}). 
To this, we study the cross-correlation coefficient between the monopole moments of the redshift-space power spectrum for the halo samples of two number densities,
\begin{align}
  \frac{P^\mathrm{S}_{\mathrm{hh},0}(k; n_1,n_2)}{\sqrt{P^\mathrm{S}_{\mathrm{hh},0}(k; n_1) P^\mathrm{S}_{\mathrm{hh},0}(k; n_2)}}
\end{align}
for the {\it Planck} cosmology at $z = 0$.
We consider the cases of $n_2 = 10^{-3.5}, 10^{-4}, 10^{-4.5}$, and $10^{-5}\,(h^{-1}\,{\rm Mpc})^{-3}$, while keeping $n_1$ fixed to $10^{-3}\,(h^{-1}\,{\rm Mpc})^{-3}$. 
Note that the halo sample of $n_2$ is a subsample of the sample of $n_1$, and 
we subtracted the shot noise from each power spectrum in the numerator and the denominator.
The figure clearly shows that the ansatz, Eq.~(\ref{eq:powerM1M2_lin}), does not hold at nonlinear scales.
As an overlap between the two samples decreases (the differences between $n_1$ and $n_2$ get larger), a deviation of the cross-correlation coefficient from unity becomes greater and starts from smaller $k$ bins. 
With the results in this figure, we conclude that it is indispensable to use the halo power spectrum of two number density bins for the emulator construction.

\section{AN OPTIMAL CHOICE OF THE NUMBER OF HIDDEN UNITS IN THE NEURAL NETWORK TRAINING}
\label{sec:architecture}

In our network architecture, we employed two hidden layers to give a large flexibility to the nonlinear mapping from the nine-dimensional input vector to the output power spectrum.
The main factor which handles the model flexibility of neural network is the number of hidden units (hereafter, we call it as $N_{\rm hidden}$).
We executed the following study to make an appropriate choice of $N_{\rm hidden}$ in our neural network.

\begin{figure}
\centering
\includegraphics[width=0.49\textwidth]{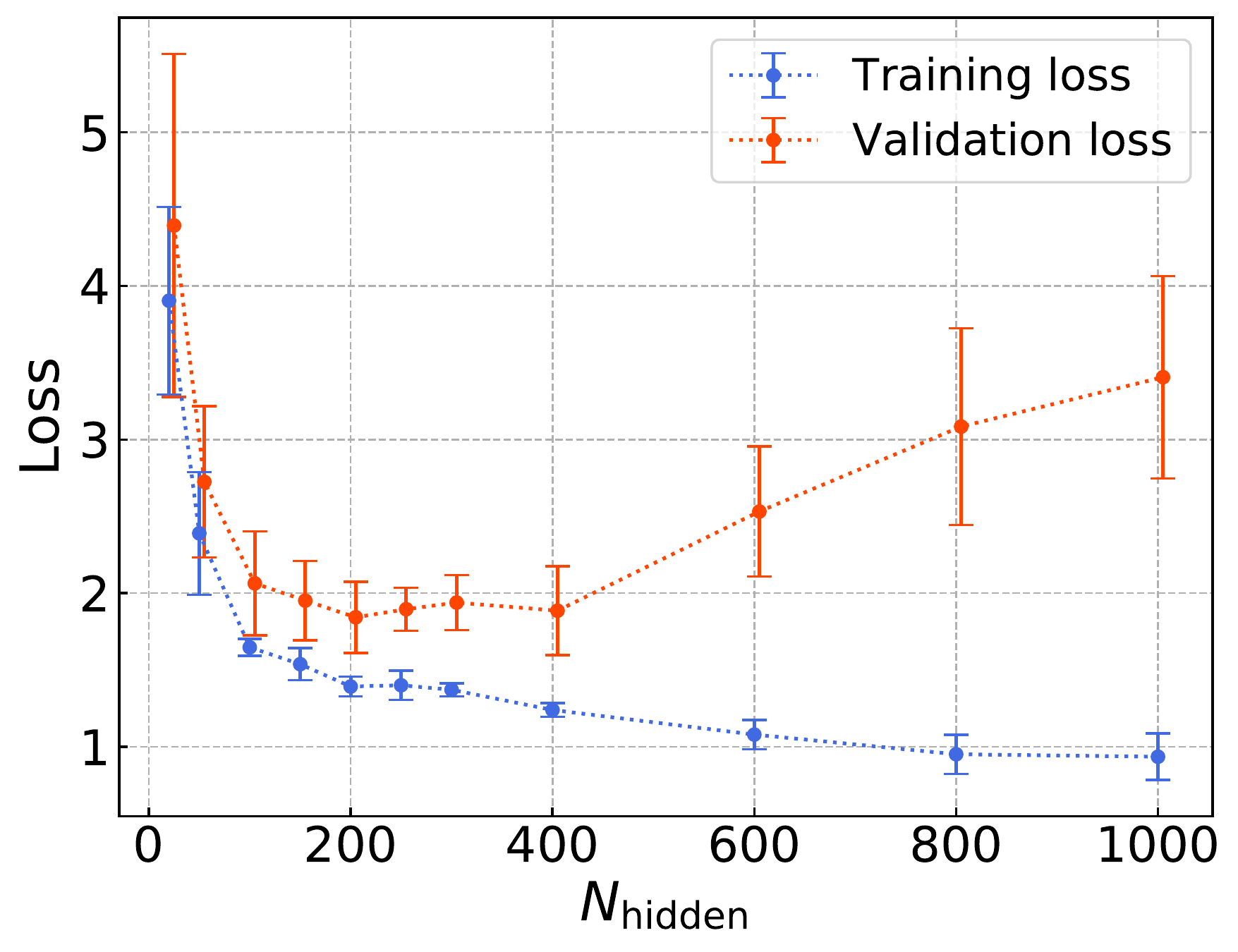}
\caption{
Shown is how the value of the loss function [Eq.~(\ref{eq:loss_func})] after the training for 1000 epochs varies with the number of hidden units, $N_{\rm hidden}$.
Blue and red symbols with error bars are the training and validation loss, respectively.
The error bars are the standard deviation among five different choices of the training/validation split (see text).
}
\label{fig:loss_func}
\end{figure}

Figure~\ref{fig:loss_func} shows how the loss function values after the training vary with $N_{\rm hidden}$.
We employ the equal $N_{\rm hidden}$ for both the two hidden layers, and run the training for 1000 epochs as we described in Sec.~\ref{subsec:training}, for the different $N_{\rm hidden}$ in the range of [20,1000].
In addition, to measure the goodness of choice of $N_{\rm hidden}$ in our network including its possible uncertainty due to the variations of dataset, we change the split of training/validation datasets; since we have five slices (slice 1--5) in the \textsc{Dark Quest} simulation suite, we can consider five different choices of the training/validation split, by choosing one of them as the validation set and remaining four slices as the training set.
In this figure, we show the mean and standard deviation of the final loss function values among the five choices of the training/validation split, for each of the training (blue) and validation (red) losses.
Note that, when we calculated the final training or validation loss, we followed the definition of Eq.~(\ref{eq:loss_func}), except that we averaged over the whole training or validation dataset, respectively.

The training loss decreases almost monotonically with the increase of $N_{\rm hidden}$, because the enhanced flexibility of the neural network enables better fittings to the training dataset.
However, this is not the case in the validation loss.
When $N_{\rm hidden}$ is low, the validation loss decreases with the increase of $N_{\rm hidden}$, similarly to the training loss.
However, as we increase $N_{\rm hidden}$ more than about 400, the validation loss also increases, which leads to a worse emulation performance.
It is due to that the neural network has too large flexibility to properly generalize to the validation data.
Since our goal is to construct an emulator that can predict not only the training dataset but also the power spectrum for new inputs, we need to suppress the validation loss and avoid such an overfitting.
The validation loss becomes the lowest when $N_{\rm hidden} \sim 200$. 
We use this value as an optimal choice of $N_{\rm hidden}$ for our emulator construction.

\section{DEPENDENCE OF THE EMULATOR ACCURACY ON THE HALO NUMBER DENSITY}
\label{sec:accuracy_number_density}

\begin{figure*}
\centering
\includegraphics[width=0.98\textwidth]{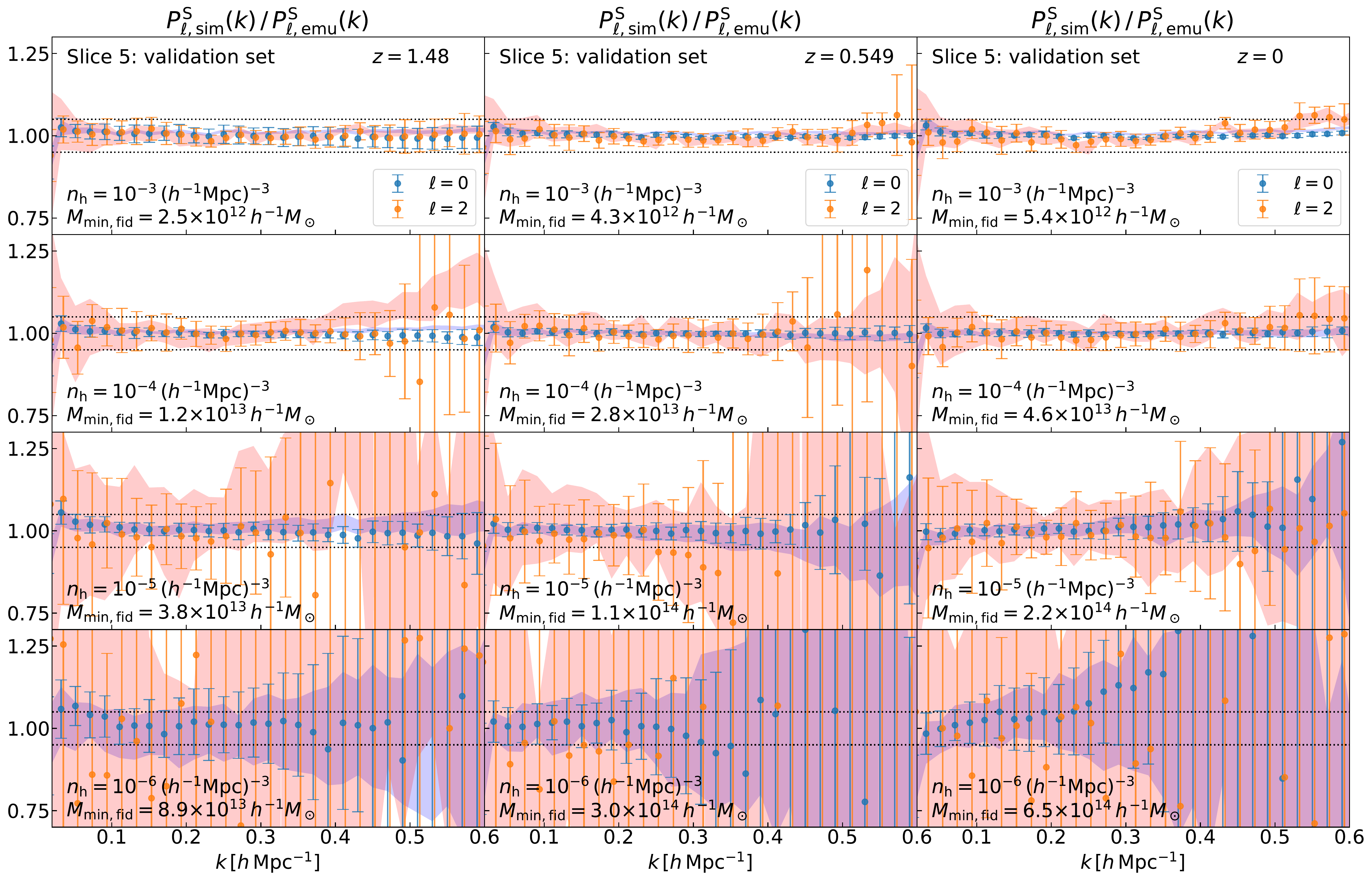}
\caption{
The accuracy of the emulator predictions for different values of halo number density.
We focus on the cases of $n_1 = n_2 = n_{\rm h}$ presented in each subplots.
The second row is identical to the lower panels in Fig.~\ref{fig:p0p2_emu_validation}.
}
\label{fig:p0p2_emu_validation_nd3-6}
\end{figure*}

\begin{figure*}
\centering
\includegraphics[width=0.98\textwidth]{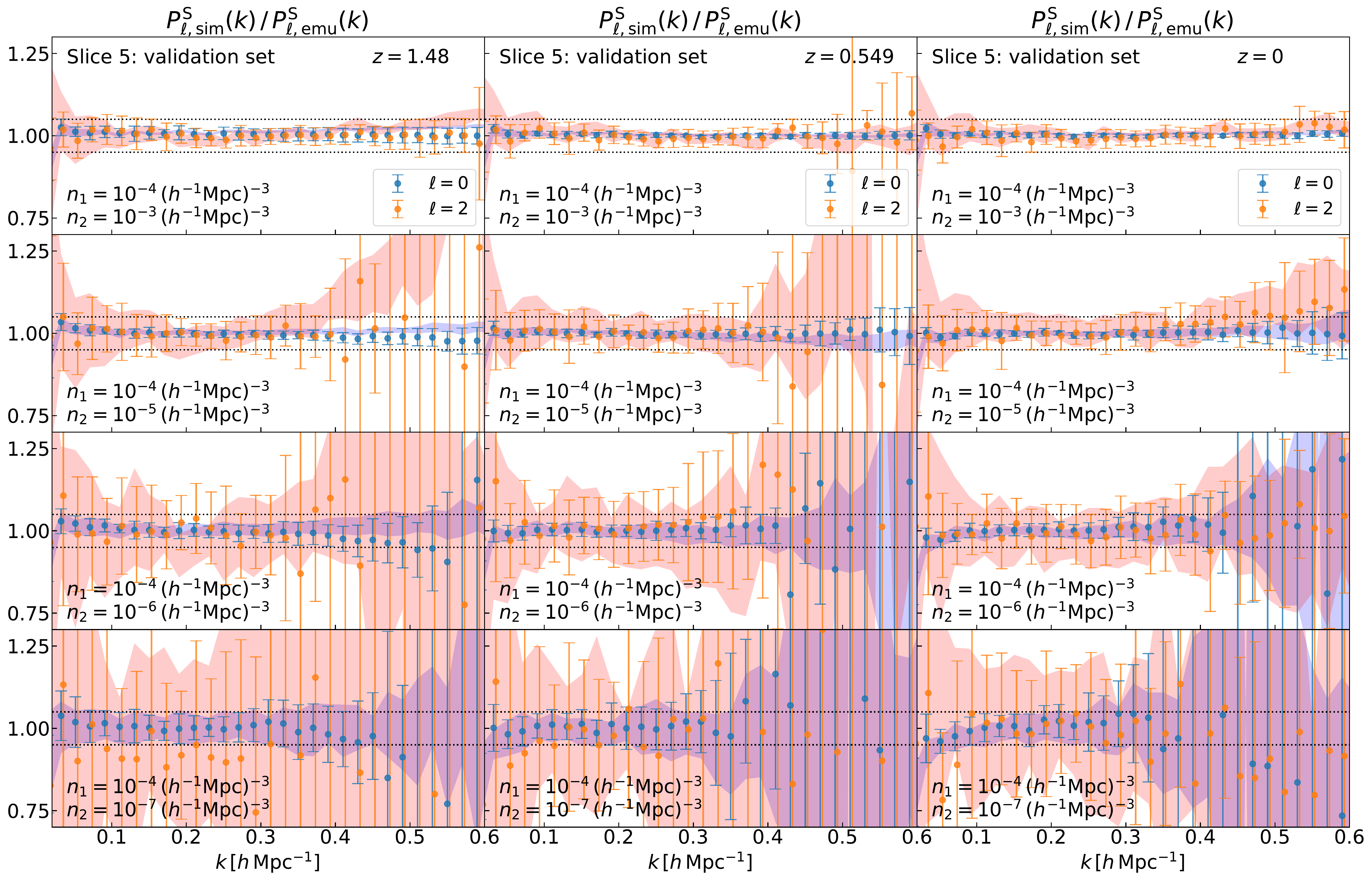}
\caption{
The same as Fig.~\ref{fig:p0p2_emu_validation_nd3-6}, but we show the cases of $n_1 \ne n_2$.
}
\label{fig:p0p2_emu_validation_hetero}
\end{figure*}

In this appendix, we study the accuracy/performance of the emulator predictions for other halo samples which we did not consider 
in the main text.
Figures~\ref{fig:p0p2_emu_validation_nd3-6} and \ref{fig:p0p2_emu_validation_hetero} show how the emulator accuracies for the monopole and quadrupole moments with the different halo samples.
In Fig.~\ref{fig:p0p2_emu_validation_nd3-6} we show the results for the power spectrum of the single number density bin, $n_{\rm h}$, where $n_{\rm h} = 10^{-3}, 10^{-4}, 10^{-5},$ and $10^{-6}\,(h^{-1}\,{\rm Mpc})^{-3}$, respectively.
Due to the severe shot noise, the accuracy for the low number density such as $n_{\rm h} = 10^{-6}\,(h^{-1}\,{\rm Mpc})^{-3}$ is much worse than that for the higher number density sample.
However, for each value of $n_{\rm h}$, the discrepancies are roughly comparable to the variance estimated from 15 realizations for the fiducial {\it Planck} cosmology, indicated by shaded regions.
This indicates that the training of the neural network reaches the limit determined from the noise levels of the training data.
The same tendency is also presented in Fig.~\ref{fig:p0p2_emu_validation_hetero}, which shows the cases in which $n_1$ and $n_2$ are different.

\section{DETAILS OF THE DEFAULT HOD MODEL}
\label{sec:hod}

In Sec.~\ref{subsec:halo_model_formalism}, we put an overall picture to implement the redshift-space galaxy power spectrum based on the halo power spectrum emulator. 
In this appendix, we provide a detailed description on the default model ingredients we use in this paper.
All the demonstrations we showed in this paper are based on this specific implementations.

We used the HOD model described by Ref.~\cite{2005ApJ...633..791Z}, which divides the galaxies into the central and satellite galaxies and treats the number distributions of them separately.
For central galaxies, we assume that the occupation number (either of 0 or 1) in a halo with mass $M$ follows a Bernoulli distribution with mean given as
\begin{align}
\label{eq:Nc}
  \avrg{N_{\rm c}} (M) = \frac{1}{2} \left[1 + {\rm erf} \left( \frac{\log_{10} M - \log_{10} M_{\rm min}}{\sigma_{\log_{10} M}} \right) \right],
\end{align}
where ${\rm erf}(x)$ denotes the error function. In other words, each halo with mass $M$ will have a central galaxy with the probability $\avrg{N_{\rm c}} (M)$. 
On the other hand, for satellite galaxies, we assume that only halos which host a central galaxies can have satellite galaxies.
For halos with a central galaxy, we assume that the number of satellite galaxies in a halo with mass $M$ follows a Poisson distribution with mean
\begin{align}
\label{eq:lambda_s}
  \lambda_\mathrm{s} (M) = \left[\frac{M- \kappa M_{\rm min}}{M_1} \right]^\alpha,
\end{align}
and therefore the mean number of satellite galaxies is
\begin{align}
\label{eq:Ns}
  \avrg{N_\mathrm{s}} (M) = \avrg{N_{\rm c}} (M) \lambda_\mathrm{s} (M).
\end{align}
These mean HOD functions are characterized by five parameters $\{ M_{\rm min}, \sigma_{\log_{10} M}, M_1, \alpha, \kappa \}$.

We also need to model the (redshift-space) position distribution of satellite galaxies inside a halo.
We assume, for simplicity, that satellite galaxies follow the spatial distribution of matter in the host halo. 
The number density profile of satellite galaxies is given as
\begin{align}
  {\cal H}(\bx; M) = \frac{\rho(\bx; M)}{M},
\end{align}
where $\rho(\bx; M)$ is the mass density profile for halo of mass $M$. 
This profile satisfies the normalization condition, $\int\!{\rm d}^3\bx~{\cal H}(\bx; M) = 1$. 
Since we can assume a spherically symmetric radial profile in the statistical average sense, 
this normalization condition reduces to $\int_0^{r_{200}}\,4\pi r^2 {\rm d}r ~{\cal H}(r; M) = 1$.
In this paper, we adopted the normalized NFW profile \cite{NFW} for the galaxy radial profile.
The NFW profile is specified by the concentration parameter $c$ in addition to the mass $M$.
For this we employed the median concentration-mass relation $c(M_{\rm 200})$ calibrated in Refs.~\cite{2015ApJ...799..108D,Diemer_2019}, through the publicly available Python toolkit \textsc{Colossus} (\url{http://www.benediktdiemer.com/code/colossus/}) \cite{Diemer_2018}.

We further take into account the RSD effect due to internal virial motions of satellite galaxies inside host halos, i.e. the FoG effect
\cite{jackson72} (also see \cite{2001MNRAS.325.1359S,2001MNRAS.321....1W} for the halo model approach of the FoG effect). 
To do this, we need to model the velocity distribution of satellite galaxies with respect to the halo center.
In this paper we employ an isotropic Gaussian distribution for the velocity distribution for simplicity, 
\begin{align}
  {\cal F}(\Delta r_\parallel; \sigma_{{\rm vir},M}) = \frac{1}{\sqrt{2\pi} 
  \frac{\sigma_{{\rm vir},M}}{aH} } \exp \left[ - \frac{(\Delta r_\parallel)^2}{2\frac{\sigma_{{\rm vir},M}^2}{a^2H^2} } \right], 
  \label{eq:Fvel}
\end{align}
where $\sigma_{\rm vir}(M)$ is the velocity dispersion for halos of $M$. 
We assume that the velocity dispersion is specified by the host halo mass as
\begin{align}
\label{eq:virial_variance}
  \sigma_{{\rm vir},M}^2 = \frac{G M}{2R_{\rm phy}},
\end{align}
where $R_{\rm phy}$ is the {\it physical} halo radius (i.e. $R_{\rm phy} = a R_{\rm 200}$). 
The distribution ${\cal F}$ denotes the distribution of the line-of-sight component of velocity, and we expressed the velocity function in terms of the positional displacement by the RSD effect due to the line-of-sight velocity component: $\Delta r_\parallel\equiv v_\parallel/(aH)$. 
The velocity function [Eq.~(\ref{eq:Fvel})] satisfies the normalization condition $\int_{-\infty}^{\infty}\!\mathrm{d}(\Delta r_\parallel)~{\cal F}(\Delta r_\parallel)=1$.

The redshift-space distribution of satellite galaxies in a given host halo is stretched by the FoG effect along the line-of-sight direction and can be expressed by a convolution of the distributions of real-space spatial distribution and the velocity function of satellite galaxies
(also see \cite{hikage12a,hikage:2013kx}),
\begin{align}
\label{eq:sspace_profile}
  {\cal H}^\mathrm{S}(\bs; M, \sigma_{{\rm vir}, M}) = \int_{-\infty}^\infty {\rm d}y~{\cal H}(\bs-y \hat{{\bf n}}; M)\,{\cal F}(y; \sigma_{{\rm vir}, M}),
\end{align}
where $\hat{{\bf n}}$ denotes the unit vector along the line-of-sight direction.
Equation~(\ref{eq:sspace_profile}) reduces to a simple multiplicative form in Fourier space:
\begin{align}
\label{eq:sspace_profile_fourier}
  \tilde{{\cal H}}^\mathrm{S}(\bk; M, \sigma_{{\rm vir}, M}) = \tilde{{\cal H}}(\bk; M)\, \tilde{{\cal F}}(k_\parallel; \sigma_{{\rm vir}, M}).
\end{align}
Note that in this work we consider only the specific model based on the NFW profile and the Gaussian velocity distribution, but actually Eq.~(\ref{eq:sspace_profile_fourier}) can be constructed from any other models of the real-space position distribution $\tilde{{\cal H}}(\bk; M)$ and displacement (velocity) distribution $\tilde{{\cal F}}(k_\parallel; \sigma_{{\rm vir}, M})$.

Finally, we showed the emulator prediction including the off-centering effect in Sec.~\ref{subsec:galaxy_power_spectrum}.
To model the off-centering effect, we introduced two ingredients: one is the off-centering probability $p_{\rm off}$ and the other is a radial profile to model the spatial distribution of off-centered galaxies in the host halo.
We assumed that each central galaxy is off-centered by the probability $p_{\rm off}$, where $p_{\rm off}$ does not depend on any other properties such as halo mass $M$.
For the spatial distribution, we assume a Gaussian radial profile for simplicity (similar to Ref.~\cite{hikage12a}),
\begin{align}
\label{eq:off-center_profile}
  {\cal P}(r_{\rm off}) {\rm d}^3 {\bf r}_{\rm off} = \frac{1}{(2\pi)^{3/2} R_{\rm off}^3 } \exp \left[ - \frac{r_{\rm off}^2}{2 R_{\rm off}^2} \right] {\rm d}^3 {\bf r}_{\rm off},
\end{align}
where $R_{\rm off}$ is a parameter that represents the typical off-centering displacement.
Using these two ingredients, the off-centered power spectrum is obtained by replacing $\avrg{N_{\rm c}}\!(M)$ in Eqs.~(\ref{eq:PS1h}) and (\ref{eq:PS2h}) with
\begin{align}
\label{eq:off-centering}
\left[(1-p_{\rm off})+p_{\rm off}\exp\left\{-\frac{1}{2}{k^2 R_{\rm off}^2}\right\} {\cal F}(k_\parallel; \sigma_{{\rm vir},M})\right] \avrg{N_{\rm c}}\!(M),
\end{align}
which is identical to Eq.~(\ref{eq:off-center}) in the main text, where $\exp\left\{-\frac{1}{2}{k^2 R_{\rm off}^2} \right\}$ is obtained as the Fourier transform of the off-centering displacement profile Eq.~(\ref{eq:off-center_profile}).

Note that in Eq.~(\ref{eq:off-centering}) we implicitly assume that the off-centered galaxies will have the random velocity relative to the halo center specified by the velocity variance [Eq.~(\ref{eq:virial_variance})].

\bibliography{lssref}

\end{document}